\documentclass[structabstract]{aa}
\usepackage{graphics}

\usepackage{txfonts}

\newlength{\dlugskr}

\def\figref#1{Fig.\,\ref{#1}}

\def\etal{{et\,al}}

\def\Msun{M$_{\sun}$}

\def\WRPN{[WR]\,PN}
\def\WRPNe{[WR]\,PNe}

\def\WRCSs{[WR]\,CSs}

\def\WELPN{WEL\,PN}
\def\WELPNe{WEL\,PNe}

\def\WELCSs{WEL\,CSs}

\def\VLPN{VL\,PN}
\def\VLPNe{VL\,PNe}

\def\oppo{O$^{++}$/O$^{+}$+O$^{++}$}
\def\hepphe{He$^{++}$/He$^{+}$+He$^{++}$}

\newcommand{\Getal}{G\'orny et~al. (2004)}
\newcommand{\Getalnb}{G\'orny et~al. 2004}
\newcommand{\bb}{{\it b\ }}
\newcommand{\dd}{{\it d\ }}

\newcommand{\kms}{~km~s$^{-1}$}

\newcommand{\msun}{\ifmmode M_{\odot} \else M$_{\odot}$\fi}
\newcommand{\rsun}{\ifmmode R_{\odot} \else R$_{\odot}$\fi}
\newcommand{\lsun}{\ifmmode L_{\odot} \else L$_{\odot}$\fi}
\newcommand{\zsun}{\ifmmode Z_{\odot} \else Z$_{\odot}$\fi}

\newcommand{\Te}{$T_{e}$}
\newcommand{\Ne}{$n_{e}$}

\newcommand{\Hb}{\ifmmode {\rm H}\beta \else H$\beta$\fi}

\newcommand{\hei}{He~{\sc i}}

\newcommand{\nii}{[N~{\sc ii}]}

\newcommand{\Oii}{[O~{\sc ii}] $\lambda$3727}
\newcommand{\oii}{[O~{\sc ii}]}
\newcommand{\Oiii}{[O~{\sc iii}] $\lambda$5007}
\newcommand{\Oiiit}{[O~{\sc iii}] $\lambda$4363}
\newcommand{\oiii}{[O~{\sc iii}]}

\newcommand{\rOiii}{[O~{\sc iii}] $\lambda$4363/5007}
\newcommand{\rNii}{[N~{\sc ii}] $\lambda$5755/6584}

\newcommand{\rSii}{[S~{\sc ii}] $\lambda$6731/6716}

\newcommand{\Np}{N$^{+}$}
\newcommand{\Npp}{N$^{++}$}

\newcommand{\Op}{O$^{+}$}
\newcommand{\Opp}{O$^{++}$}

\begin{document}

\title{Planetary nebulae in the direction of the Galactic bulge:\\
        On nebulae with emission-line central stars
\thanks{
 Based on observations made at the Cerro Tololo Interamerican 
 Observatory and the European Southern Observatory}
\fnmsep
\thanks{
 Tables 2 and 3 and Figures 21 and 22
 are only available in electronic form at http://www.aanda.org}
       }
\subtitle{}

\titlerunning{Planetary nebulae in Galactic bulge}

\author{G\'{o}rny, S. K.
       \inst{1}
\and Chiappini, C.
       \inst{2,3}
\and Stasi\'nska, G.
       \inst{4}
\and
       Cuisinier, F.
       \inst{5}
}

\offprints{S. K. G\'{o}rny}

\institute{Copernicus Astronomical Center, Rabia\'nska 8,
       PL-87-100 Toru\'n, Poland \\
       \email{skg@ncac.torun.pl}
\and
       Geneva Observatory,
       Geneva University,
       51 Chemin des Mailletes,
       CH-1290 Sauverny, Switzerland\\
       \email{Cristina.Chiappini@unige.ch}
\and
       Osservatorio Astronomico di Trieste,
       OAT-INAF,
       Via Tiepolo 11,
       Trieste, Italy
\and
       LUTH, Observatoire de Paris, CNRS, Universit\'e Paris Diderot;
       Place Jules Janssen 92190 Meudon, France\\
\email{grazyna.stasinska@obspm.fr}
\and
       GEMAC, Observat\'orio do Valongo/UFRJ,
       Ladeira do Pedro Ant{\^{o}}nio 43, 20.080-090 Rio de Janeiro, Brazil \\
       \email{francois@ov.ufrj.br}
}

\date{Received ???; accepted ???}

\abstract 
{} 
{We present a homogeneous set of spectroscopic measurements secured with
4-meter class telescopes for a sample of 90 planetary nebulae (PNe) located
in the direction of the Galactic bulge.} 
{We derive their plasma parameters and chemical abundances. For half of the
objects this is done for the first time. We discuss the accuracy of
these data and compare it with other recently published samples. We 
analyze various properties of PNe with emission-line central stars in the 
Galactic bulge. }
{Investigating the spectra we found that 7 of those PNe are ionized by
Wolf-Rayet ([WR]) type stars of the very late (VL) spectral class [WC\,11]
and 8 by weak emission-line (WEL) stars. From the analysis we conclude that
the PN central stars of WEL, VL and remaining [WR] types form three,
evolutionary unconnected forms of enhanced mass-loss among central stars of
PNe. [WR]~PNe seem to be intrinsically brighter than other PNe. Overall,
we find no statistically significant evidence that the chemical
composition of PNe with emission-line central stars is different from that of
the remaining Galactic bulge PNe.}
{}
\maketitle

\section{Introduction}

Planetary nebulae (PNe) are a short evolutionary phase in the life of low
and intermediate mass stars occurring after they leave the Asymptotic Giant
Branch (AGB). This basic fact was established  decades ago
(Shklovsky 1956, Paczy\'nski 1971) yet the details of processes leading to
the creation of the nebula and its subsequent evolution remain unclear. It
is not known for example if the central stars (CSs) of PNe are predominantly
hydrogen or helium burning. In fact, only for a small subclass of them, with
spectra similar to massive Wolf-Rayet stars, do we know what is powering
their evolution: since such stars are practically hydrogen free, they must
be burning helium. For the remaining CSs it is believed that
they are burning hydrogen in a shell and in fact most of the available
evolutionary models assume hydrogen as a fuel of the nuclear reactions at
these stages of evolution (see G\'orny~\& Tylenda 2000).

The atmospheres of Wolf-Rayet type CSs are peculiar and very different from
most CSs. Their outermost layers are mostly composed of helium, carbon and
oxygen (see review by Werner \& Herwig 2006). They are also characterized by
enhanced mass-loss which triggers the occurrence of prominent stellar
emission bands of C, O and He. The origin of their name comes from the
appearance of their stellar spectra as they
closely resemble those of genuine massive Wolf-Rayet population I stars of
the WC spectral subclass. The similarity is so close that the same
classification scheme can be used (Crowther et~al. 1998, Acker \& Neiner
2003)\footnote{
  In this work we will continue to use  the older classification scheme with
  spectral classes ranging from coolest [WC\,11] to hottest [WC 2] stars
  because they are easier to apply to the central stars with lower quality
  spectra - see \Getal\ for more detailed justification.}. 
To mark the difference with population I WR stars, a notation between 
brackets is used: [WR] or [WC].

The general question of the origin of Wolf-Rayet type central stars
at the center of some PNe is one of the open problems in the PN
field. But \WRPNe\footnote{
 In the rest of the paper, we use the notation \WRPNe\ 
 for PNe with [WR] central stars and \WRCSs\ referring to such stars.}
have other secrets of their own. For example, it is not understood
why in the Milky Way there are so few CSs with intermediate-class [WC]
spectral types. Those that are known are located predominantly in the bulge
of our Galaxy while there is clearly an underpopulation of such objects in
the disk (G\'orny 2001). The picture of Galactic bulge/disk dichotomy of
\WRPNe\ seemed to be disturbed when \Getal\ reported on a large number of
PNe central stars classified as [WC\,11] in the bulge. Before that,
\WRPNe\ of this type were known only in the Galactic disk. In addition,
the number of PNe with "weak emission-line" (WEL) central stars (a class
introduced by Tylenda et~al. 1993) found by \Getal\ in the
bulge was also surprisingly high.

In this paper we present new spectra of 90 PNe in the Galactic
bulge direction. The goal of our observational program was to increase the
sample of bulge PNe with homeogenous abundance determinations and to
discover new emission-line CSs. Combined with high quality literature data
we have gathered spectroscopic data of 245 PNe, most of them with a
high probability of physically belonging to the bulge of the Milky Way. The
remaining PNe most probably belong to the inner disk of our Galaxy. In this
work, we use the spectroscopic information combined with other data to
discuss the evolutionary status of Galactic bulge PNe with different types
of emission-line CSs as compared to PNe with normal nuclei.

This paper is organized as follows. In Section 2 we present the observations
and describe the reduction procedures. In Section 3 the data quality is
assessed. In Section 4 we describe the method applied to derive the plasma
parameters and abundances.  In Section 5 for common objects we compare our
results with those of other authors.  Section 6 presents our newly
discovered PNe with emission-line CSs including a discussion on their
classification, rate of occurrence and selection effects, both in the bulge
and inner disk samples. In Section 7 various properties of these objects are
described and compared to normal PNe. A summary of our results and our
conclusions can be found in Section 8.

\section{Observations and reduction}

We present results of spectroscopic observations of 90 PNe secured
during three runs on two different 4-meter class telescopes by C.\,Chiappini
and F.\,Cuisinier. The observations were performed in July 2001 and
April/May 2002 at the Cerro Tololo Interamerican Observatory (CTIO) using
the RC-spectrograph attached to the 4-meter telescope and in July 2002 with
the 3.6 meter telescope and the EFOSC\,II instrument of the European
Southern Observatory (ESO). The log of observations is presented in Table~1.

\begin{table*}
\caption{Log of observations.}
\begin{tabular}{
                     l
                     l @{\hspace{0.10cm}}
                     c @{\hspace{0.10cm}}
                     c @{\hspace{0.20cm}}
                     c @{\hspace{0.20cm}}
                     c @{\hspace{0.30cm}}
                     c }
\hline
   Date &
   Observatory &
   Telescope &
   Range &
   Resolution &
   Observed PNe \\
\hline
   2001, July      & CTIO & 4m + RC-spectrograph & ~~3900\--7580\AA\ & $\sim$1000 & 35 \\
   2002, April/May & CTIO & 4m + RC-spectrograph & ~~3700\--7390\AA\ & $\sim$1000 & 37 \\
   2002, July      & ESO  & 3.5m + EFOSC\,II     & ~~3600\--7350\AA\ & $\sim$2000 & 23 \\
\hline
\end{tabular}
\end{table*}

\begin{figure}
\resizebox{0.95\hsize}{!}{\includegraphics{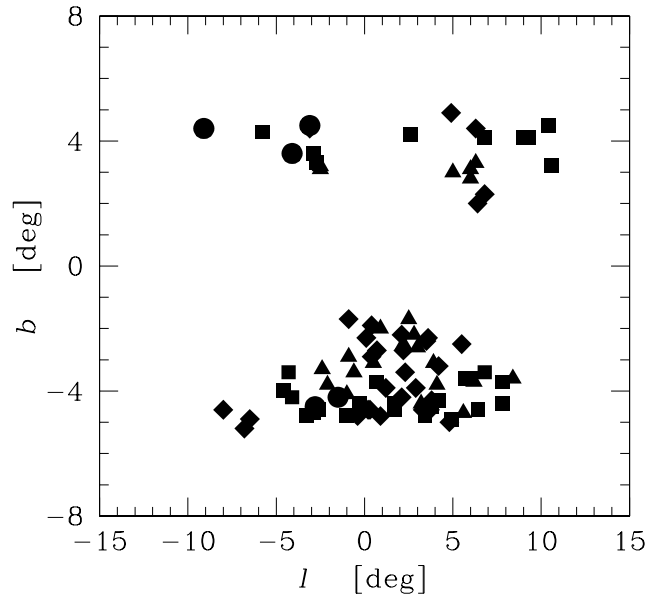}}
\caption[]{
  Distribution of the observed PNe in Galactic coordinates.  Objects observed
  during different runs are marked: CTIO 2001 -- squares, CTIO 2002 --
  diamonds, ESO 2002 -- triangles. Objects observed during both seasons at
  CTIO are marked with circles.
}
\label{l_b}
\end{figure}

In Fig.~\ref{l_b} we show the sky distribution of the observed objects in
Galactic coordinates. Practically all these PNe can be regarded as belonging
to the Galactic bulge as they satisfy the standard criteria (Stasi\'nska
\& Tylenda 1994), namely: a) they are located within 10 degrees from the
center of the Galaxy, b) have diameters smaller than 20\arcsec\ and c) known
radio fluxes at 5GHz smaller than 100\,mJy.

For the majority of PNe in our sample the presented spectroscopic
observations allowed us to derive accurate plasma parameters and chemical
abundances for the first time. For the remaining objects such data had already
been published elsewhere, e.g. 24 PNe are in common with \Getal\ and 9 with
Wang \& Liu (2007). There are also 44 objects in common with Exter \etal.
(2004) however in this case information on some crucial diagnostic lines is
often available only in our spectra.

A series of spectra with different exposures were taken for each
observed PN. The times ranged from a few seconds to 1 hour, to secure
unsaturated detections of the strongest nebular lines as well as good signal
for the weak but important features. On-sky projected slit apertures of 4
arcsec for PN observations and 10 arcsec in the case of standard star spectra
were used. Each night at least three different standard stars (Feige\,58,
Feige\,110, EG\,274, GD\,108 or LT\,6248) were observed together with
appropriate calibration lamp spectra, dome and sky flats.

The spectra obtained at CTIO have a spectral resolution R=1000 and span from
3900\AA\ to 7580\AA\ for observations taken in year 2001 and from
3700\AA\ to 7390\AA\ in 2002. The spectra obtained at ESO have a spectral
resolution of about R=2000 and span from 3600\AA\ to 7350\AA. In both cases,
due to the fact that no order blocking filters were used, the spectra of
standard stars can be contaminated in their red part with light coming from
the second order blue wing. The effect on PN spectra is much smaller
since most of the radiation is emitted in narrow lines.  In  some
PN observations, the use of broad-band filters is not desirable if one
is interested in registering blue lines, like the \Oii\ line. However
it is possible to use transmission curves as derived from
different standard stars of different effective temperatures to
disentangle the effects of second order contamination. In Appendix\,A we
describe our method to account for second order contamination during data
reduction and find that it adds only a few percent of 
uncertainty in the line fluxes.

Procedures from the long-slit spectral package of MIDAS\footnote{MIDAS is
developed and maintained by the European Southern Observatory.} were used to
reduce and calibrate the spectra. These involved bias subtraction, flatfield
correction, atmospheric extinction correction, and wavelength and flux
calibration.

The 1-dimensional spectra of standard stars were obtained in the usual manner
by summing over the appropriate rows of a sky-subtracted frame. In the case
of PNe spectra a multi-step method had to be adopted. This was due to
the fact that many lines of the observed nebulae are weak and that the
objects are located in crowded Galactic bulge fields with many nearby
contaminating stars. For this reason before extracting 1-D spectra we
removed all the underlying background sources leaving only nebular emission
lines. This involved both sky continuum emission and telluric and
interstellar lines (perpendicular to the dispersion direction) and all the
stellar continuum components (spanning along the dispersion). The
complicated system of contaminating features could not be satisfactorily
removed in just one step. A separate fitting to both directions had to be
performed starting from test removing of sky background components, followed
by removing spectra of stars and finally performing a proper sky
subtraction. In some cases this process had to be repeated in an iterative
way.

The final step was to detect the nebular lines in the frame. The procedure
we used considered that a signal can be attributed to a nebular line only if
it was stronger than 2 sigmas of the averaged background noise and at least
half of its neighboring pixels were also found above that level. In this way
spatial contours of all secure nebular lines were established while the
remaining background could be set to zero. After summing the frame in
the direction perpendicular to the dispersion, a 1-D spectrum of the PN was
obtained. The applied procedure helps to maximize the S/N of weak nebular
features as fewer noise-dominated pixels are integrated.

Intensities of the lines were measured from the 1-D spectra by
employing the REWIA package\footnote{
  J.Borkowski; {\tt www.ncac.torun.pl/cgi-bin/rewia2html}}
adopting Gaussian profiles and performing multi-Gaussian fits when necessary.

The line intensities have been corrected for extinction using an iterative
procedure, adopting the extinction law of Seaton (1979) in order to
reproduce the theoretical case B Balmer lines ratios at the electron
temperature and density derived for the object.  For the CTIO and ESO
observations in 2002 the H$\alpha$/H$\beta$ ratio was used while for the
CTIO 2001 observations only the H$\alpha$/H$\gamma$ could be used\footnote{
In the CTIO 2001 run, the H$\beta$ line was registered on hot pixels of one
of the bad CCD columns and could not be reliably measured.}. Therefore the
line intensities for the latter subsample of observations have in fact been
derived with respect to H$\alpha$ and only later recalculated to the
expected H$\beta$ intensity.

In the case of 2002 CTIO and ESO observations the basic reddening procedure
described above did not always give the theoretically expected ratios of
H$\gamma$/H$\beta$ and/or H$\delta$/H$\beta$. This is not unusual and can be
attributed to many factors like deviations from the adopted extinction law or
flux calibration problems (we have tried to evaluate the importance of such
effects, see below). It is crucial however that the intensities of the
\Oiiit\ and \Oii\ lines are reddening-corrected in the best possible way.
For this reason we have applied an additional correction procedure (similar
to the one described in G\'orny et~al. 2004) to bring the H$\gamma$/H$\beta$
and/or H$\delta$/H$\beta$ ratios to their theoretically expected values and
then used a proportional correction to all the nearby lines. The principal
refinement to the recipe described in G\'orny et~al. (2004) was that we took
due consideration of the possible random errors involved in line
measurements and applied the procedure only if the deviations were
substantial compared to expected inaccuracies. We thus avoided translating
uncertainties from individual measurements of (sometimes rather weak)
H$\delta$ and H$\gamma$ lines onto other lines.

In Table~2$^{\,}$\footnote{Table 2 is available in electronic form
only.} we present the dereddened intensities of all important lines on the 
scale of H$\beta$=100. The lines additionally corrected in the way described above
are marked with "c" in the table.

\section{Evaluation of data quality}

In the ESO 2002 observations the spectral resolving power of $\approx$2000
resulted in a substantial blending of the H$\alpha$ line at 6563\AA\ with
the \nii\ line at 6548\AA. Since the \nii\ $\lambda$6584/6548 ratio is given
by atomic physics to be 3.05, we obtained the H$\alpha$ intensity by
subtracting from the measured sum of H$\alpha$ and \nii\ $\lambda$6548
intensities the value of the \nii\ $\lambda$6584 intensity divided by 3.05. 
The cases when this approach was necessary are marked with "b" in Table~2.
The accuracy of the H$\alpha$ line intensities obtained in such a way is
estimated to be around 10\%.

The same mark "b" is placed for cases with strong blending of lines of the
\rSii\ doublet. The accuracy of the inferred ratio of these two lines can
probably be as low as 40\% in the most difficult cases marked additionally
with a colon (:b) in Table~2.

The accuracies of all the remaining line measurements are estimated to be of
roughly 5\%. This includes all direct sources of errors like photon
shot noise as well as CCD readout, bias and sky background induced
noises. The standard star measurement and interstellar/atmospheric
extinction related uncertainties further increase the possible errors. In
cases marked with a colon in Table~2 the uncertainty is around 20\% and in
the rare cases of extremely weak lines or lines contaminated with sky
features or field stars (marked with semicolon) this can be as high as 40\%.

\begin{figure}
\resizebox{.95\hsize}{!}{\includegraphics{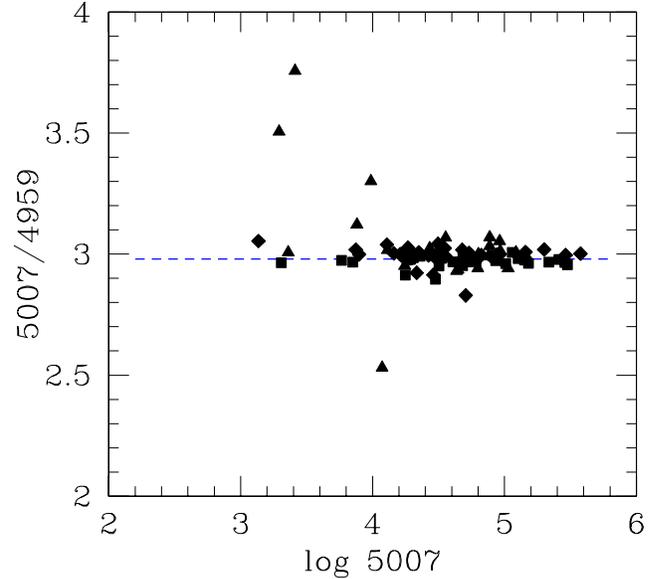}}
\caption[]{
  Intensity ratio of the [O~{\sc iii}] $\lambda$5007 and 4959 lines as the
  function of the flux in [O~{\sc iii}] $\lambda$5007 (arbitrary
  units). Objects observed 
  during
  different runs are marked: CTIO 2001 -- squares, CTIO 2002 -- diamonds, ESO
  2002 -- triangles. Dashed line represents theoretical value.
}
\label{oiii}
\end{figure}

\begin{figure}
\resizebox{.95\hsize}{!}{\includegraphics{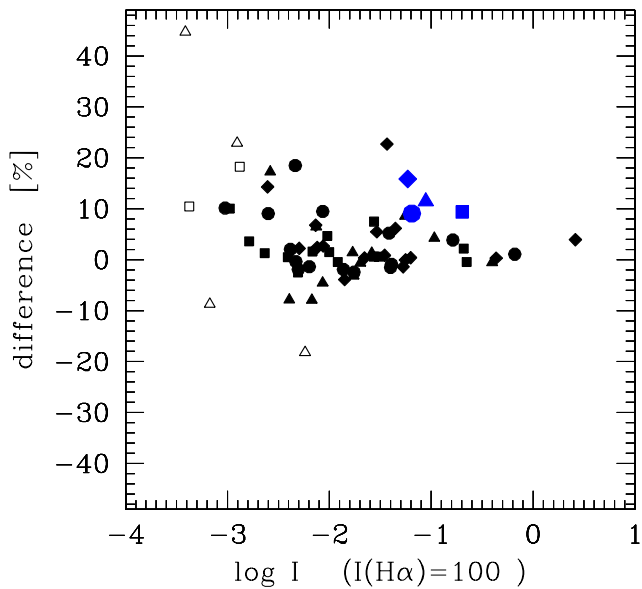}}
\caption[]{
  Relative differences in line measurements (reddening uncorrected) from two
  different spectra at CTIO in 2001 and in 2002 as a function of line
  intensity (log\,I(H$\alpha$)=2). Different symbols mark observations of
  four PNe: H\,2-1 -- triangles; H\,1-9 -- squares; M\,2-11 -- diamonds;
  H\,1-46 -- circles.  The data for \oii\ $\lambda$7325 line are marked with
  larger blue symbols. Open symbols mark line measurements judged as 
  uncertain.
}
\label{ctio2x}
\end{figure}

Because the ratios of some lines should be practically constant and are
known from atomic physics they can be used to evaluate the quality of
spectroscopic observations. One of them is the  \oiii\
$\lambda$5007/4959 line ratio which should be 2.98 (Storey \& Zeippen
2000). Note that these lines are usually so bright they easily saturate and
therefore have to be extracted from the shortest exposure frames. The
measured values for our data are plotted in \figref{oiii}.  The dispersion
of the observed ratios for the strong and intermediate brightness lines is
in agreement with the estimated 5\% error of the line intensity
measurements\footnote{ The most outlying points in \figref{oiii} are
evolved, faint nebulae: H\,1-64, H\,2-22, Pe\,2-12 and H\,2-30. The first
three are low excitation PNe without emission-line CSs. The latter PN is
high excitation object but no nebular parameters could be calculated due to
the quality of the spectra.}.

We have tried also to use  the already mentioned \nii\ $\lambda$6584/6548
ratios derived from our spectra. In this case the deviations from the
theoretical value turned out to be much larger. We have checked, however,
that the main reason for the discrepancy is due to contamination of \nii\
$\lambda$6548 line from the wing of the strong adjacent H$\alpha$ line. It
therefore turns out that the \nii\ $\lambda$6584/6548 ratio is simply not
applicable as a tracer of line measurement quality for our observations.

A final test of the accuracy of line measurements is shown in
\figref{ctio2x}. This plot presents relative differences of all lines of 4
PNe (H\,2-1, H\,1-9, M\,2-11 and H1-46) registered during CTIO observations
in 2001 and repeated in 2002. In this case the differences reflect the
combined effect of more factors since during the two nights e.g. standard star
selection could be different and the atmospheric conditions or detailed
instrument settings are certainly not identical. As can be seen in
\figref{ctio2x}, most of the differences in line measurements are again
within 5\% and only very few exceed 10\%. Many of the outlying points
correspond to lines already acknowledged to be uncertain while inspecting
the spectra and measuring the line intensities (open symbols). It is
important to note that the measurement of the \oii\ $\lambda$7325 lines seems
to be rather uncertain (shown with larger symbols) even though their
intensities are relatively high -- at the level of 1/10 of the H$\alpha$
intensity. This is due to the difficulty in establishing the instrument
response function with a high confidence at both extreme red and
blue ends of the spectral range\footnote{
   Compare e.g. the different shapes of
   response functions from individual standard stars at these wavelengths shown
   in Fig.~A2 in the Appendix.
}. An additional argument for a systematic rather than statistical character
of these errors is that they are biased in the same direction in all four
PNe presented in \figref{ctio2x}. With CTIO 2001 observations the \oii\
$\lambda$7325 intensities seem always larger than in CTIO 2002. For this reason
we have adopted 10\% as a default error for \oii\
$\lambda$7325 measurements in further calculations.

\section{Plasma parameters and chemical abundances}

We use the classical empirical method to derive the plasma parameters.
First, the electron densities are deduced from the \rSii\ ratio and electron
temperatures from the \rOiii\ and/or \rNii\ ratios. These are used to refine
the inferred reddening correction as described above. The chemical
abundances are derived with the code ABELION as in G\'orny et~al. (2004),
but the atomic data have been updated using sources listed in Stasi\'nska
(2005) as well as Tayal (2007) for \oii\ and Porter et al. (2007) for \hei.

The [N~{\sc ii}] $\lambda$5755 and  [O~{\sc ii}]
$\lambda$$\lambda$7320,7330 lines can be affected by recombination from
\Npp\ and \Opp\ ions. This was taken into account by using the
expressions given in Liu et al. (2000), the \rNii\ temperature and assuming
that \Npp/H=\Opp/H $\times$ \Np/\Op. This has a  negligible
effect on the computed abundances. The real effect could be larger if, as
suggested by Liu (2006 and references therein), the recombination lines were
actually coming from a much cooler zone. In the most extreme cases from Wang
\& Liu (2007), where the observational data allow a better correction for
the effect of recombination, the resulting abundances are modified by a few
percent at most.

Thus the recombination contribution cannot solve the known problem of a
frequent difference in O$^+$ ionic abundance derived from $\lambda$7325 as
compared to $\lambda$3727. The discrepancy between these two O$^+$ values
varies from object to object. For some PNe there is almost perfect agreement
while for other PNe the difference exceeds a factor of 2. We have noticed
however that the abundance ratios O$^+$($\lambda$3727)/O$^+$($\lambda$7325)
that we derive seem to cluster around a value characteristic for a given
observing run and PN environment (e.g. bulge or inner-disk objects). We have
therefore used the relevant median ratio to correct the
O$^+$($\lambda$7325) abundance down when necessary to obtain an O$^+$
estimation scaled to the value expected from the O$^+$($\lambda$3727) line. For
example, in the case of CTIO 2002 when the blue O$^+$ line was not observed
due to high extinction, a factor of 0.66 was used to correct the other line. The
same factor was applied to CTIO 2001 observations that all missed the \Oii\
line since it was outside the observed range. Thanks to this procedure, the
results are not biased by the selection of data sources (whether the
$\lambda$3727 could be registered in a particular observing run) or the line
being unobserved for some PNe (e.g. due to high extinction). In a last step
if there were O$^+$ estimates from both lines a final value for a given PN
was calculated by taking an average weighted by their respective
uncertainties.

The uncertainties in abundance ratios and other derived parameters were
obtained by propagating uncertainties in the observed emission line
intensities using Monte-Carlo simulations. It should be noted that our
approach does not take into account such sources of possible errors like
variations in extinction law or unknown structure of the nebulae. This
should be compensated to some degree by our conservative assumption of line
intensity errors of at least 5\% even for the strongest lines. In order to
avoid dealing with uncertain values we remove from consideration any
parameter for which the two-sigma error from the Monte-Carlo simulation is
larger than 0.3\,dex. In cases where we will include more uncertain
parameters, this will be indicated.

\begin{figure}
\resizebox{.95\hsize}{!}{\includegraphics{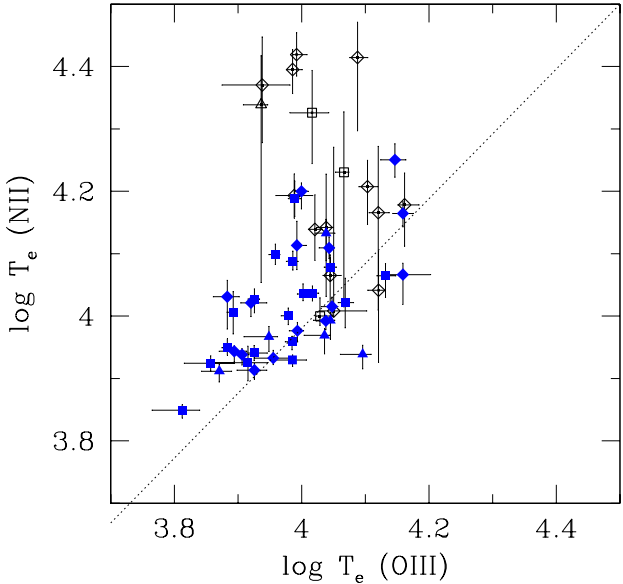}}
\caption[]{
  The electron temperature derived from \rOiii\ versus the electron
  temperature derived from \rNii\ (CTIO observations in 2001 --
  squares, CTIO 2002 -- diamonds, ESO 2002 -- triangles). Open black
  symbols mark cases with \Te(N~{\sc ii}) judged uncertain and replaced with
  \Te(O~{\sc iii}) in abundance calculations.
}
\label{to_tn}
\end{figure}

Similarly to G\'orny et al. (2004) we use \Te(N~{\sc ii}), the temperature
derived from \rNii\ for lines from ions with low ionization potential
(S$^+$, N$^+$, O$^+$) and \Te(O~{\sc iii}), the temperature derived from
\rOiii\ for hydrogen and lines from other ions with intermediate and high
ionization potentials. If neither of these temperatures is available (only 6
cases) we refrain from abundance determinations. In \figref{to_tn}\ we plot
\Te(O~{\sc iii}) versus \Te(N~{\sc ii}) for our PN data.  The observed
behavior of these temperatures is quite similar to the one shown in Fig.~2
of \Getal. The only difference is that in the new sample there are no PNe
with measured values of \Te(O~{\sc iii}) larger than 16000K. As in \Getal\
in about 25\% of cases \Te(N~{\sc ii}) was very uncertain and we used the
\Te(O~{\sc iii}) for all the ions.

After computing ionic abundances, the elemental abundances are obtained
using the ionization correction factors (ICFs) from Kingsburgh \& Barlow
(1994). In the present work we derive additionally the abundance of
chlorine, adopting the ICF from Liu \etal. (2000). 
Table~3$^{\,}$\footnote{Table 3 is available in electronic form only.} lists
the plasma diagnostics and ionic and elemental abundances ordered by PN\,G
numbers.  In this table, there are three rows for each object, and a fourth
row used to separate them. The first row gives the values of parameters
computed from the nominal values of the observational data. The second and
third row give the upper and lower limits respectively of these parameters.
Column (1) of Table~3 gives the PNG number; Column (2) gives the usual name
of the PN; Column (3) gives the electron density deduced from \rSii, columns
(4) and (5) give the electron temperature deduced from \rOiii\ and \rNii\
respectively (the value of \Te(N~{\sc ii}) is in parenthesis if \Te(O ~{\sc
iii}) was chosen for all ions). Column (6) gives the He/H ratio, columns
(7) to (12) the N/H, O/H, Ne/H, S/H, Ar/H, Cl/H ratios, respectively. Column
(13) gives the logarithmic extinction C at \Hb\ derived from the spectra.

\section{Comparison with other samples}

We now compare the quality of our results with previously published data.
For this purpose we derived the plasma parameters and abundances using
exactly the same procedures, assumptions and atomic data for the objects of
the sample introduced here and taking the data of PNe originally observed by
G\'orny et~al. (2004), objects observed by Escudero \& Costa (2001) and
Escudero \etal. (2004) and also those of Wang \& Liu (2007). In the latter
case we used only the optical data to be consistent with what was available
from our own observations.  Finally, the same method was applied to objects
from Exter et al. (2004).

\addtocounter{table}{2}
\begin{table*} 
\begin{minipage}[t]{1.65\columnwidth}
\caption{
 Differences in dex on plasma parameters and
 chemical abundances adopting data from this work
 versus data taken from the literature. The number of
 analyzed PNe in common is given in parenthesis.}
\centering
\renewcommand{\footnoterule}{}
\begin{tabular}{ l @{\hspace{0.70cm}}
                 c @{\hspace{0.55cm}}
                 c @{\hspace{0.55cm}}
                 c @{\hspace{0.55cm}}
                 c @{\hspace{0.65cm}}
                 c @{\hspace{0.65cm}}
                 c @{\hspace{0.65cm}}
                 c }
\hline
    sample\footnote{Sample {\it G} includes objects in common with
           original PNe observations of G\'orny et al. (2004);
           sample {\it ECM} -- Escudero \& Costa (2001) or Escudero et al. (2004);
           sample {\it WL} -- Wang \& Liu (2007)
           and {\it EBW} -- Exter et al. (2004).} &
    $\Delta$ log\,\Te(N~{\sc ii}) &
    $\Delta$ log\,\Te(O~{\sc iii}) &
    $\Delta$ log\,\Ne(S~{\sc ii}) &
    $\Delta$ log\,O/H &
    $\Delta$ log\,N/O &
    $\Delta$ log\,S/Ar \\
\hline
    {\it G}   & 0.01 ~~(~8) & 0.02 ~~(~8) & 0.13 ~~(~9) & 0.04 ~~(~9) & 0.32 ~~(~9) & 0.20 ~~(~5) \\
    {\it ECM}  & 0.02 ~~(~9) & 0.02 ~~(~9) & 0.29 ~~(10) & 0.10 ~~(12) & 0.17 ~~(12) & 0.11 ~~(~7) \\
    {\it WL}   & 0.02 ~~(~8) & 0.01 ~~(~8) & 0.06 ~~(~8) & 0.04 ~~(~8) & 0.18 ~~(~8) & 0.07 ~~(~4) \\
    {\it EBW}  & 0.08 ~~(10) & 0.04 ~~(10) & 0.13 ~~(22) & 0.19 ~~(20) & 0.27 ~~(20) & 0.15 ~~(10) \\
\hline 
\end{tabular}
\end{minipage}
\end{table*}

First we analyse the electronic temperatures. Columns 2 and 3 of Table~4
present the median of the differences of derived temperatures for PNe in
common between this work and other samples. As it can be seen, the agreement
between the different authors is good, typically well below 0.1 dex. The
differences are therefore of the same order as individual errors derived
with the Monte-Carlo method, suggesting that our error estimates were
reasonable. Larger differences are found with respect to the PNe in common
with Exter et al. (2004). Inspecting the individual cases of PNe
with \Te(N~{\sc ii}) much larger than \Te(O~{\sc iii}) we have noticed that
usually this property is confirmed by data of different observers.

The derived electron densities from \rSii\ ratio for objects in common
between different samples are compared in column 4 of Table~4. As can be
seen the agreement is also usually good. However, for objects with weaker
lines we have checked that the deviations are larger, especially with
respect to the Exter \etal. (2004) sample.

The typical differences in O/H between the computations using our data and
literature data are within 0.1 dex (column 5 of Table~4). Only comparing
with data from Exter \etal. (2004) did we find a typical difference of almost
0.2 dex. What is more important however, is that we  noticed that using
measurements from Exter \etal. (2004) would make the observed O/H
distribution of bulge PNe much broader simply due to the lower quality of
that data. In particular, analyzing PNe in common with our list of targets
one finds in Exter \etal. (2004) three cases with log\,O/H$>$9 -- a value
that otherwise seems a strong upper cut off for derived oxygen abundances of PNe
in the Galactic bulge direction (compare with Fig.\,11 of G\'orny et~al.2004
and a discussion on possible flattening of the O/H gradient or see
\figref{hist_oh} below).

In columns 6 and 7 of Table~4 we present differences in derived N/O and
S/Ar. In the latter cases the uncertainties can be larger, particularly when
the lines are extremely weak. In addition, the S/Ar ratios of Exter \etal.
(2004) are systematically lower than what we find.

From the results presented in this Section we can conclude that the quality
of our data is comparable to that found in the best recent literature
sources. On the other hand, the comparison with Exter et al. (2004) shows our
data to be of superior quality, especially for \Te\ and O/H. This is
probably due to the indirect flux calibration procedure adopted by Exter
\etal. In addition, due to the apparently lower sensitivity of their
observations, temperature diagnostic lines are not available for about 40\%
of the PNe included in Exter et al. (2004), making the data unusable for
abundance determinations.

\section{Discovery of new emission-line central stars} 
\subsection{Spectra of the newly discovered objects} 

With our new uniform high quality spectra of 90 PNe seen in the direction of
the Galactic bulge we were able to perform a search for new emission-line
CSs. This was motivated by the previous search for such objects towards the
Galactic bulge, which revealed a large number of very late type
\WRCSs\ (see Fig.\,6 of G\'orny et~al. 2004) and several new \WELCSs. As
mentioned in Section 1, this was unexpected, since no [WC\,11] CSs
were known previously in this zone. It seemed therefore that the
distribution of spectral [WC] types among objects located in the Galactic
bulge was different from that in the Galactic disk, but not in the way
assumed by G\'orny (2001).

\begin{figure}
\resizebox{\hsize}{!}{\includegraphics{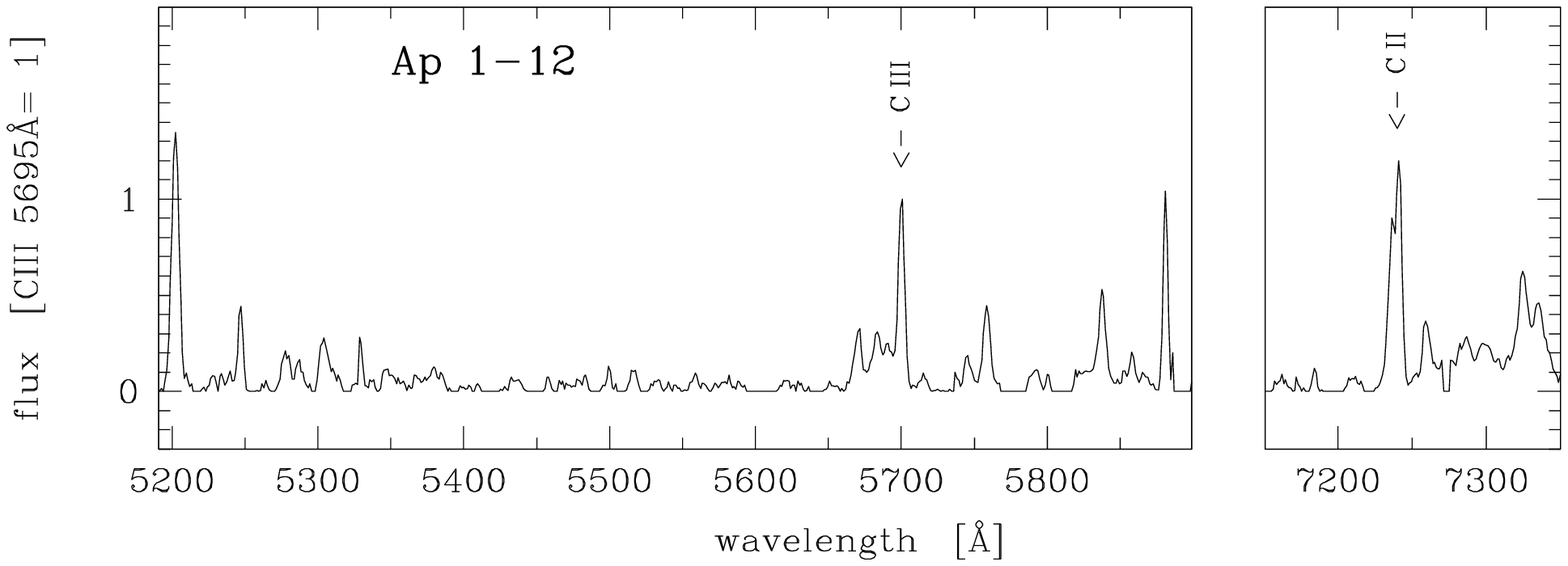}}
\resizebox{\hsize}{!}{\includegraphics{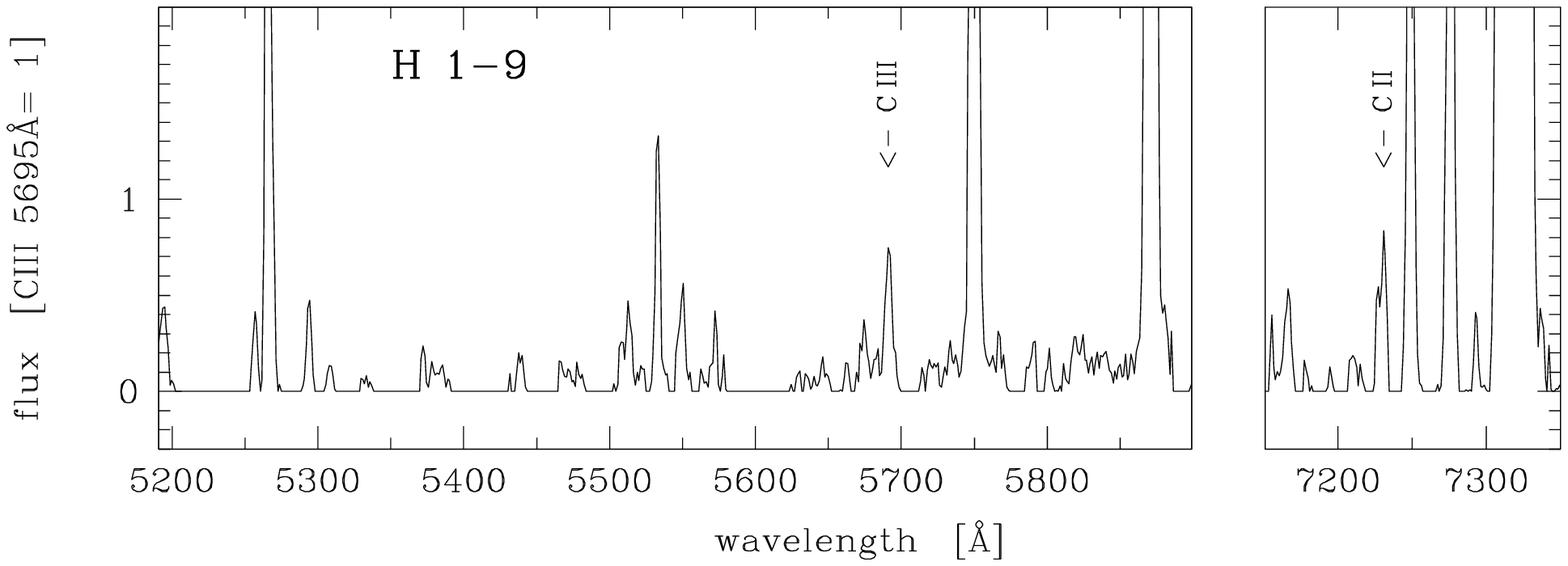}}
\resizebox{\hsize}{!}{\includegraphics{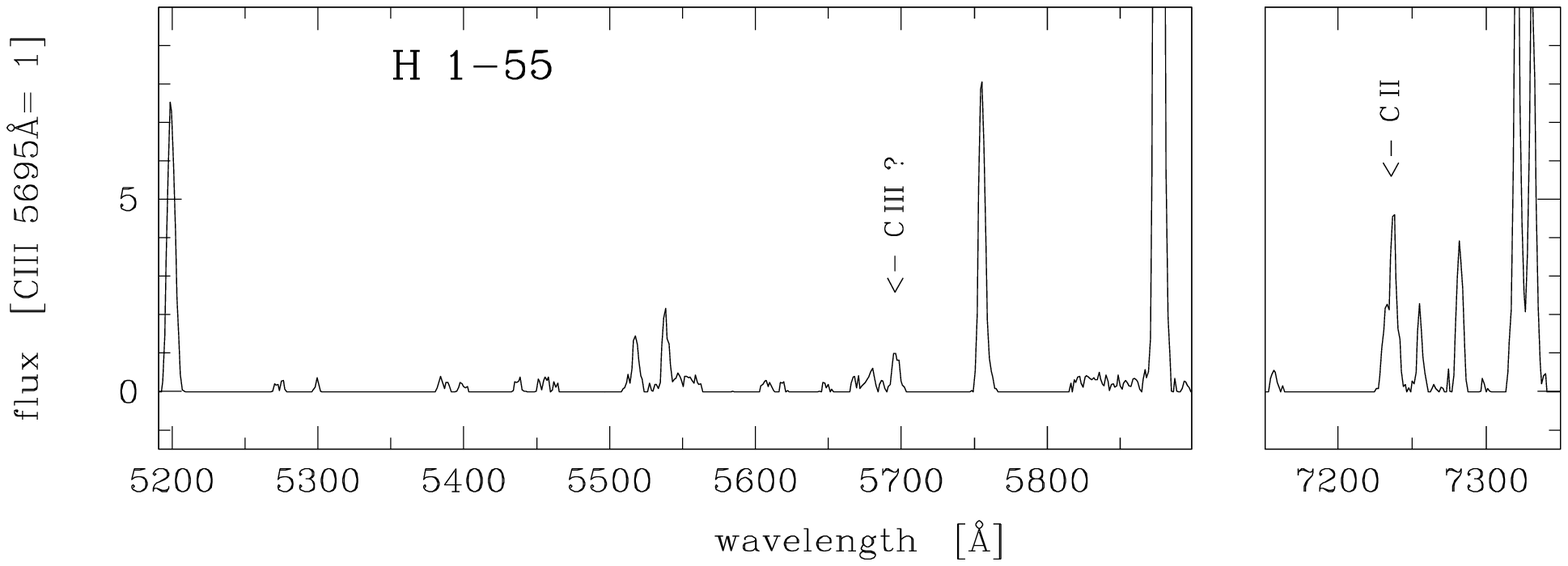}}
\resizebox{\hsize}{!}{\includegraphics{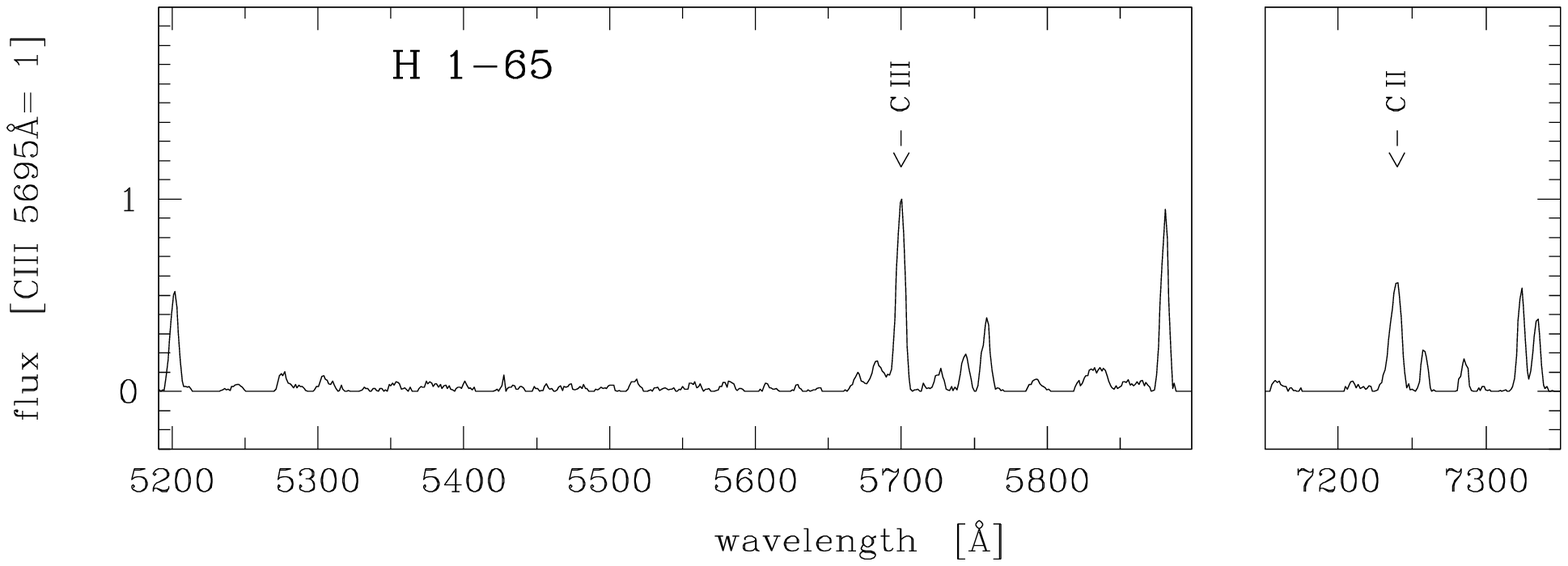}}
\resizebox{\hsize}{!}{\includegraphics{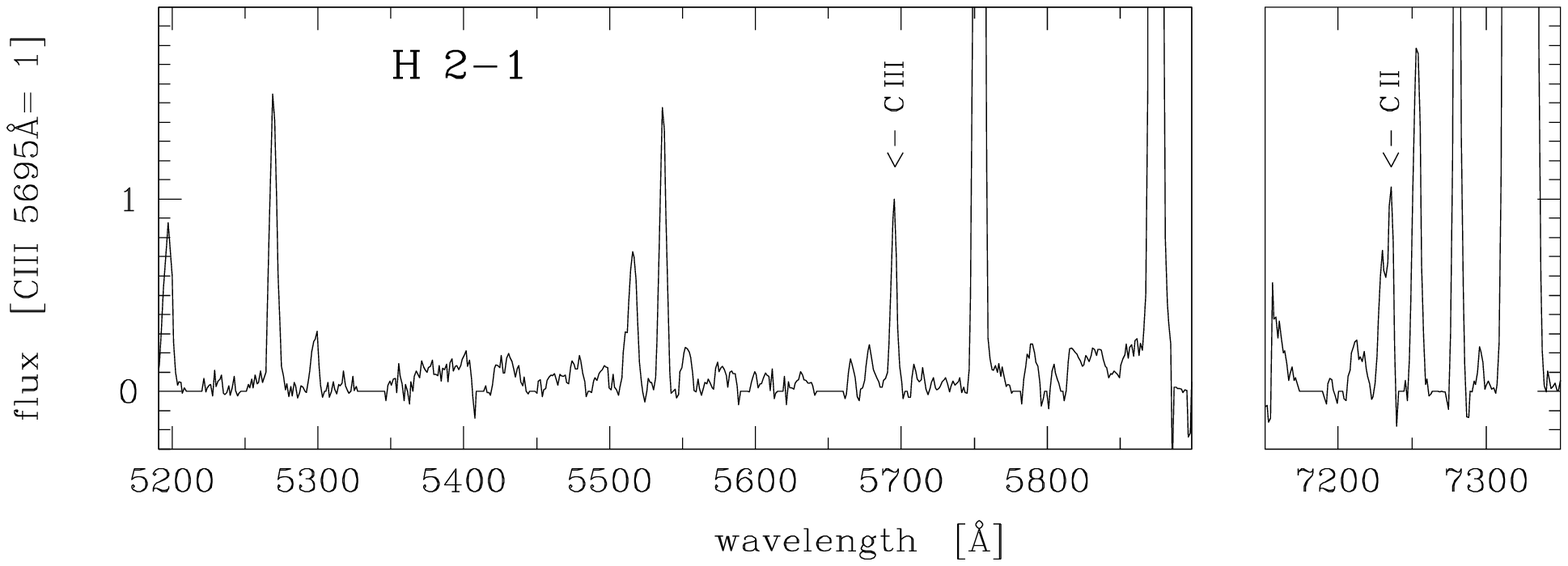}}
\resizebox{\hsize}{!}{\includegraphics{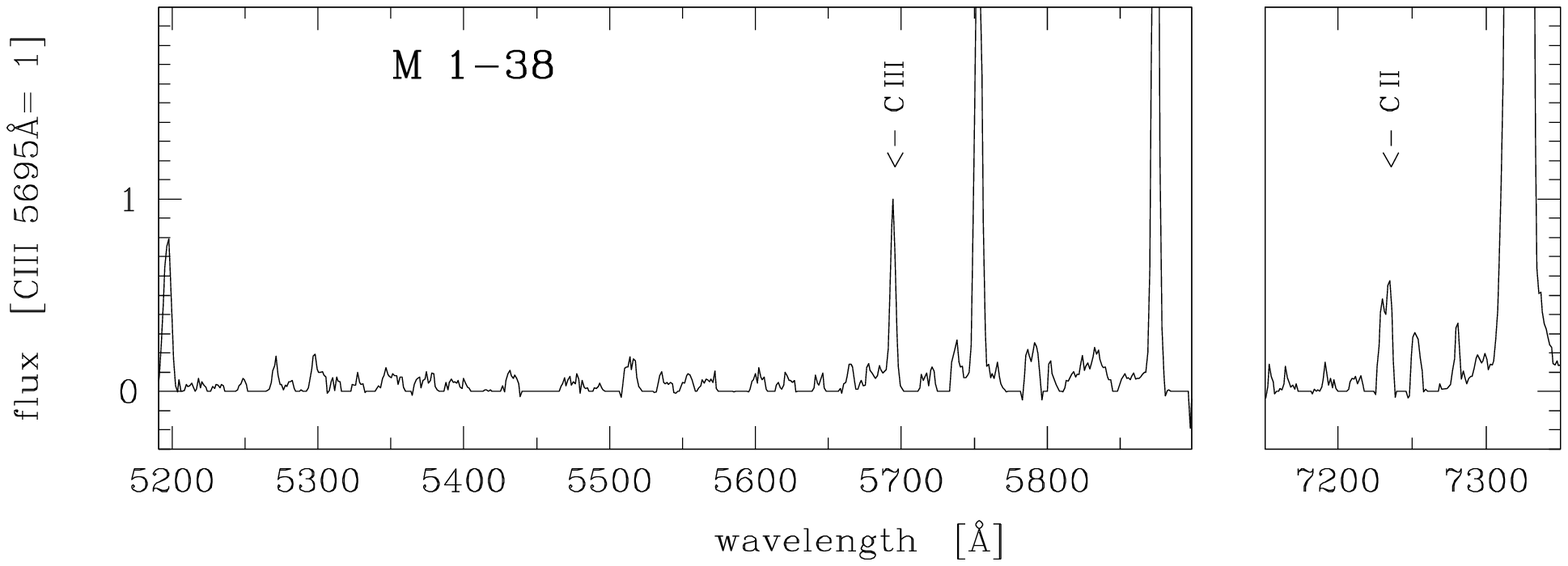}}
\resizebox{\hsize}{!}{\includegraphics{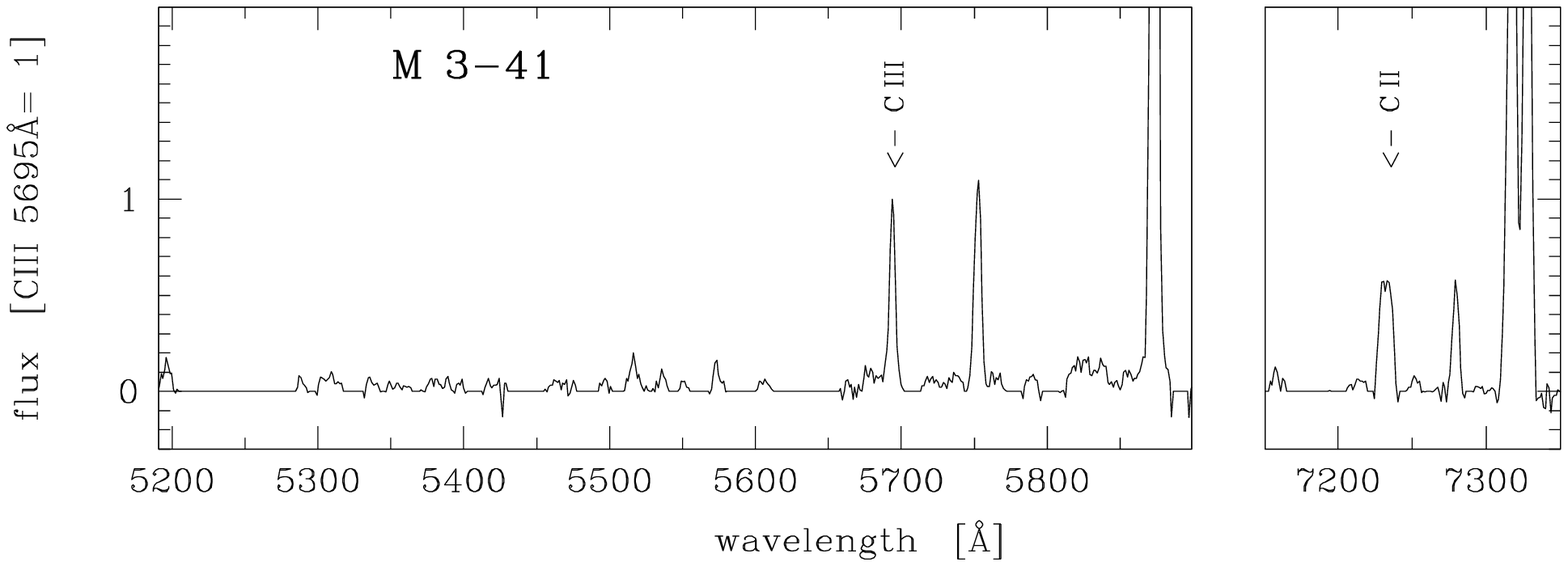}}
\caption[]{
 Spectra of the new \VLPNe\ (very late type [WC\,11]-like
 emission-line central stars PNe).
}
\label{spectra_VL}
\end{figure}

In the present work we discovered a further 7 CSs with very late emission-line
spectra\footnote{ 
 One of them, H\,1-55 should be confirmed when better spectra are available.
} (hereafter called VL).
With the classical classification scheme (Hu \& Bibo 1990) they would be
described as [WC\,11]. The spectra are presented in \figref{spectra_VL}.
Both C~{\sc iii} $\lambda$5695 and C~{\sc ii} $\lambda$7235 lines can be
identified while the C~{\sc iv} $\lambda$5805 feature is absent. The list of
stellar emission lines we looked for was the same as in \Getal.

As in the case of objects discovered by \Getal\ the nebulae surrounding
[WC\,11] stars in the bulge are all of very low ionization with spectra
dominated by the \nii\ doublet and hydrogen Balmer series.  Only for two of
these PNe (H\,2-1 and H\,1-9) are the central stars apparently hot enough,
leading to an \Oiii\ intensity comparable to that of H$\beta$.  The
same was found by G\'orny et al. (2004) for M\,3-17.  For all the other PNe
with [WC\,11] CSs in our sample, the \Oiii\ line (often the strongest
nebular line in normal PNe) is either undetectable or at a level of a few
percent of H$\beta$. This means that there are no conditions for higher
ionization potential ions in the nebula and at least the C~{\sc iii}
$\lambda$5695 line has to be of stellar origin. Some nebular contribution to
the recombination C~{\sc ii} $\lambda$7235 cannot however be totally ruled
out. The objects with the highest probability of such contribution are
H\,1-9 and Ap\,1-12. It is interesting that other extremely low ionization
PNe in the bulge do show the C~{\sc ii} recombination line (the C~{\sc iii}
bands are not observed) though in most cases the emission clearly comes  from
a spatially extended region and is therefore of nebular nature.

\begin{figure}
\centering
\resizebox{.94\hsize}{!}{\includegraphics{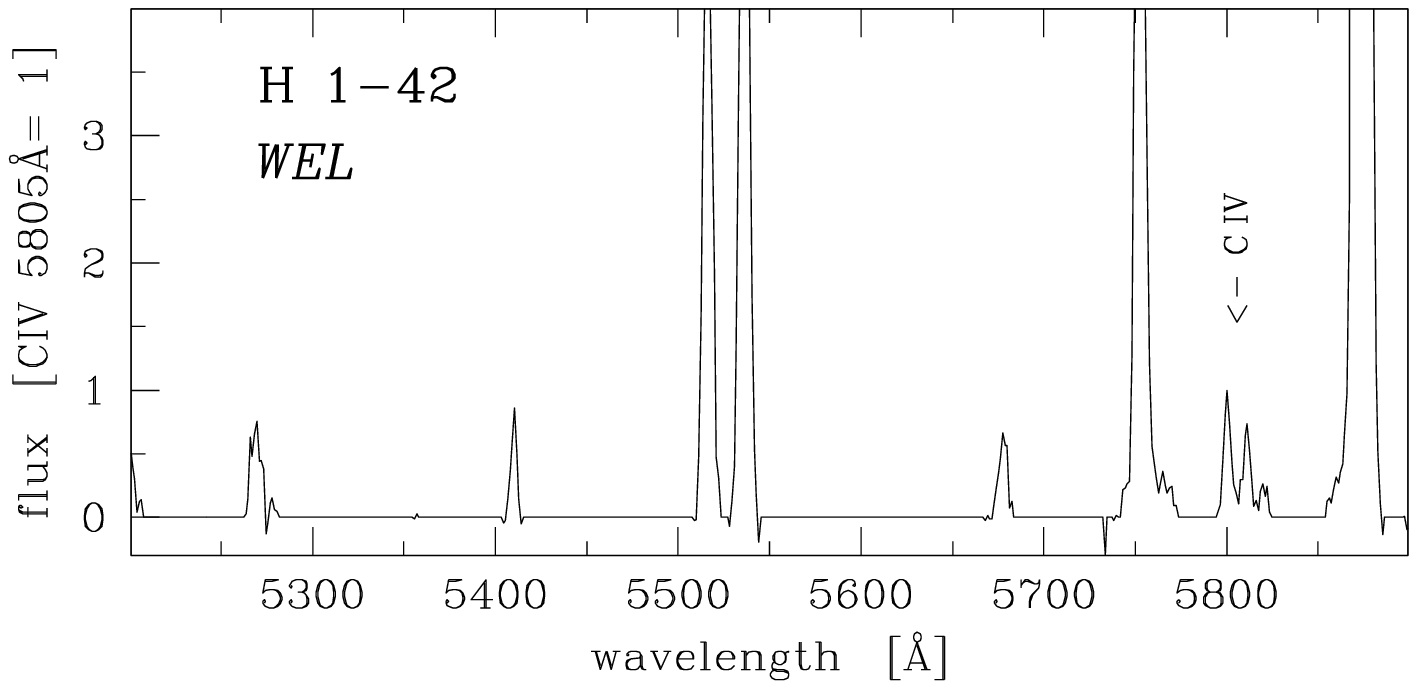}}
\resizebox{.94\hsize}{!}{\includegraphics{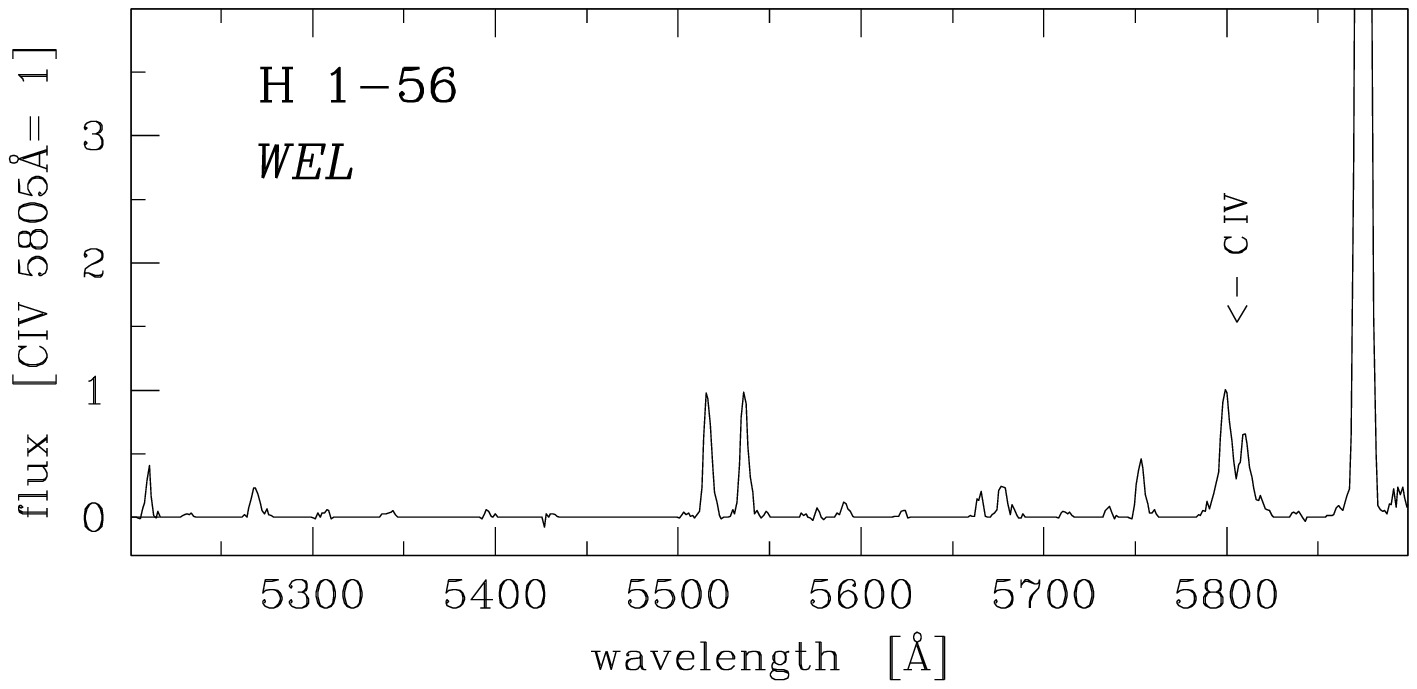}}
\resizebox{.94\hsize}{!}{\includegraphics{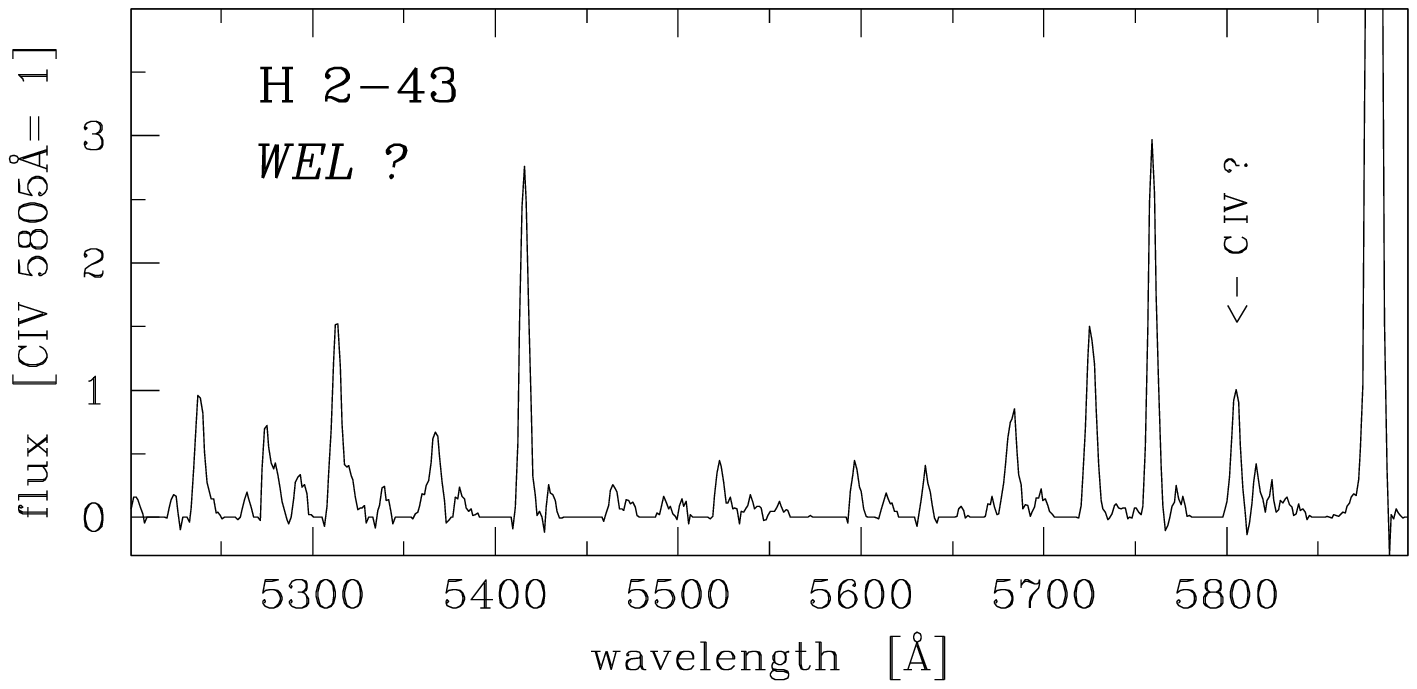}}
\resizebox{.94\hsize}{!}{\includegraphics{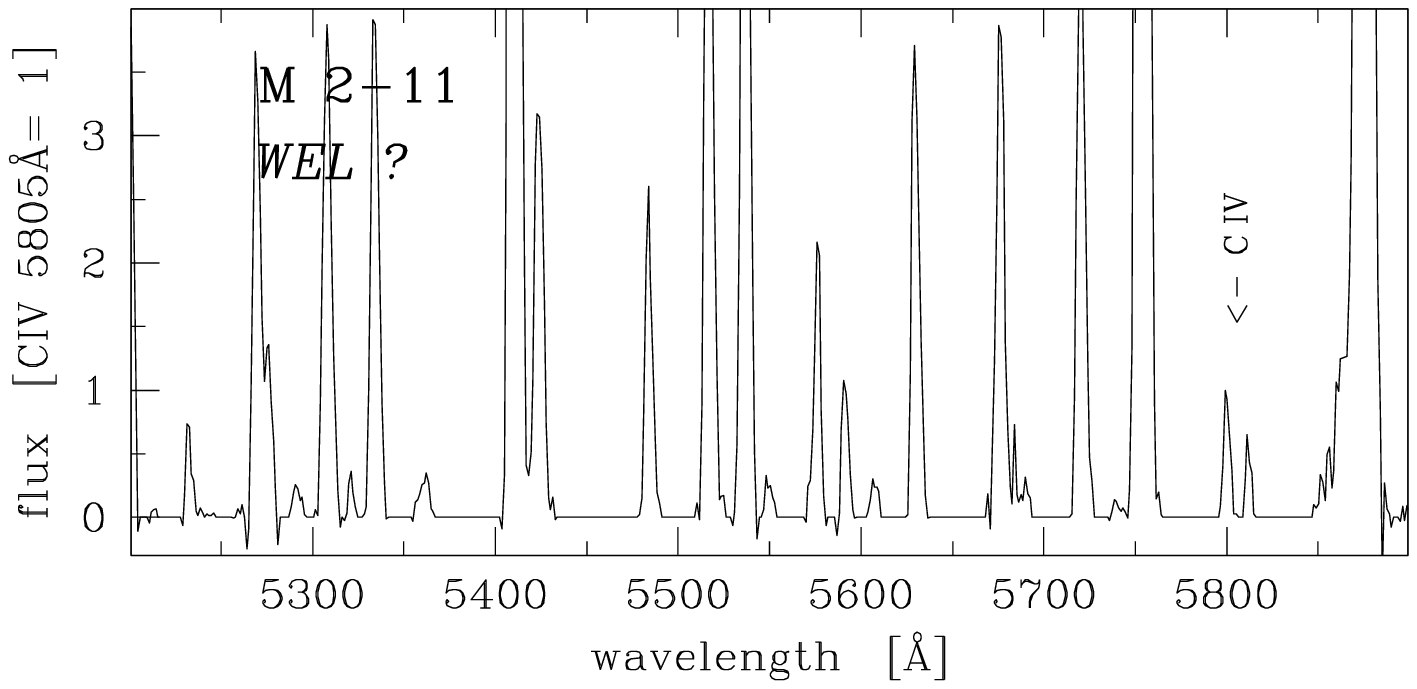}}
\resizebox{.94\hsize}{!}{\includegraphics{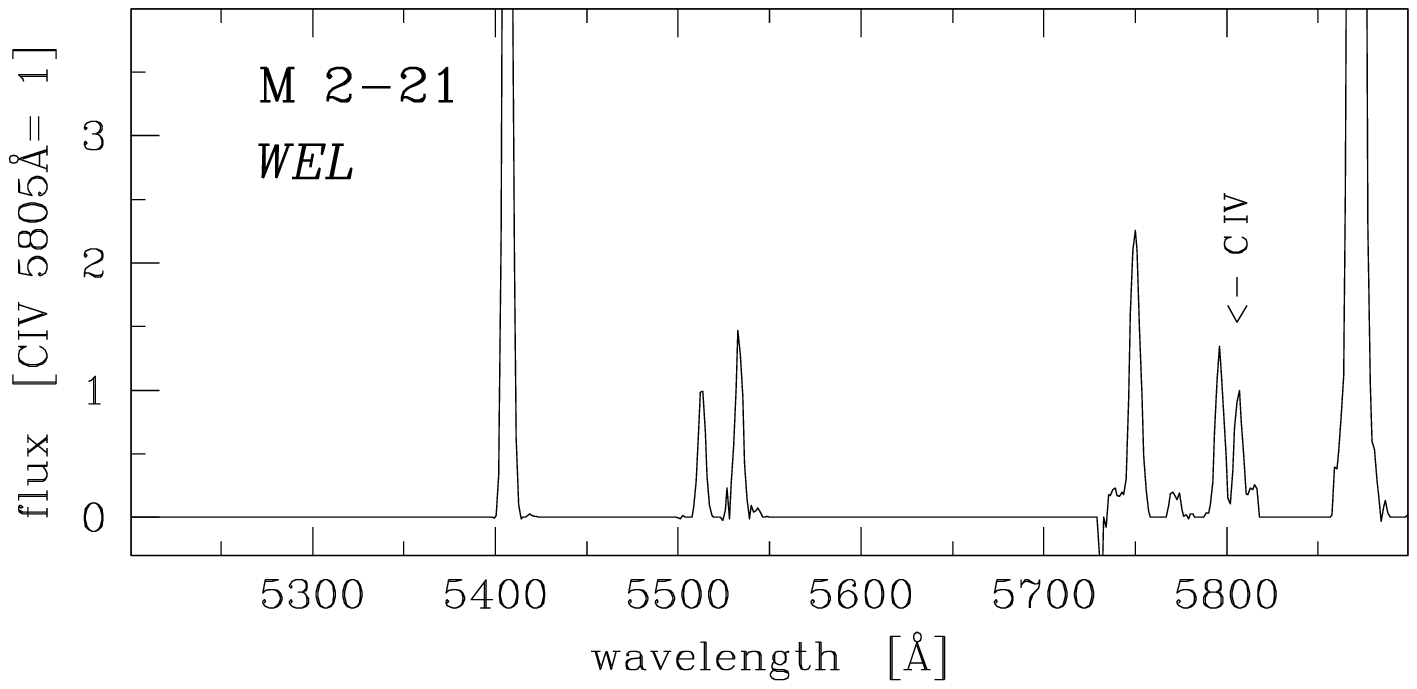}}
\resizebox{.94\hsize}{!}{\includegraphics{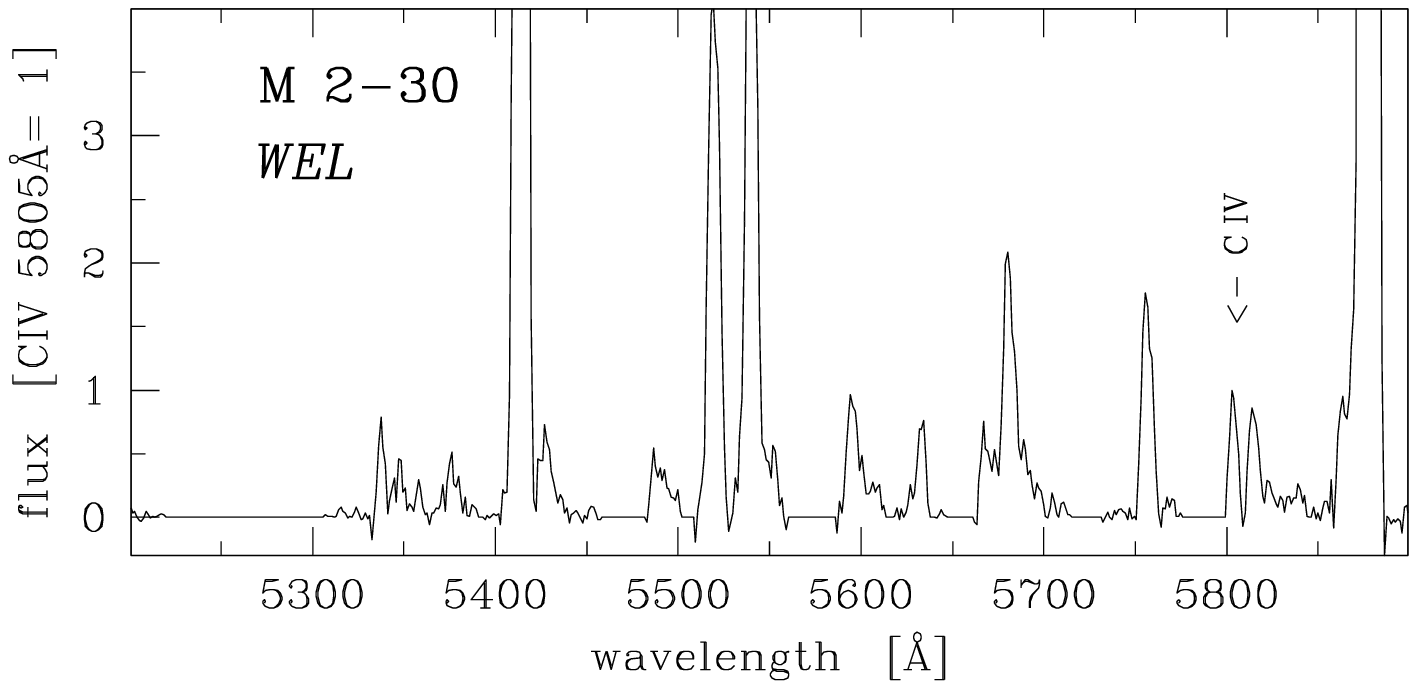}}
\resizebox{.94\hsize}{!}{\includegraphics{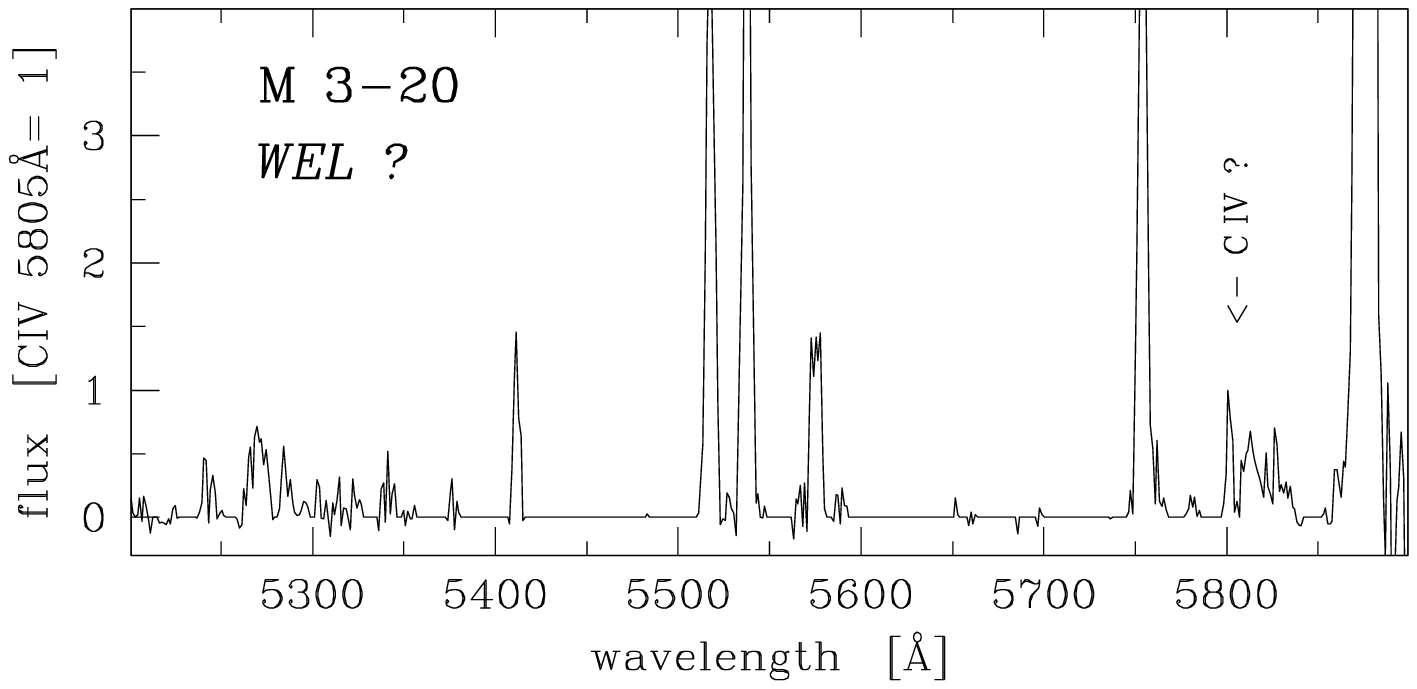}}
\resizebox{.94\hsize}{!}{\includegraphics{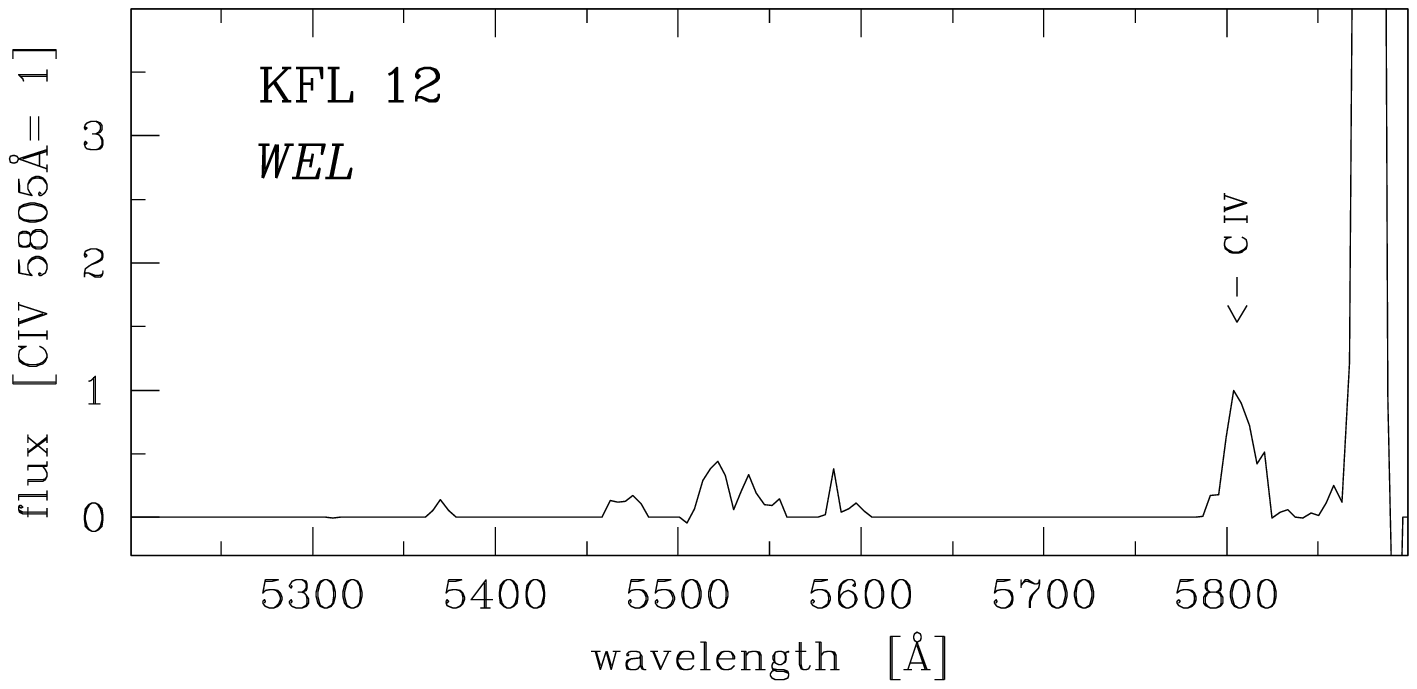}}
\caption[]{
 Spectra of the newly discovered \WELPNe\ (weak emission-line
 central stars PNe).
}
\label{spectra_wels}
\end{figure}

We also identified 8 new \WELPNe\footnote{
  Additional three \WELPNe\ pertaining to the Galactic disk can be identified
  by examination of measured lines tabulated by Wang~\& Liu (2007): M\,1-20,
  IC\,4699 \& IC\,4846. However, IC\,4846 has coordinates too far-off
  to be regarded an inner-disk member and was not included in the analyzed
  sample.
}.
According to the definition of Tylenda \etal. (1993), the spectra of these 
objects usually present only the C~{\sc iv} emission at 5805\AA\ and 
this feature is anyhow much weaker and narrower than in \WRPNe. The spectra of 
the new WEL CSs are shown in \figref{spectra_wels}. Two of these objects were 
previously inspected by G\'orny et al. (2004) but no emission lines were 
detected. On the other hand none of the previously known [WR] or \WELCSs\ 
escaped identification\footnote{
  The only exception is M\,2-34 (PN\,G007.8-03.7) which has a doubtful
  spectral classification and was not confirmed as [WR] type
  central star by G\'orny et~al. (2004)
}.

\subsection{Number of objects}

With the aim of performing a statistical analysis of the population of PNe with
emission-line CSs we combined our present sample with the ones of Cuisinier
\etal. (2000), Escudero \& Costa (2001), Escudero et al. (2004), \Getal\ and
Wang \& Liu (2007). These PNe can be divided into two distinct populations:
one composed of the objects pertaining physically to the Galactic bulge and a
second with most of the objects related to the Galactic inner-disk. To
distinguish the members of the first group, hereafter referred to as the
{\it b} subsample, we used the same criteria as in
\Getal. The objects that do not fulfill these criteria are likely to be
located in the disk (the {\it d} subsample). Since all of the latter have
been chosen at small angular distances from the center of the Galaxy, the
majority of them probably belongs to the inner-disk population. It has been
shown by \Getal\ that although the abovementioned selection criteria do not
contain any reference to kinematical properties, a clear distinction can be
observed also in radial velocities of the members of the \bb and \dd subsamples
(see Fig.\,12 and 13 of G\'orny et~al. 2004). Of course, as the criteria are
of a statistical nature, individual objects assigned to one of the groups
may in reality belong to a different population. This can be also the case
of some PNe with emission-line CSs as discussed in the next section.

In total our \bb subsample  consists of 180 PNe and our \dd subsample
consists of 65 PNe. In subsample \bb there are 25 PNe with WEL CSs, 14 PNe
with VL CSs and 9 \WRPNe\ of earlier types (from now on simply referred to
as \WRPNe). In the \dd sample we have 12 \WELPNe, only one confirmed PN with
a VL CS and 4 other objects with earlier type \WRCSs. There are no other
presently known \WRPNe\ physically pertaining to the Galactic bulge and not
included in our {\it b} sample. However, the situation is different with our
{\it d} sample as 11 probable inner-disk \WRPNe\ not belonging to our sample
are known to exist (see Table~4 of \Getalnb). This means that there are
important selection effects one should take into account and that one should
not compare the frequency of occurrence of [WR]~PNe using only the PNe of
our \bb and \dd subsamples.

\begin{figure}
\resizebox{0.95\hsize}{!}{\includegraphics{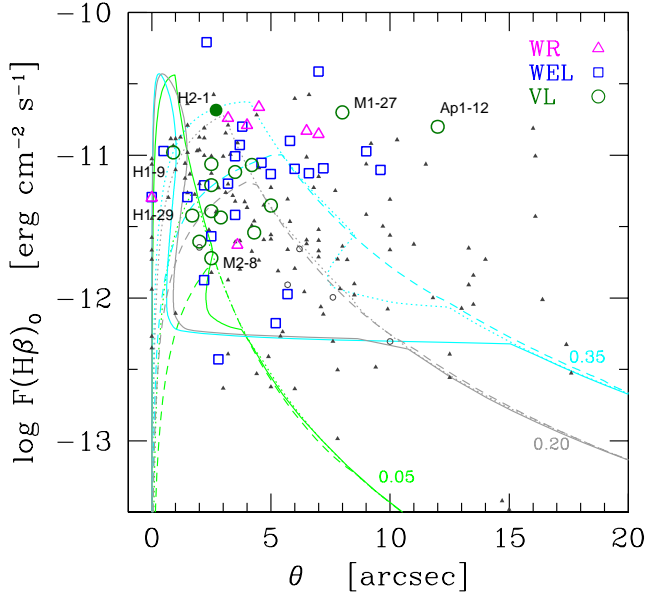}}
\caption[]{
  The relation of apparent diameter and reddening corrected H$\beta$ flux for
  Galactic bulge planetary nebulae. Magenta triangles mark \WRPNe,
  blue squares \WELPNe\ 
  and green circles mark \VLPNe\ ([WC\,11]-like spectra).
  Normal PNe with no emission-line central stars are
  presented with small circles (lowest ionization) and small triangles (all
  remaining). The lines present model calculations for central stars of
  0.57\Msun\ (dashed lines), 0.60\Msun\ (dotted) and 0.64\Msun\ (solid) and
  are labeled by the adopted total nebular masses of the nebula: 
  0.05, 0.20 and 0.35\Msun.
}
\label{wr_diam_fhb}
\end{figure}

\subsection{Selection effects}

In \figref{wr_diam_fhb} we show the reddening corrected H$\beta$ fluxes of
the PNe from our subsample \bb as a function of their apparent diameters. As
all of the nebulae under consideration should be located at approximately
8.5\,kpc, the distance-dependence of both parameters can be ignored in this
case.  PNe with emission-line CSs are represented by large symbols. The PNe
with CSs that do not exhibit emission lines are marked with small black
symbols. We will refer to them as {\it normal} or {\it other} PNe throughout
the rest of this text.

It is striking that the \WRPNe\ are among the brightest PNe of the
subsample. Extrapolating this property of the \bb\ sample to other samples
of PNe, we can infer that the estimated occurrence frequency of \WRPNe\ can
be severely biased in magnitude limited samples. This could explain why, in
external galaxies, the rate of occurrence of \WRPNe\ is much higher than in
our Galaxy (e.g. in the Sagittarius dwarf galaxy, see Zijlstra et~al. 2006).

There are some outliers. One of them is M\,2-8 (PN\,G\,352.1+05.1) a [WR] PN
with log\,F(H$\beta$)$_0$=$-$11.6 and a central star classified as [WC\,2-3].
Its angular position is 9.4~arcsec away from the center of Galaxy so it is
very close to the border value of 10~arcsec usually adopted to define bulge
objects. The physical distance from the Sun can be evaluated with
statistical methods (see next Section) as 10.7 to 13.3 kpc. It is therefore
not excluded that this PN is actually located behind the bulge region. The
radial velocity V$_{lsr}$\footnote{
 All the radial velocities used in this paper have been taken from Durand et
 al. (1998) and corrected for solar motion using the formulae of Beaulieu et 
 al. (2000)}
of only 2.8\kms\ is also not in contradiction with
the hypothesis that M\,2-8 more likely belongs to the disk population (see
Fig.\,12 and 13 of \Getalnb)

The other nebulae outlying their group location in \figref{wr_diam_fhb} are
M\,1-27 (PN\,G356.5-02.3) and Ap\,1-12 (PNG\,003.3-04.6) of VL type. In this
case both the large observed diameter and H$\beta$ brightness do not fit the
rest of VL PNe and seem too large. The distance estimation gives 4 to about
6 kpc for both of them -- almost half of what is usually adopted as a
distance to the Galactic bulge. One could therefore suspect that in reality
these PNe are disk objects located in the foreground of the bulge. The
radial velocities of M\,1-27 is V$_{lsr}$=-53\kms\ and not in contradiction
with the object belonging either to the disk or to the bulge system. In the
case of Ap\,1-12 the V$_{lsr}$=173\kms\ clearly indicates however that the
PN should be kinematically associated with the bulge system.

If having a high brightness nebula is a general property of \WRPNe\ in different
galactic systems it would explain why so many previously known \WRPNe\ are
missing in our \dd\ subsample. The reason is that most of the targets
included in the present analysis (except for observations by Cuisinier et~al.
2000 and Wang \& Liu 2007) were secured in programs deliberately
concentrated on obtaining spectra for the previously unobserved or poorly
observed PNe that are intrinsically fainter. Given the fact that \WRPNe\ are
probably the brightest PNe in a given population, this will certainly
introduce a bias in the estimated occurrence frequency. In the samples
limited to only the brightest PNe the relative number of \WRPNe\ could be
overestimated. In the case of our \dd\ subsample, on the contrary, many
\WRPNe\ have been excluded.

\figref{wr_diam_fhb} shows that \VLPNe\ are generally at least a factor 2
weaker than \WRPNe.  This is a first indication that these classes should be
analyzed separately. The number of \WELPNe\ and that of
\VLPNe\ can easily be underestimated in samples limited to only the
brightest PNe. However in the case of our $d$ sample, since it is biased
towards low luminosity PNe as mentioned above, the fraction of \WELPNe\ and
\VLPNe\ is expected to be larger than in a complete inner-disk sample.

\subsection{Rate of occurrence}

Assuming that the abovementioned selection effects are nonetheless small
for VL PNe we can attempt a comparison of their rate of occurrence in the
Galactic bulge and inner-disk. The 14 VL~PNe from the \bb sample represent
7.8\% of the total number of PNe in this sample, whereas the single
confirmed \VLPN\ in our $d$ sample suggests 1.5\% in the inner-disk.
However, if we consider that Ap\,1-12 and/or M\,1-27 are also in fact
located in the disk rather than bulge then the difference will substantially
diminish. Given the small absolute numbers involved, it is therefore not
certain that there is a statistically meaningful difference in the rate of
occurrence of VL PNe between these two Galactic populations. On the other
hand, the observational selection effects do not have to work in favor
of detecting relatively more VL PNe in the \dd sample (see discussion
in \Getalnb).

Analyzing the spectra of VL PNe, an easily noticeable property is that they
are associated with PNe of very low excitation. Eight out of fourteen VL PNe
in the \bb sample have \Oiii\ lines either not detected or
fainter than 1\% of H$\beta$. In only one object, H\,1-9, this line is
stronger than H$\beta$. We can therefore check if the relatively high number
of VL PNe found in the Galactic bulge is not related to the fact that they
are associated with a particular population of very low-ionization PNe
possibly more frequent in this region. In our \bb sample there are 21
(12.1\%) PNe with \Oiii\ $<$ H$\beta$, whereas in the \dd sample there are 5
such objects (8.5\%). The relative population of very low-ionization PNe in
both subsamples is therefore similar. Unfortunately, again the
absolute numbers are too small to judge if the fact that VL CSs have been
found in  67\% of low ionization PNe in the \bb subsample is statistically
significant. An analysis of a much larger number of inner-disk PNe would be
necessary to conclude if the rate of occurrence of VL~PNe in the disk
(presently derived as $\sim$20\% to $\sim$40\%) is indeed much lower than in
the bulge.

Finally, assuming that the selection effects are reasonably small also for
\WELPNe,  we can evaluate their rates in both subsamples. There are 25
(or 14.5\%) such objects in the \bb subsample and 10 (16.9\%) of them in
\dd subsample. Obviously there is no difference in the frequency of
occurrence of \WELPNe\ in both groups despite the fact that the Galactic bulge
and inner disk PNe should originate from different stellar populations and
different epochs of Galactic evolution.

In conclusion, as can be seen from the above considerations, the answer to
the question of how frequent are the different types of emission-line CSs
remains open, until additional similarly large and uniform samples of
deep PNe spectra are collected. In particular, the analysis of PNe in the
Milky Way inner-disk should shed more light on the connection between VL PNe
and low excitation nebulae (G\'orny, in preparation). Since the number of
low excitation PNe is relatively easy to establish (e.g. using existing
surveys) it would help to investigate if the presence of the VL PNe group is
characteristic only of the metallicity/age conditions of the bulge
population. It cannot be totally excluded that the surprisingly large number
of discoveries of VL PNe by \Getal\ and this work is simply due to the fact
that they were not expected to exist and therefore not searched for by surveys
of other Galactic regions.

\section{Properties of the bulge PNe with emission-line central stars }

\begin{figure}
\resizebox{0.95\hsize}{!}{\includegraphics{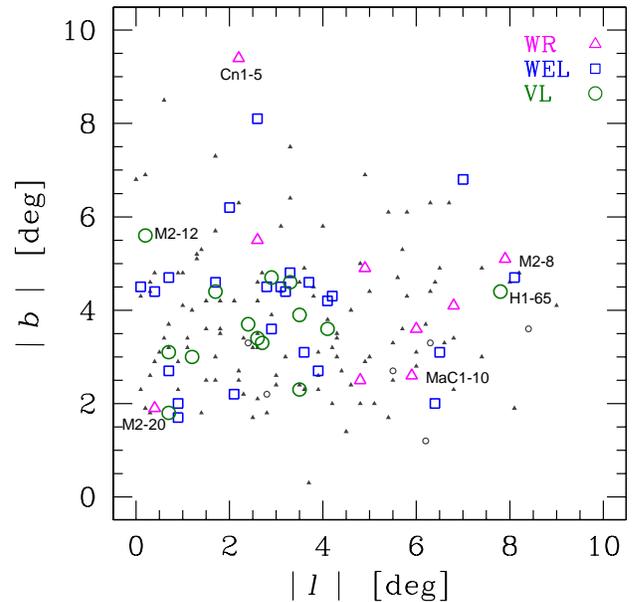}}
\caption[]{
  Locations of different types of bulge PNe in Galactic coordinates (absolute
  values). The meaning of the symbols is the same as in \figref{wr_diam_fhb}.
}
\label{wr_lb}
\end{figure}

The aim of this section is to compare the properties of PNe with
emission-line CSs with the "normal" ones. We will restrict our analysis to
PNe with emission-line CSs of our \bb subsample, i.e. pertaining
to the Galactic bulge. This sample is more numerous, statistically
more meaningful and likely more complete and homogeneous than the much
smaller \dd subsample we have collected.

\subsection{Spatial distribution}

We begin with the simplest property i.e. the spatial distribution on the sky
in the Galactic coordinates {\it l,b} (\figref{wr_lb}). Because of the small
number of objects we present the absolute values of the coordinates assuming
a perfect point-symmetry of the bulge PNe distribution. It is clear from
\figref{wr_lb} that the locations of \WRPNe, \WELPNe\ and \VLPNe\ differ,
suggesting that they may originate from different populations of stars.

More than a half of bulge [WR]\,PNe are located at longitudes
larger than 4.5 degrees from the Galactic center, whereas there are very few
\WELPNe\ and \VLPNe\ at such locations. From simple geometric considerations
the \WRPNe\ seem therefore to form the most external system and it would be
tempting to consider the \WRPNe\ as  members of the inner disk
population and not physically pertaining to the bulge. Indeed, the object
with the largest angular separation from the center is M\,2-8, already
suspected to belong to the Galactic disk based on its relatively low
H$\beta$ flux and low radial velocity. The other \WRPNe\ seem however not in
contradiction in this respect with their bulge association. On the other
hand,  the distances of \WRPNe\ derived with the
Shklovsky statistical method suggest they could be located at 4.3\,kpc i.e.
only half-way to the bulge. We strongly argue however (see the next
Section) that these distances are not reliable because the common
assumptions, e.g. on the typical nebular mass, are not met by \WRPNe.

Most of the bulge \WRCSs\ fall within the range [WC4] -- [WC6] of spectral
classes that are uncommon among the disk \WRPNe\ (see e.g. Fig.\,1 of
G\'orny 2001). It is a particular feature of this group, and it seems very
unlikely that these objects would be Galactic disk members only by chance
located in the direction of Galactic bulge. Nonetheless, the spectral type
cannot be used as a parameter discriminating \WRPNe\ pertaining to the bulge
from those in the Galactic inner-disk. In the latter system there are also
four objects with similar [WC] type (see Table~1 of \Getalnb).
Interestingly, the Shklovksy distances of these objects are, as should be
expected, convincingly smaller than the bulge
\WRPNe\ distances mentioned above.

For the VL PNe in \figref{wr_lb} the distribution is completely different.
They are practically all within 4.5 degrees from the Galactic center with
the exception of H\,1-65 PNe\footnote{
  Note that H\,1-65 has been classified as \WELPN\ by Acker \& Neiner
  (2003) nonetheless our spectrum in \figref{spectra_VL} clearly shows 
  the C\,{\sc iii} emission at 5695\AA\ and the absence of C\,{\sc iv} 
  5805\AA\ that is present in other \WELPNe.
}. It can be noticed also in \figref{wr_lb} that the low
ionization PNe not classified as \VLPNe\ (small open circles) are 
not distributed like the \VLPNe\  but rather like the \WRPNe.

\WELPNe\ show a distribution similar to that of \VLPNe, suggesting that both
groups can be physically closer to the center of the Milky Way and originate
from a different stellar population or epoch of the bulge history than the
\WRPNe.

\begin{figure}
\resizebox{0.95\hsize}{!}{\includegraphics{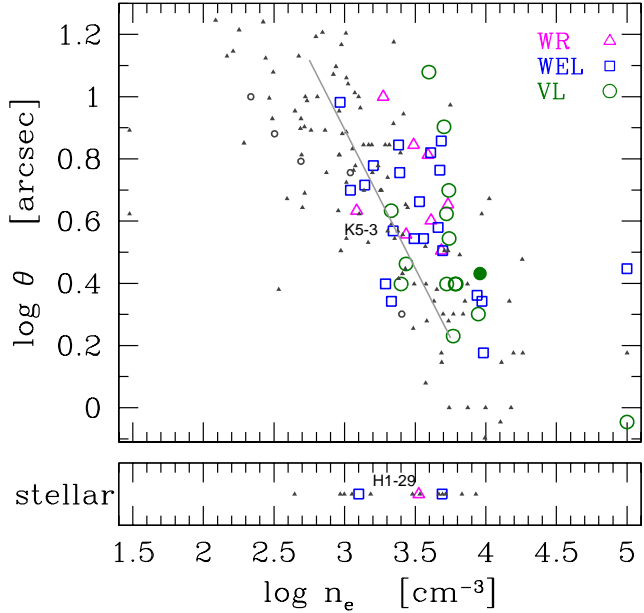}}
\caption[]{
  Apparent diameter vs. electron density for different types of bulge PNe.
  Bottom panel is for PNe too small to be reliably measured. The meaning of
  the symbols is the same as in \figref{wr_diam_fhb}. Filled symbol marks
  inner-disk VL PN.
}
\label{ne_diam}
\end{figure}

\subsection{Masses of the nebulae}

Figure \ref{ne_diam}\ presents the nebular diameter as a function of
electron density for the \bb sample PNe. The solid line shown in this plot
is a linear fit to the distribution of normal PNe. As can be seen, most 
PNe with emission-line CSs, irrespective of the particular spectral type,
are located to the right of this line. This means that the ionized matter in
these nebulae is denser than in the average PNe. If all the nebulae would be
uniformly filled with gas and ionized at the same percentage of their volume
that would also mean that the nebulae around emission-line CSs are more
massive.

The hypothesis that the nebulae around \WRCSs\ are more massive than the
normal PNe was proposed by G\'orny (1996) as a result of the analysis of
statistical distances to these objects. Taking the average distance
calculated with the Shklovsky method it was suggested that the \WRPNe\ are
closer by a factor of 2 to the observer than the other Galactic PNe. As that
investigation concerned 350 PNe with known types of the central star
(appropriate spectroscopic observations available) there seemed to be no
observational selection effect responsible for this behavior nor physical
reason why \WRPNe\ should be closer to the Sun. G\'orny (1996) concluded
that the apparent difference can be explained if the nebulae around \WRCSs\
are almost 5 times more massive than the typical nebula or/and if their
filling factor parameters are considerably lower. The second possibility is
quite likely as the morphological images of \WRPNe\ reveal a larger number
of sub-structures (G\'orny 2001) while their expansion velocity profiles are
characterized by large scale turbulence component sometimes found also in
\WELPNe\ but seldom in normal PNe (Gesicki et~al. 2006).

Here, we will adopt a similar reasoning to that of G\'orny (1996). Using
distances derived with statistical methods is equivalent to assuming that
all the nebulae can be represented by a simple model of hydrogen gas
occupying uniformly a sphere with a certain filling factor (like in the
Shklovsky distance method) or that this gas has a constant density equal to
that derived from spectroscopic observations (as in the distance method
proposed by Barlow 1987). The mass of the gas is assumed to be the same for
all objects. Since we are dealing with Galactic bulge PNe, if the model
assumptions are correct and hold for all nebulae then using either method
the derived average distance for each subsample should be the same. We
assume a Galactic bulge distance of 8.5\,kpc from the Sun.

Starting with the method proposed by Barlow (1987), making the common
assumption that the typical ionized mass of the nebula is 0.2\Msun\ and
taking the densities as derived from our observations of \rSii\ lines one
obtains a distance scale that is too long for the normal PNe with non
emission-line CSs. The adopted Galactic bulge distance would be obtained if
the mass of the average nebula assumed in the model is 0.13\Msun\footnote{
The typical
  nebular mass would be $\sim$0.19\Msun\ if we consider only 33 normal PNe
  with CSs hotter than log\,T$_{\star}$$>$4.5 assuming such nebulae are density
  bounded.}.
If we now use the Shklovsky method to derive the same distance adopting the
ionized nebular mass of 0.13\Msun\ it is necessary to assume also that the
average filling factor is $\epsilon$=0.75.

\begin{figure}
\resizebox{0.95\hsize}{!}{\includegraphics{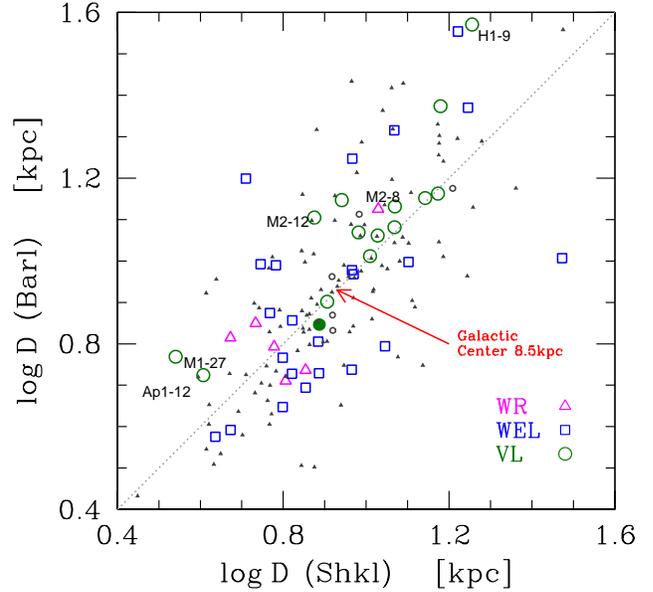}}
\caption[]{
  Distances derived with the Shklovsky method vs. method of Barlow (1987) for
  different types of bulge PNe. The meaning of the symbols is the same as in
  \figref{wr_diam_fhb} and \figref{ne_diam}.
}
\label{wr_dist}
\end{figure}

The distances for all the Galactic bulge PNe of our \bb\ sample computed
using both methods with the abovementioned assumptions on nebular parameters
are plotted in \figref{wr_dist}. One can see in this plot that the distances
of \WRPNe\ calculated with such a nebular model are too short compared to
the adopted bulge distance (arrow), meaning that their true ionized masses must
be larger. The only [WR]\,PN with a distance exceeding 8.5kpc is M\,2-8,
declared before as a probable background object not truly associated with
the bulge. On the other hand the distances to VL PNe are generally too large,
again except for M\,1-17 and Ap\,1-12 already suspected rather to be 
foreground disk objects based on their large relative brightness. Excluding
these two nebulae to obtain a mean distance of 8.5 kpc for VL PNe, their
average filling factor should be closer to $\epsilon$=0.5 and their ionized
masses as small as M$_{ion}$=0.05\Msun. Taking into account that VL PNe are
very low ionization nebulae, they can be only partially ionized and in that
case their total nebular mass could be larger (but see the discussion on
evolutionary status below). For the 5 \WRPNe\ in
\figref{wr_dist}\ with short distances one obtains M$_{ion}$=0.34\Msun. As
far as the \WELPNe\ are concerned they do not seem to distinguish themselves from the
normal PNe in
\figref{wr_dist}. Finally, because the derivation of typical ionized masses or
the filling factor by reversing statistical distance methods is very
uncertain, the quoted numbers should be regarded as a rough estimate giving
only an indication of the scale of possible differences between different
groups of PNe.

As already discussed, in \figref{wr_diam_fhb} one can compare the locations
of the sample \bb\ PNe in the diameter versus flux F(H$\beta$)$_0$ plane with a
set of theoretical tracks. These tracks have been calculated assuming that
the central star is radiating like a black-body and evolving according to
the Bl\"ocker (1995) models. Three different stellar masses are shown:
0.57\Msun\ (long dashed lines), 0.6\Msun\ (dotted) and 0.64\Msun\ (solid).
The model of the surrounding nebula was a simple constant-density sphere
uniformly filled with gas and expanding with a constant velocity. Such
simple PNe models are not realistic as e.g. they do not take into account the
interaction between the stellar winds and the nebula.  They are however
sufficient for our illustrative purposes. Three different sets of models are
presented in the plot for each stellar mass assuming a total nebular mass of
M$_{neb}$=0.05, 0.20 and 0.35\Msun\ (see labels). The adopted expansion
velocity was set to 20km/s and the filling factor
$\epsilon$=0.75 in all the cases.

The model calculations presented in \figref{wr_diam_fhb} indicate what kinds
of nebulae can produce the bright \WRPNe. They cannot be directly compared
as the \WRCSs\ are helium-burning objects whereas the presented
models are based on evolutionary tracks of hydrogen-burning stars. It can be
deduced however that  more massive nebulae can easily be the brightest
objects in a given population even if they surround intermediate mass
CSs. The high-mass stars (M$_*$$>$0.64\Msun) cannot produce
objects like bulge \WRPNe\ because they evolve too quickly. The time-scale
of their evolution would have to be slowed-down considerably compared with
what the stellar Bl\"ocker (1995) models predict. This is however not very
likely, knowing that the \WRCSs\ have high mass loss. Intensive mass
loss will rather speed up the evolution bringing the star faster to high
temperatures due to the stripping of the external layers of the
star\footnote{
  Compare with G\'orny et~al. (1994) to see how an enhanced mass-loss is
  reflected in evolutionary diagrams for H-burning central stars or G\'orny \&
  Tylenda (2000) for He-burning models.
}.

Concerning the \WELPNe\ it is not certain (as for most of the normal nebulae
as well) if their CSs are powered by burning helium or hydrogen.
From their locations in \figref{wr_diam_fhb} it may be suspected they
represent a diversity of possible parameter combinations and can have
CSs of different masses and/or very different nebulae. The \VLPNe\
on the other hand are probably more uniform and more obviously distinct from
\WRPNe.

\subsection{Evolutionary status}

\begin{figure}
\resizebox{0.95\hsize}{!}{\includegraphics{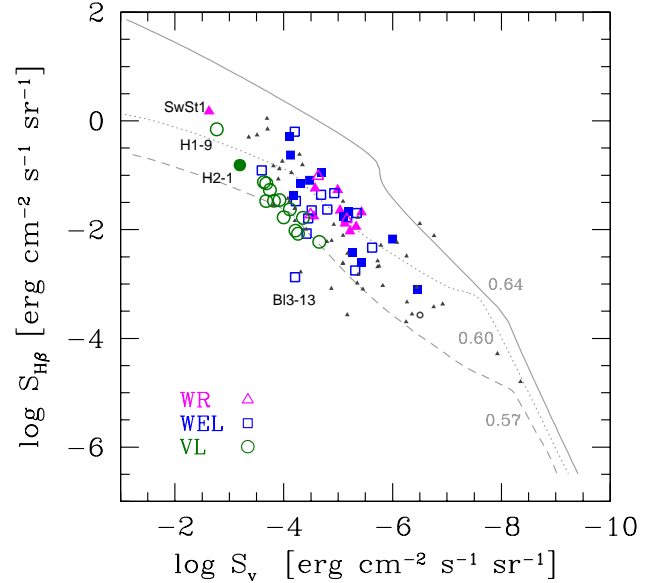}}
\caption[]{
  Surface brightness S$_{H\beta}$ versus S$_V$ for different types of bulge
  and inner-disk PNe. The bulge PNe are marked with large open symbols:
  magenta triangles -- \WRPNe, blue squares -- \WELPNe,
  green circles -- \VLPNe\ ([WC\,11]-like spectra). 
  Large filled symbols of the equivalent shape
  mark inner-disk WEL and \WRPNe. Small symbols mark
  normal bulge PNe with the notation as in \figref{wr_diam_fhb}.
  The lines present model calculations for central stars of
  0.57, 0.60 and 0.64\Msun\ adopting a simple nebular model with M$_{neb}$=0.20\msun,
  $\epsilon$=0.75 and V$_{exp}$=20km/s.
}
\label{wr_ss}
\end{figure}

Figure \ref{wr_ss}\ presents the locations of sample \bb PNe (open symbols)
in the S$_{H\beta}$
versus S$_V$ plot, S$_{H\beta}$ being the nebular H$\beta$ surface brightness and S$_V$ a similarly defined parameter 
based on the stellar flux in the visual V band and on the nebular diameter
$\theta$ (G\'orny \etal. 1997):
\[
  S_{\mbox{v}} = F_{\mbox{v}} / (\pi \theta ^2).
\]
This parameter is very useful for distinguishing
stars at different stages of post-AGB evolution. Both S$_{H\beta}$ and S$_V$
have the advantage of being distance independent therefore we can also plot in
\figref{wr_ss}\ the locations of \WRPNe\ and \WELPNe\ from our
subsample \dd (filled symbols). The evolutionary tracks have been calculated
with a simple model of normal bulge PNe with M$_{neb}$=0.20\msun, filling
factor $\epsilon$=0.75 and expansion V$_{exp}$=20km/s for central stars
of 0.57, 0.60 and 0.64\Msun\ evolving in agreement with
Bl\"ocker (1995) models.

It is seen in \figref{wr_ss}\ that the locations of \WRPNe\ and \VLPNe\ are
clearly separated. The \VLPNe\ seem to be originating from stars of lower
masses than the [WR] (but again the exact values of plotted parameters
depend on selected nebular model and details of central star evolution - see
G\'orny \& Tylenda 2000 for some examples). The lines presenting model
calculations allow us to follow the typical evolution of a PN and its central
star that is going from young high surface brightness objects to old,
dispersed and low surface brightness nebulae. None of them links
regions occupied by the VL and \WRPNe.  The \VLPNe\ locations in this frame
form a very narrow strip suggesting that the masses of their CSs may be
very similar and that one can in fact observe an evolutionary sequence
within this group. The direction of their evolution would be in agreement
with model predictions but clearly not towards \WRPNe.

There seems to be only one VL object in the present \bb subsample that could
evolve into a [WR]-type star: H\,1-9. Its location in \figref{wr_ss}\ is
close to the position of the long-known, unusual very late type object
SwSt\,1 and other Galactic disk [WC11] or [WC10] central stars (see e.g.
Fig.\,5 of G\'orny 2001 for comparison). As can be learned from
\figref{wr_diam_fhb}, H\,1-9 is possibly the brightest true bulge VL PN in
our sample and at the same time an object with the smallest observed
diameter in this group. In fact, the known disk [WC11] PNe are characterized
by small angular diameters, with the clear exception of K2-16 and possibly
PM1-188 -- see G\'orny \& Tylenda (2000) and Pe\~na (2005). It is
interesting that the location of the inner-disk VL object H\,2-1 is also
rather close to the abovementioned region in \figref{wr_ss}. It is noteworthy that
the locations of the other inner-disk-suspected M\,1-27 and Ap\,1-12 are
within the bulk of the region occupied by VL PNe in this plane (not marked).

One can also notice in \figref{wr_ss}\ that the positions occupied by the
\VLPNe\ are clearly separated from regions where the \WELPNe\ are found
(compared to \figref{wr_diam_fhb}\ with a significant number of \VLPNe\ and
\WELPNe\ sharing similar locations). It can therefore be argued that
although \VLPNe\ have central stars of lower temperatures they probably
form a separate class of objects and will evolve neither into \WRPNe\
nor \WELPNe.

\subsection{Nebular properties}

\begin{figure}
\resizebox{0.95\hsize}{!}{\includegraphics{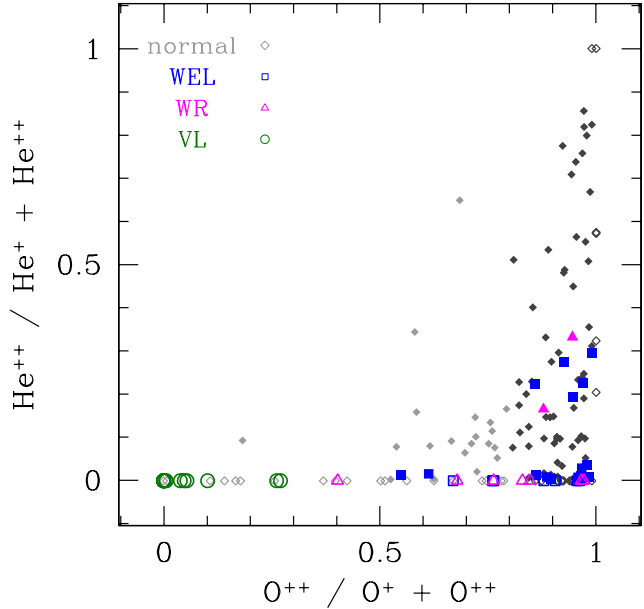}}
\caption[]{
  Ionization level of different types of bulge PNe as indicated by the
  relative abundance of O$^{++}$ ions versus relative abundance of He$^{++}$
  ions. \WRPNe\ are marked with magenta triangles,  \WELPNe\ blue
  squares and \VLPNe\ ([WC\,11]-like spectra) with green circles.
  Normal PNe are marked with diamonds: black mark
  objects with high level of ionization (\oppo$>$0.8), grey mark remaining
  objects. Open symbols indicate PNe with lines from relevant ions below the
  detection limit.
}
\label{oppo_hepphe}
\end{figure}

In this section the general properties of PNe surrounding emission-line and
non-emission line CSs in the Galactic bulge will be presented. We
start with the detected level of ionization of the nebular gas. For this
purpose in \figref{oppo_hepphe} we present the ionic ratio \oppo\ versus
\hepphe\ for all PNe from our \bb\ sample.
As can be noticed, if judged only by the relative numbers of doubly to
singly ionized oxygen atoms the \WELPNe\ would appear among the PNe of
highest ionization with almost all of them having
\oppo$>$0.85. Surprisingly, on the other hand only a few \WELPNe\ have any
significant amounts of doubly ionized helium and, on average, the
He$^{++}$/He$^{+}$+He$^{++}$ ratio is lower than in normal PNe. This could
be due to a difference in the spectral energy distribution of the ionizing
radiation field between \WELCSs\ and normal CSs i.e. those without enhanced
mass loss. It has to be noted that this is not only an interesting fact
distinguishing the CSs. It can have some important consequences e.g. for the
determination of chemical abundances since in many cases the use of ICFs is
necessary to correct for unobserved ionization stages. The assumption that
ionizing radiation fields of the CSs are similar is important, but as can be
seen clearly not totally justified in this case.

For \WRPNe\ the interpretation of \figref{oppo_hepphe} is more difficult
since in the spectra of these objects one may expect also the presence of
stellar recombination lines of helium. This is a consequence of the strong
winds from these stars and adds additional uncertainty to the
interpretation. Nevertheless, in two out of nine bulge \WRPNe\ in our sample
a measurable nebular He$^{++}$ line has undoubtedly been identified.  This
is a comparable ratio to what we noticed for \WELPNe\ and in agreement with
the fact that e.g. judging from their positions in \figref{wr_ss} the
range of temperatures of their CSs and their present phase in the
evolution of the nebula should on average be quite similar for the two
groups.

\begin{figure*}
\resizebox{0.475\hsize}{!}{\includegraphics{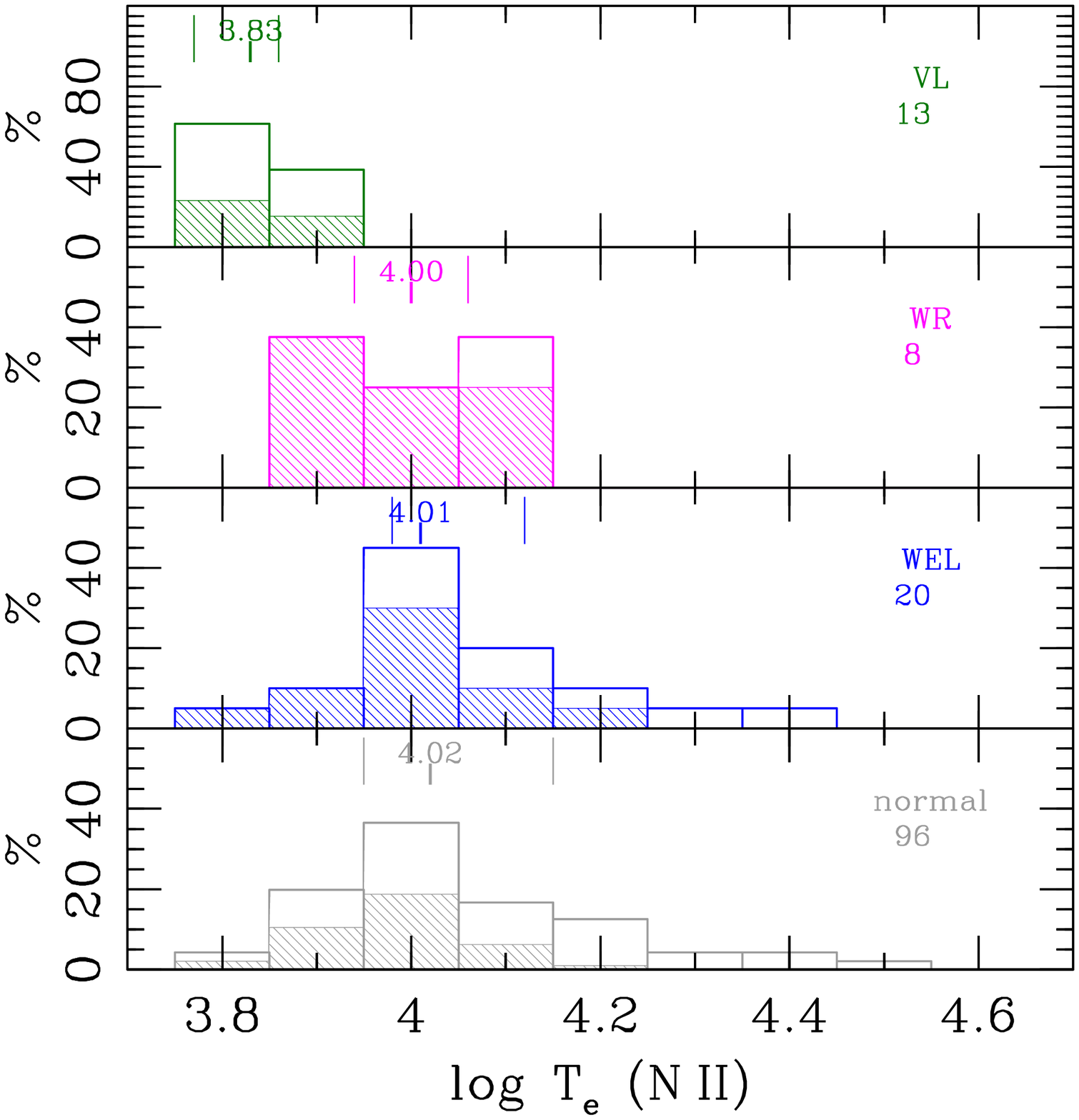}}
\resizebox{0.475\hsize}{!}{\includegraphics{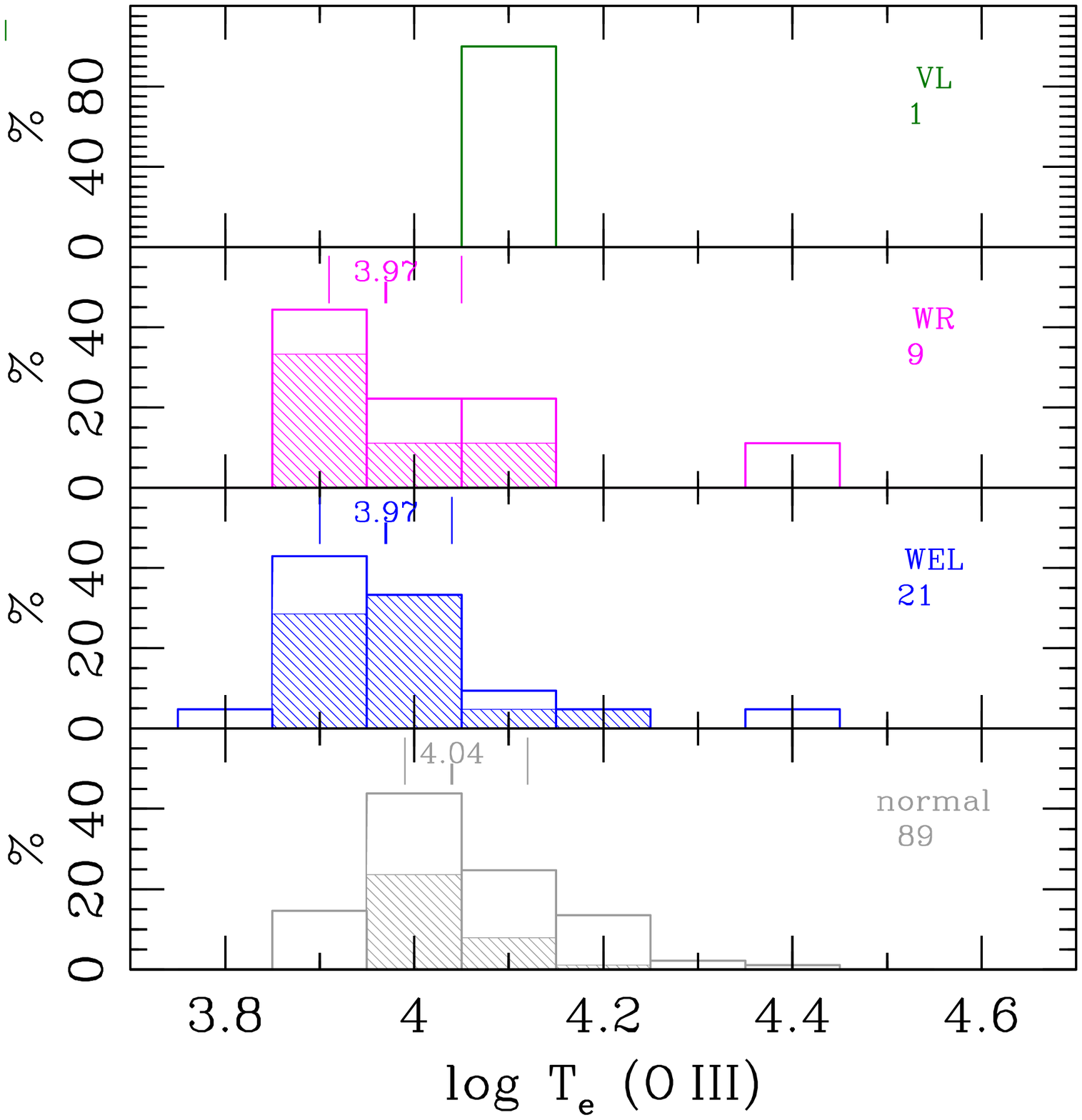}}
\caption[]{
  Distributions of electron temperatures derived from \rNii\ (left) and
  \rOiii\ (right) line ratios for Galactic bulge VL (i.e. [WC\,11]-like),
  [WR], WEL and normal
  PNe  -- green, magenta, blue and grey histograms from top to bottom.
  The shaded bars in each histogram represent 
  the best quality determinations.
  (Objects with errors $>$0.3\,dex are not included.) The
  median values, the 25 and 75 percentiles are marked with three short
  vertical lines above each histogram.  Total numbers of objects used 
  are shown in the right-hand parts of the panels below sample names.
}
\label{hist_T}
\end{figure*}

In \figref{ne_diam}\  we presented the derived electron densities of the
observed nebulae. It was quite apparent from that figure that taking nebulae
of a given nebular diameter (that can be regarded as an indicator of
evolutionary advancement) all the nebulae with emission-line CSs seemed to
be denser than their normal counterparts.

Another important parameter of the plasma - electron temperatures are
presented in \figref{hist_T}. It displays the distributions of \Te(N~{\sc
ii}) and \Te(O~{\sc iii}) temperatures derived respectively from \nii\ lines and from \oiii\
line ratios for all the four groups of the \bb\ sample PNe. The inspection
of histograms in the right panel reveals that in the case of \WRPNe\
and \WELPNe\ the distributions of \Te(O~{\sc iii}) temperature differ from what is
derived for normal PNe. There seem to be more objects with lower
temperatures in the former two groups.  For the \WELPNe\ the
Kolmogorov-Smirnov and Wilcoxon nonparametric tests confirm that this
difference is statistically significant while in the case of \WRPNe\ it may
still be a result of random effects due to the small number of objects in the
sample.

In \figref{hist_T}\ (left panel) it can be seen that the \Te(N~{\sc
ii}) of \VLPNe\ tend to be substantially lower than for other PNe. The
median value of log\,\Te(N~{\sc ii}) is 3.83 for \VLPNe\ as compared to
4.02 for the normal PNe. On the other hand, the \Te(N~{\sc ii}) temperatures
of \WRPNe\ and \WELPNe\ show near perfect similarity to the distribution of
the normal PNe.

\begin{figure}
\resizebox{0.95\hsize}{!}{\includegraphics{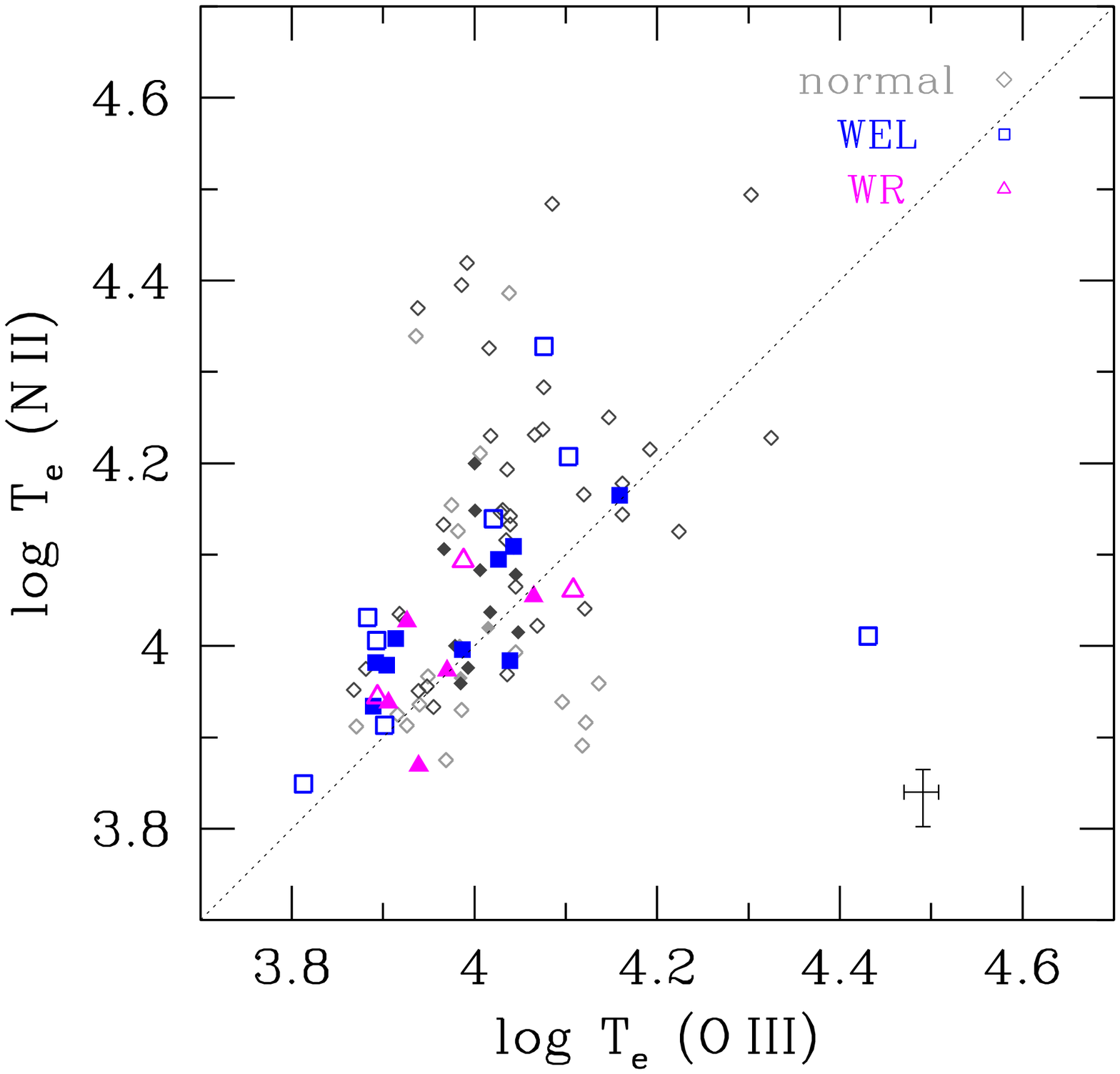}}
\caption[]{
  The electron temperature derived from \rOiii\ versus the electron
  temperature derived from \rNii\ line ratios for Galactic bulge \WRPNe\
  (magenta triangles), \WELPNe\ (blue squares) and normal PNe 
  (diamond, high ionization -
  black and low ionization - grey, defined as for \figref{oppo_hepphe}).
  Filled symbols mark data of best quality with derived errors below the median
  error of PNe in the combined bulge sample (indicated with error-bars cross).
}
\label{wr_to_tn}
\end{figure}

Figure \ref{wr_to_tn} compares \Te(N~{\sc ii}) and \Te(O~{\sc iii}) for PNe
where both electron temperatures could be derived. In general we see from
this plot that both values are relatively well correlated as could be
expected from the similarity of histograms in the bottom left and right panel of
\figref{wr_to_tn}. Despite that, there is a tendency for higher
\Te(N~{\sc ii}) with respect to \Te(O~{\sc iii}) values for PNe with higher
levels of oxygen ionization (black diamonds, \oppo$>$0.8 -- see
\figref{oppo_hepphe}). The WEL PNe show  similarity with this group
as practically all plotted objects display \Te(N~{\sc ii})$>$\Te(O~{\sc
iii}). For the \WRPNe\ the situation is not clear as most of them seem to be
located much closer to the one-to-one relation plotted with the dotted line in
\figref{wr_to_tn}.

In the case of VL PNe the \Te(O~{\sc iii}) is available for only one PN as due
to their extremely low ionization the oxygen is only singly ionized or lines
are too faint to be measured. The locations of grey symbols in
\figref{wr_to_tn} indicate that the lower values of \Te(N~{\sc ii}) are a
general tendency of low ionization PNe. Nonetheless, the distributions
observed in the left panel of \figref{hist_T} clearly demonstrate that
\Te(N~{\sc ii}) temperatures of VL PNe are lower than in any other group of
bulge PNe.

\subsection{Chemical abundances}

\subsubsection{The O/H ratio}

\begin{figure}
\resizebox{0.95\hsize}{!}{\includegraphics{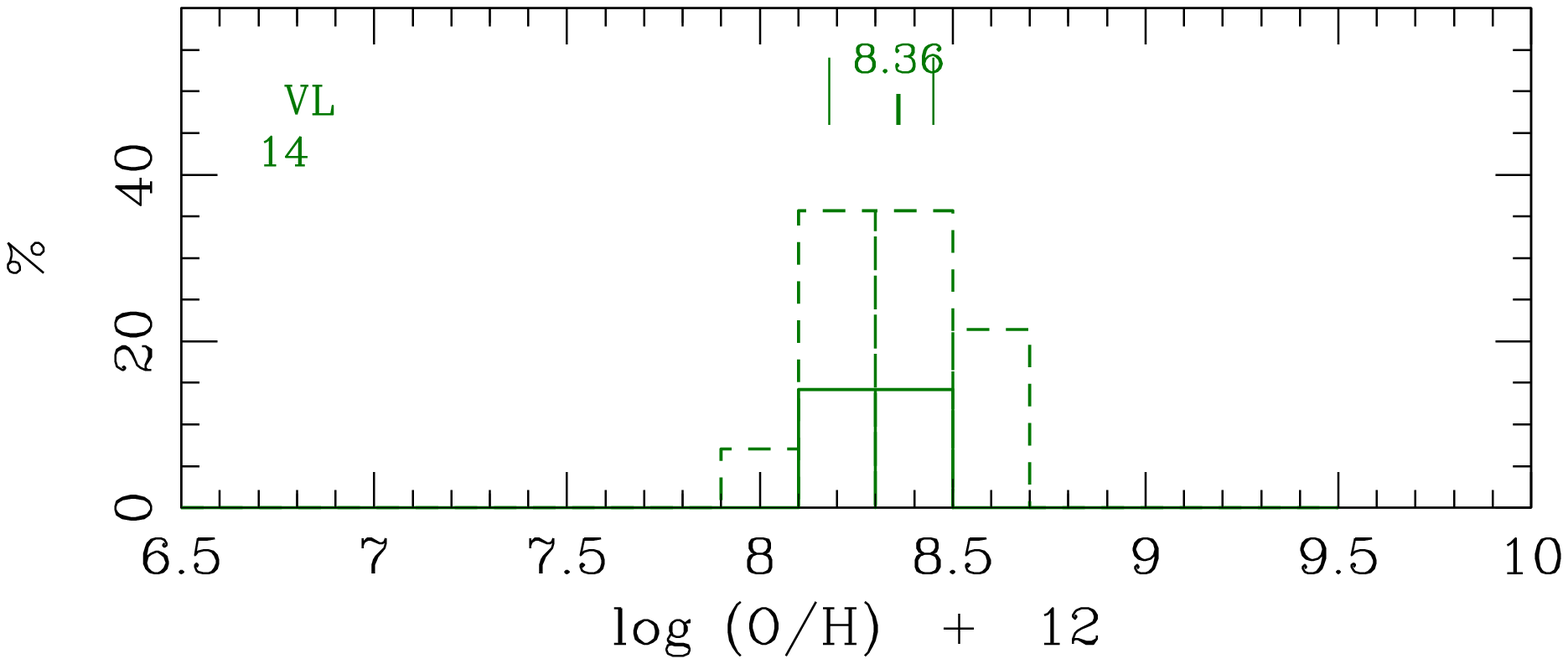}}

\resizebox{0.95\hsize}{!}{\includegraphics{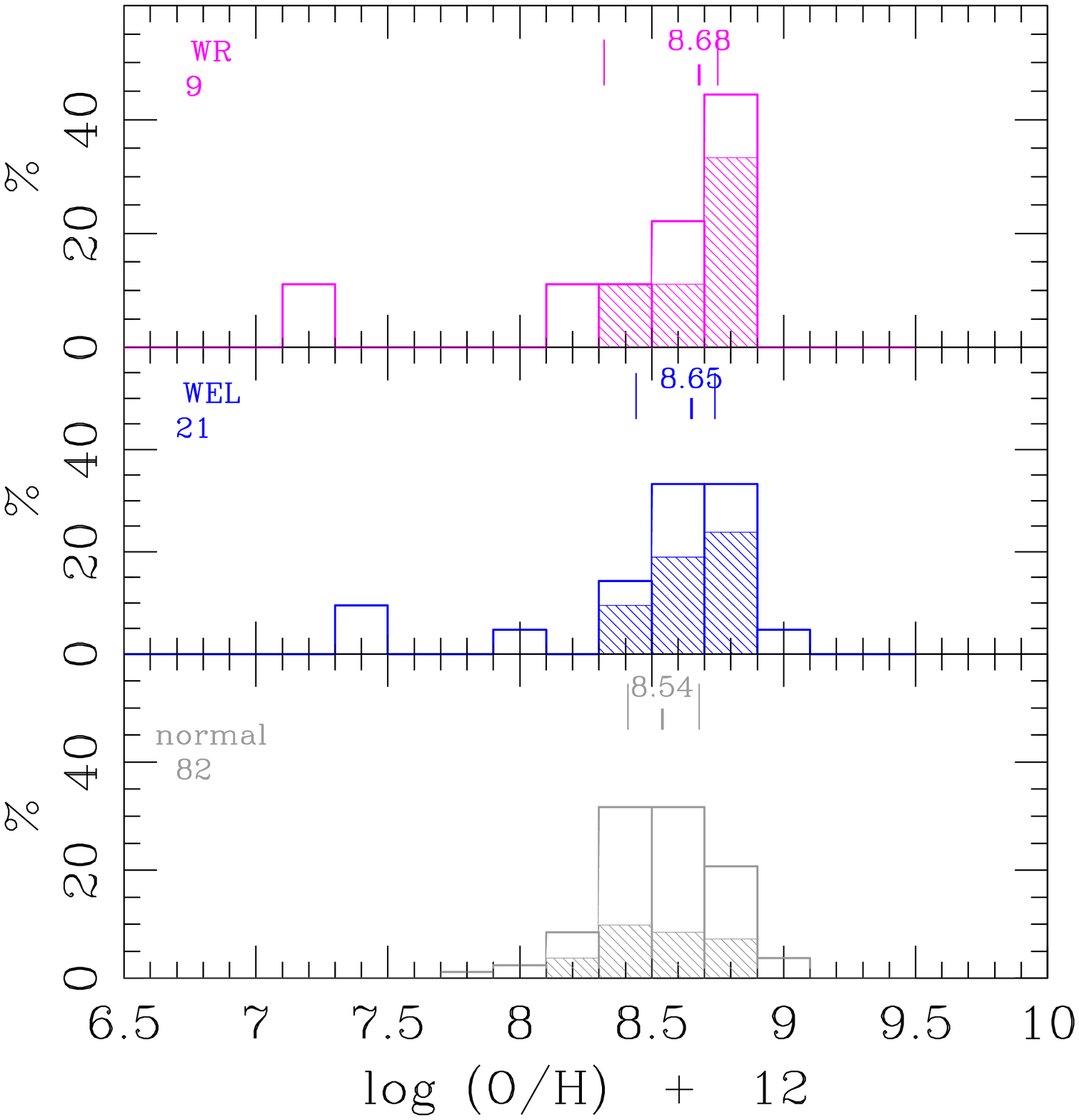}}
\caption[]{
  Distributions of O/H abundance ratio for Galactic bulge \VLPNe\
  ([WC\,11]-like spectra), \WRPNe, \WELPNe\ and
  normal PNe. The same notation as in \figref{hist_T} except for the dashed
  histogram of VL PNe that presents the distribution including all available data
  while the solid line histogram marks the part derived with data of
  satisfactory quality (errors $<$0.3\,dex).
}
\label{hist_oh}
\end{figure}

In \figref{hist_oh} we present the histograms of the oxygen abundance.
This element is believed to be generally unmodified during the life of
intermediate mass stars,  reflecting the primordial composition of the
matter from which the star was born (see Stasi\'nska 2007, Chiappini et al.
2009).  The four histograms present the distributions for the normal, [WR],
WEL and VL PNe subsamples analyzed here. The hatched parts of histograms
represent the component of the distribution derived from the data of best
quality -- the individual errors are smaller than the median of errors of a
given parameter in the  \bb subsample (i.e. non-emission and the emission-line
central star PNe combined). The remaining objects have errors not larger
that 0.3 dex as results with larger uncertainties have been rejected. An
exception has been made in the case of the distribution of VL PNe (dashed
line, top panel of \figref{hist_oh}) where we have used all the available
data regardless of this condition.

\begin{figure*}
\resizebox{0.70\hsize}{!}{\includegraphics{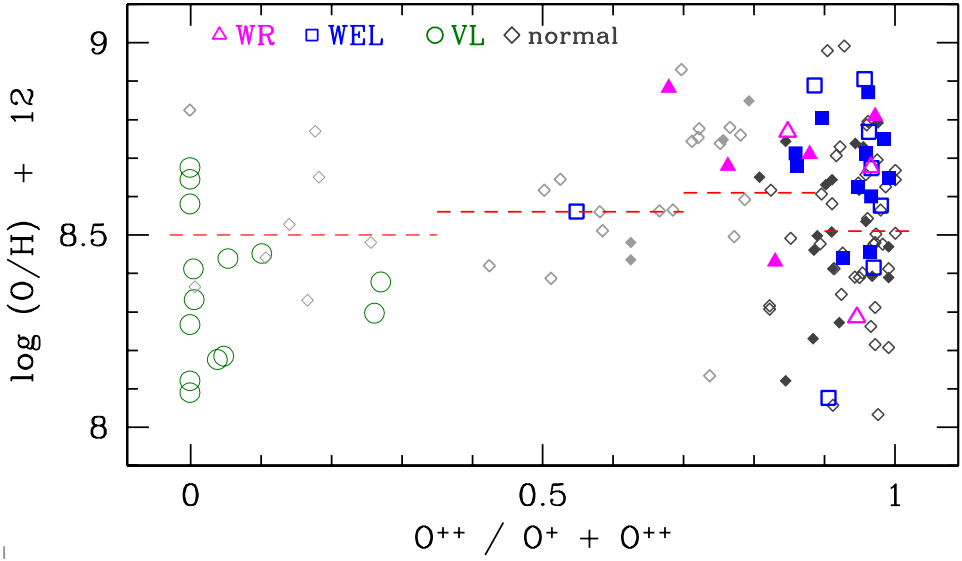}}
\caption[]{
  The relation of O/H abundance ratio with ionization parameter \oppo\ for
  Galactic bulge PNe. Different symbols mark: \WRPNe\ -- magenta 
  triangles; \WELPNe\ --
  blue squares; \VLPNe\ ([WC\,11]-like spectra) -- green 
  circles; normal PNe -- black or grey diamonds. Filled
  symbols mark objects with best quality data. Objects with derived O/H
  quality beyond adopted rejection limit (errors$>$0.3\,dex) are represented
  with open thin-line symbols. Red dashed lines mark the median O/H values
  for normal PNe divided into four bins of \oppo\ parameter: $<$0.35;
  0.35--0.7; 0.7--0.9 and $>$0.9. (Note, three low-metallicity PNe 
  fall below the
  presented range of O/H - see Section 7.5.4 and \figref{oppo_abund},
  available on-line.)
}
\label{oppo_oh_det}
\end{figure*}

The oxygen abundances of VL PNe seem lower than in any other group of
PNe\footnote{
  Due to the small number of objects the difference cannot be
  confirmed by Kolmogorov-Smirnov and Wilcoxon nonparametric tests to be
  statistically significant.
}. One can raise the question of whether the extreme low ionization of these objects
does not contribute to this fact as some atoms may remain neutral and
unobserved. In \figref{oppo_oh_det} one can see that other nebulae with
equally low oxygen ionization level (\oppo$<$0.35) apparently have  higher
O/H ratios than VL PNe. This comparison suffers however from the large
errors of the derived O/H ratio for these objects.

The O/H abundance in \WRPNe\ and \WELPNe\ also appears different from normal
nebulae. There are hardly any objects with log\,O/H$>$9 in either group
but the median O/H values are approximately 0.1 dex larger for
\WRPNe\ and \WELPNe. The difference is small and in reality not confirmed by
nonparametric tests. However, the statistically confirmed fact that the
\Te(O~{\sc iii}) temperatures are lower in \WRPNe\ and \WELPNe\ (see above) is in
accordance with a higher oxygen abundance in these objects as it could lead to
more efficient cooling of the nebular plasma.

Nonetheless, there are some possible systematic effects that have to be
considered. From \figref{oppo_oh_det} it seems evident that a comparison
limited only to the PNe with \oppo$>$0.9 would
leave no doubt that the derived O/H ratios for \WELPNe\ are higher than for the
normal PNe (note their locations above the dashed line marking the median
value). One has to be aware though of the effects brought in by the use
of ICFs. For the O/H ratio the ICF should correct for the possible 
presence of ions more charged than  O$^{2+}$ and has the form:

\begin{center}
ICF=((He$^+$ + He$^{2+}$) / He$^{+}$) $^{2/3}$.
\end{center}

For the \WELPNe, as well as for the \WRPNe, due to the fact that there is
very little or no He$^{2+}$ detected, the ICF will be close to or equal to 1. This
means that no correction is necessary as lines of all oxygen ions are accessible
in optical spectra. For the other nebulae however any systematic error in
the form of ICF applied will have a direct impact on the derived O abundance
and therefore on the conclusions we infer about possible differences between
normal PNe and subtypes with emission-line CSs.

An additional factor that may add some uncertainty  is the already mentioned
(Section 4) problem of a difference between O$^+$ abundances derived from
\oii\ $\lambda$3727 and \oii\ $\lambda$7325. For the total oxygen abundance
in WEL PNe this problem may be ignored as most of the oxygen is in the form
of doubly ionized O$^{2+}$. But for the PNe with lower ionization it is
important especially if the electron temperatures are low presenting
favorable conditions for a significant contribution from recombination.

\begin{figure}
\resizebox{0.95\hsize}{!}{\includegraphics{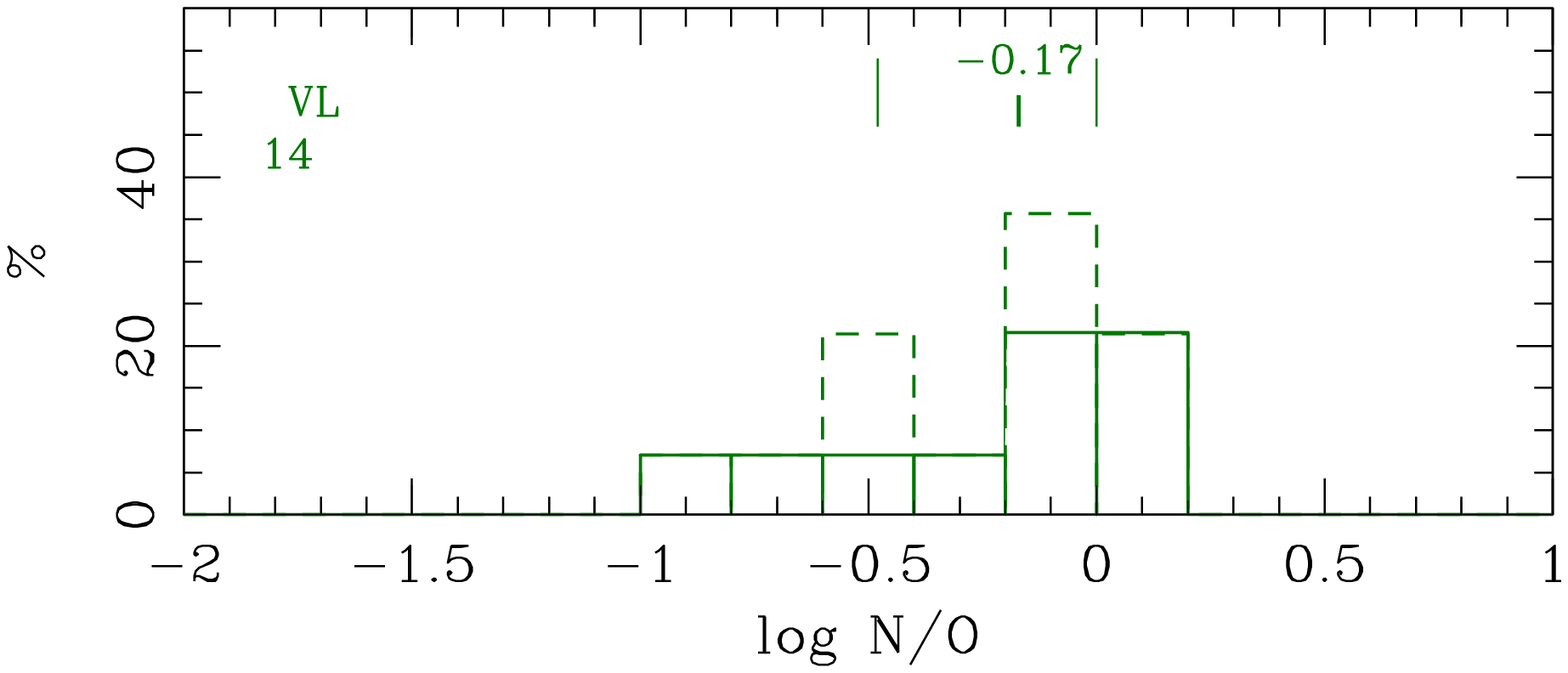}}

\resizebox{0.95\hsize}{!}{\includegraphics{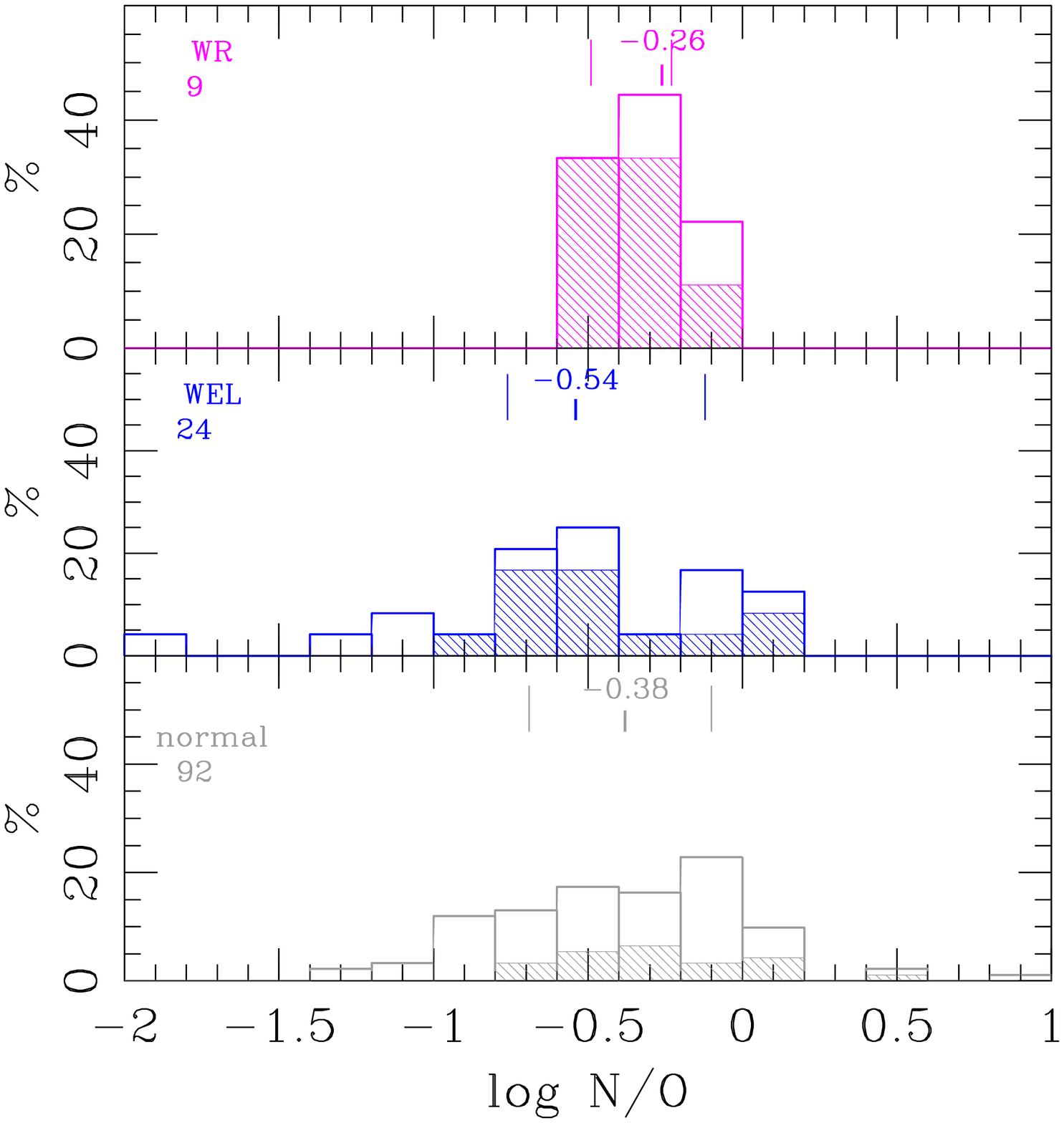}}
\caption[]{
  Distributions of N/O abundance ratio for Galactic bulge \VLPNe, 
  \WRPNe, \WELPNe\ and
  normal PNe. The same notation as in \figref{hist_oh}.
}
\label{hist_no}
\end{figure}

\subsubsection{The N/O and He/H abundance ratios}

In \figref{hist_no} the distributions of log\,N/O for the four subsamples of
bulge PNe are presented. The nonparametric tests do not indicate strong,
statistically significant differences between them. The derived median N/O
value for WEL PNe is smaller than for normal PNe but the distributions have
similar width and range. It should be noted also that this abundance ratio
is derived assuming N/O=N$^+$/O$^+$, therefore the uncertainty in the O$^+$
could directly influence the nitrogen abundances.

\begin{figure}[t]
\resizebox{0.95\hsize}{!}{\includegraphics{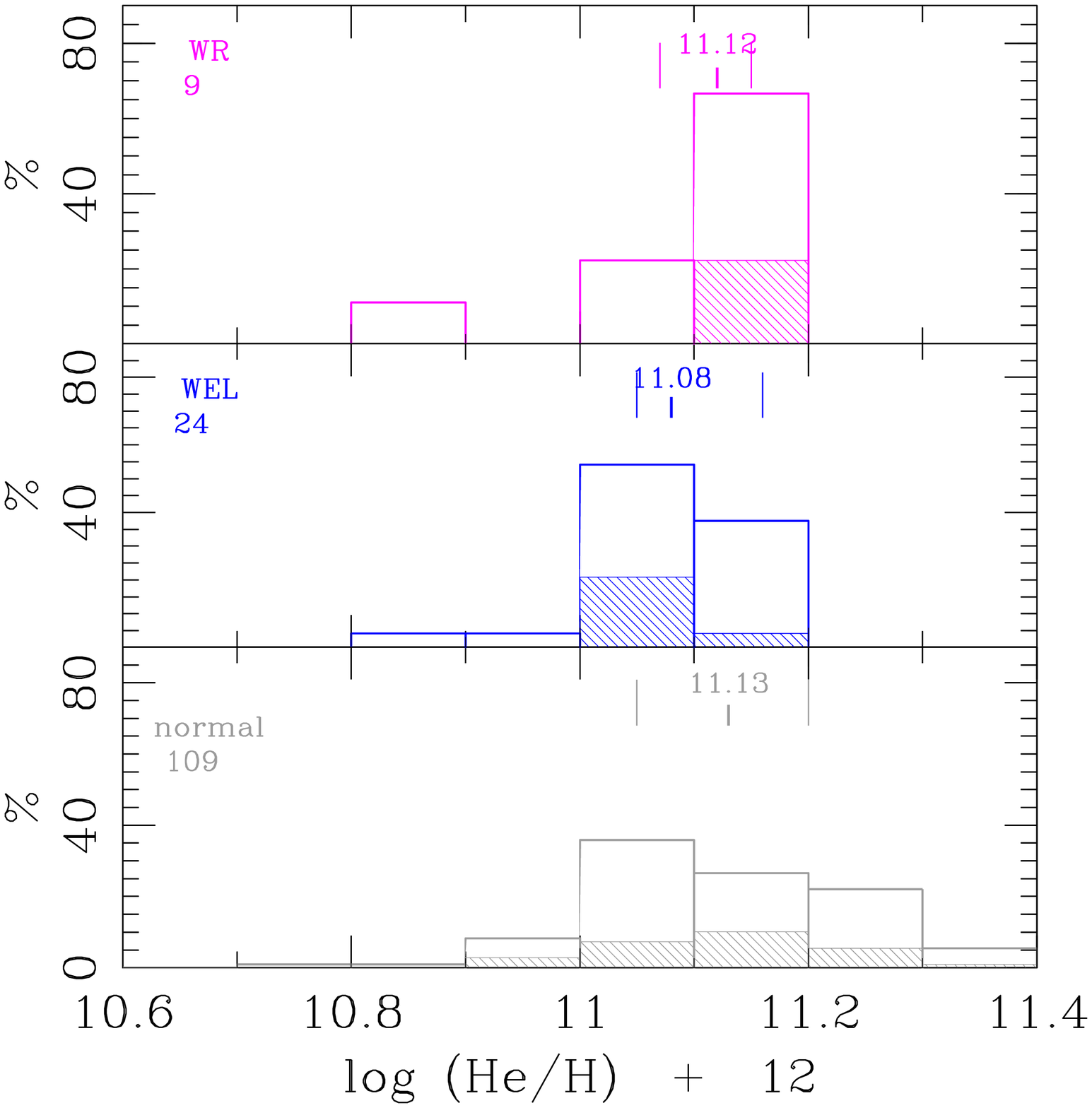}}
\caption[]{
  Distribution of He/H abundance ratio for Galactic bulge [WR], WEL and
  normal PNe. The same notation as in \figref{hist_T}.
}
\label{hist_heh}
\end{figure}

\begin{figure}[t]
\resizebox{0.95\hsize}{!}{\includegraphics{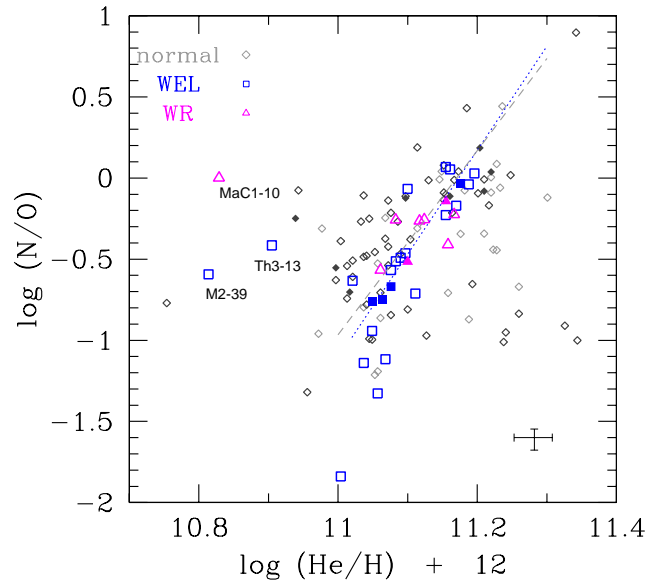}}
\caption[]{
  N/O versus He/H for Galactic bulge [WR], WEL and normal PNe. The same
  notation of symbols as in \figref{wr_to_tn}. Lines are fits obtained by
  using WEL (short-dashed) and normal PNe (long-dashed) for objects with
  log\,N/O$>$$-$1 and He/H above Solar.
}
\label{diag_heh_no}
\end{figure}

Figure \ref{hist_heh} presents the distributions of the He/H ratio. In reality
the computed He abundances do not take into account the possible presence of
neutral helium. For this reason, the He/H abundance ratios for objects with
\oppo$<$0.4 have been excluded from further considerations as being
significantly underestimated. All VL PNe fall into this category and are
therefore absent in \figref{hist_heh}.

Applying  nonparametric tests one finds out that the distributions presented
in \figref{hist_heh} are not distinguishable in a statistically meaningful
way. It can be noticed however that neither the \WRPNe\ (but the overall
number of objects is small) nor the WEL PNe have helium abundances higher
than log\,He/H+12=11.2 whereas for almost 1/3rd of the normal PNe the helium
abundances have been found to be higher than this value. This is surprising,
since \WRCSs\ have very strong He-rich winds, and one would expect this to
be reflected in the nebulae.

Both helium and nitrogen are expected to be produced in more massive
progenitors of PNe. In \figref{diag_heh_no} we present a relation of the
abundances of both elements for [WR], WEL and normal PNe. For the latter two
groups we  also plot a linear fit to the locations of objects with
log\,N/O$>$$-$1 (but rejecting PNe with presumably underestimated
log\,He/H+12$<$10.93, i.e. below solar). As can be noticed for WEL PNe there
is a tight correlation of He and N enrichment in the parent star (dotted
line). Surprisingly, though the dispersion of points is much larger, an
almost identical correlation is derived for the normal PNe. The [WR] PNe also
seem to follow a similar relation as the WEL PNe although the number of
nebulae is too small to derive firm conclusions.

\begin{figure*}
\resizebox{0.475\hsize}{!}{\includegraphics{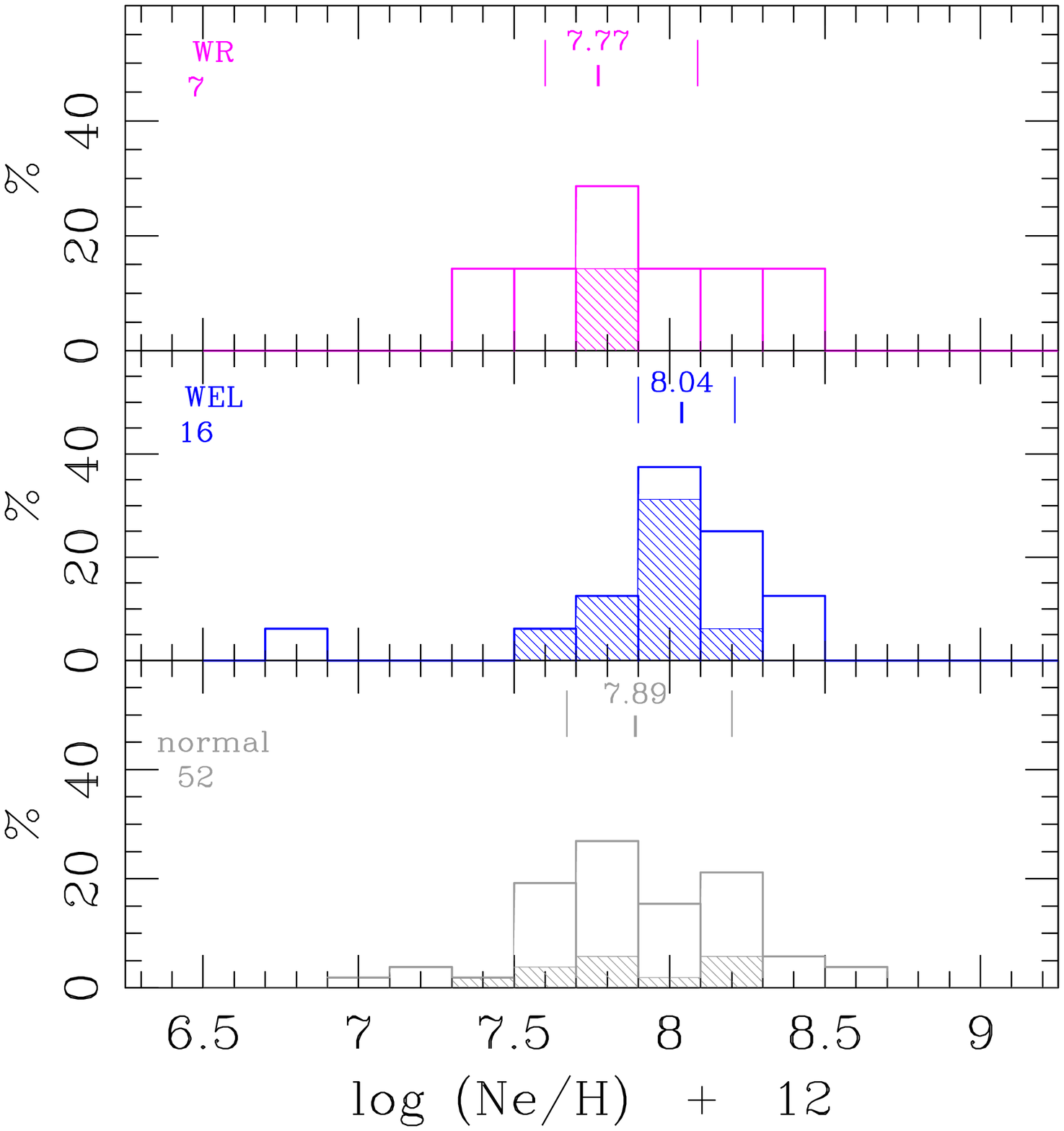}}
\resizebox{0.475\hsize}{!}{\includegraphics{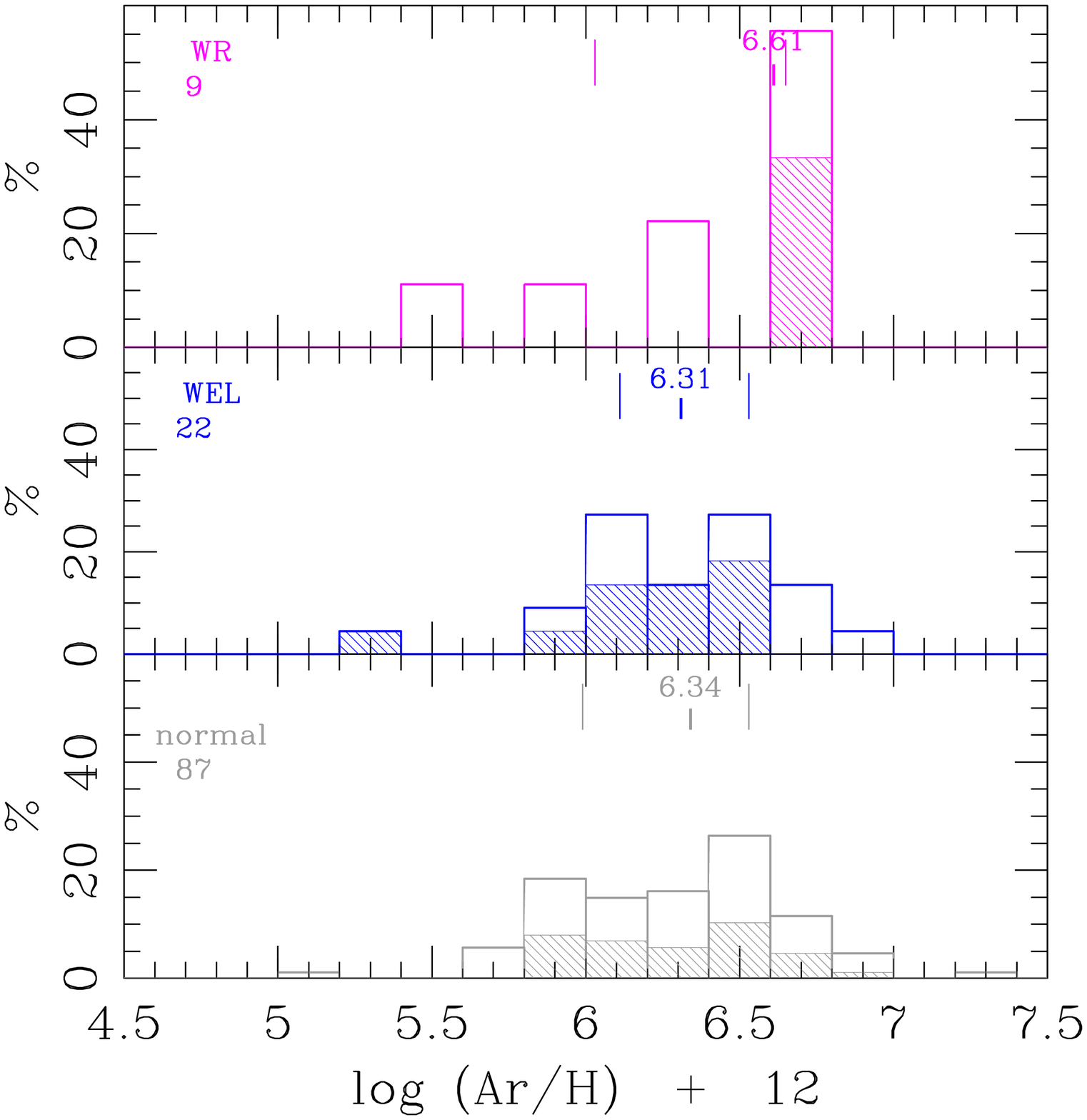}}

\resizebox{0.475\hsize}{!}{\includegraphics{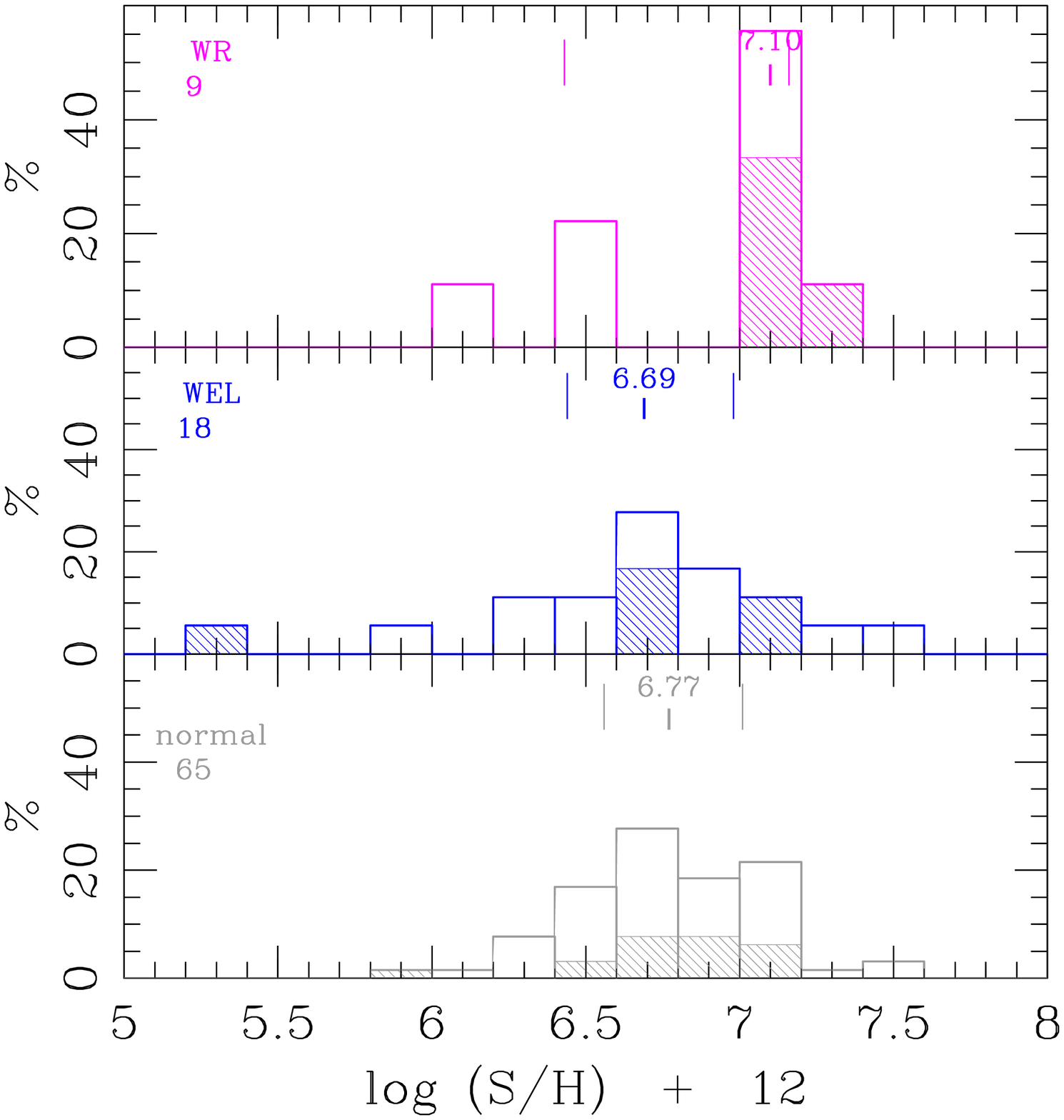}}
\resizebox{0.475\hsize}{!}{\includegraphics{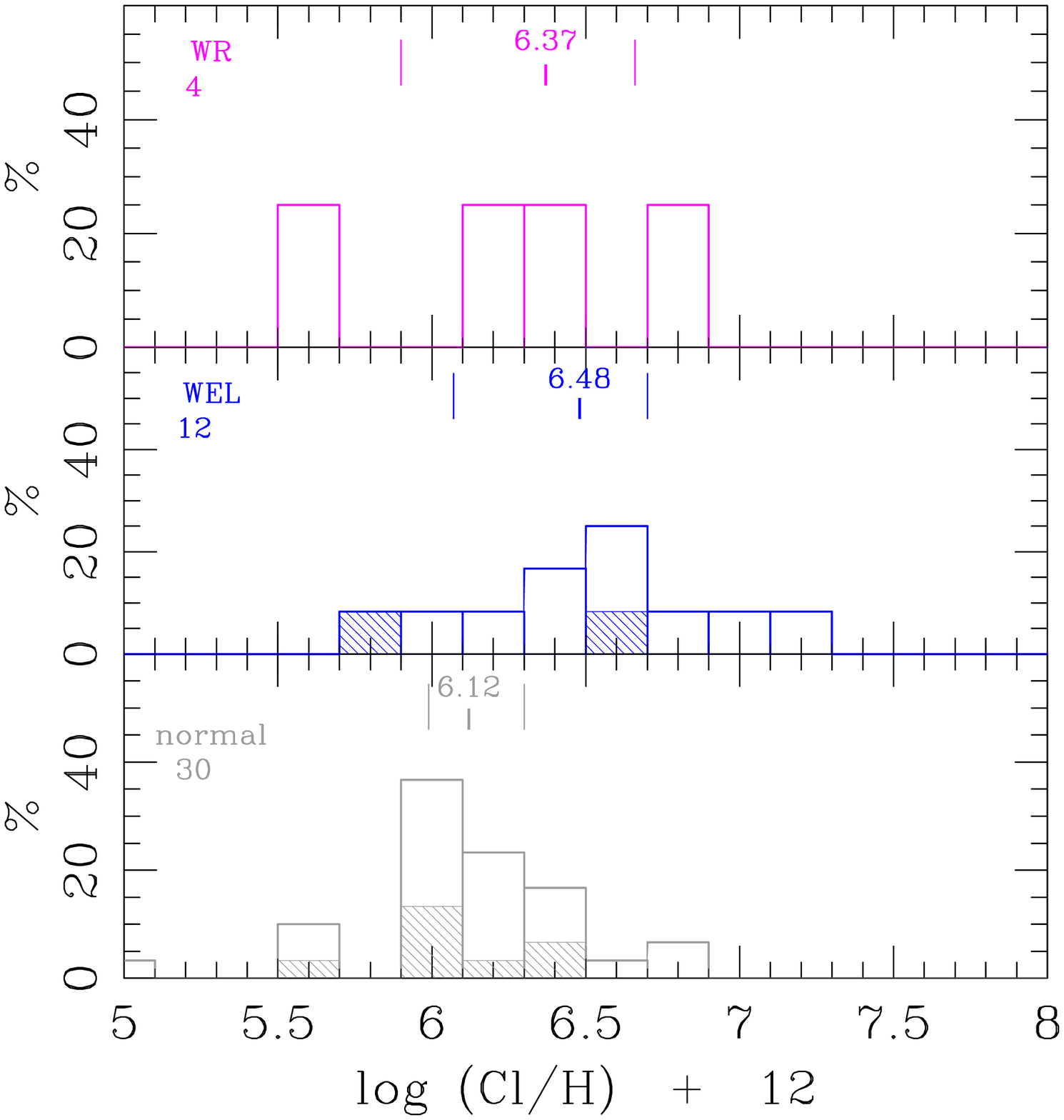}}
\caption[]{
  Distributions of neon, argon, sulfur and chlorine abundance for Galactic
  bulge [WR], WEL and normal PNe. The same notation as in \figref{hist_T}.
}
\label{hist_other_h}
\end{figure*}

\subsubsection{Abundances of other elements}

Figure \ref{hist_other_h} presents the distributions of neon, argon, sulfur
and chlorine abundances for the Galactic bulge [WR], WEL and normal PNe. The
VL PNe have been omitted because of their very low ionization, which leads
to uncertain abundances as the ICFs used are not applicable. To minimize ICF
related effects for other types of objects only nebulae with \oppo$>$0.4
have been used to derive the distributions of Ne, Ar and S abundances and
objects with \oppo$>$0.8 in the case of Cl abundance\footnote{
  Figures \ref{oppo_abund} and \ref{hepphe_abund} available on-line present
  the relations of important abundance ratios of all the bulge PNe as a
  function of \oppo( and \hepphe)
  and can be analyzed for effects of imperfect ICF correction.
}.

\onlfig{21}{
\begin{figure*}
\resizebox{0.32\hsize}{!}{\includegraphics{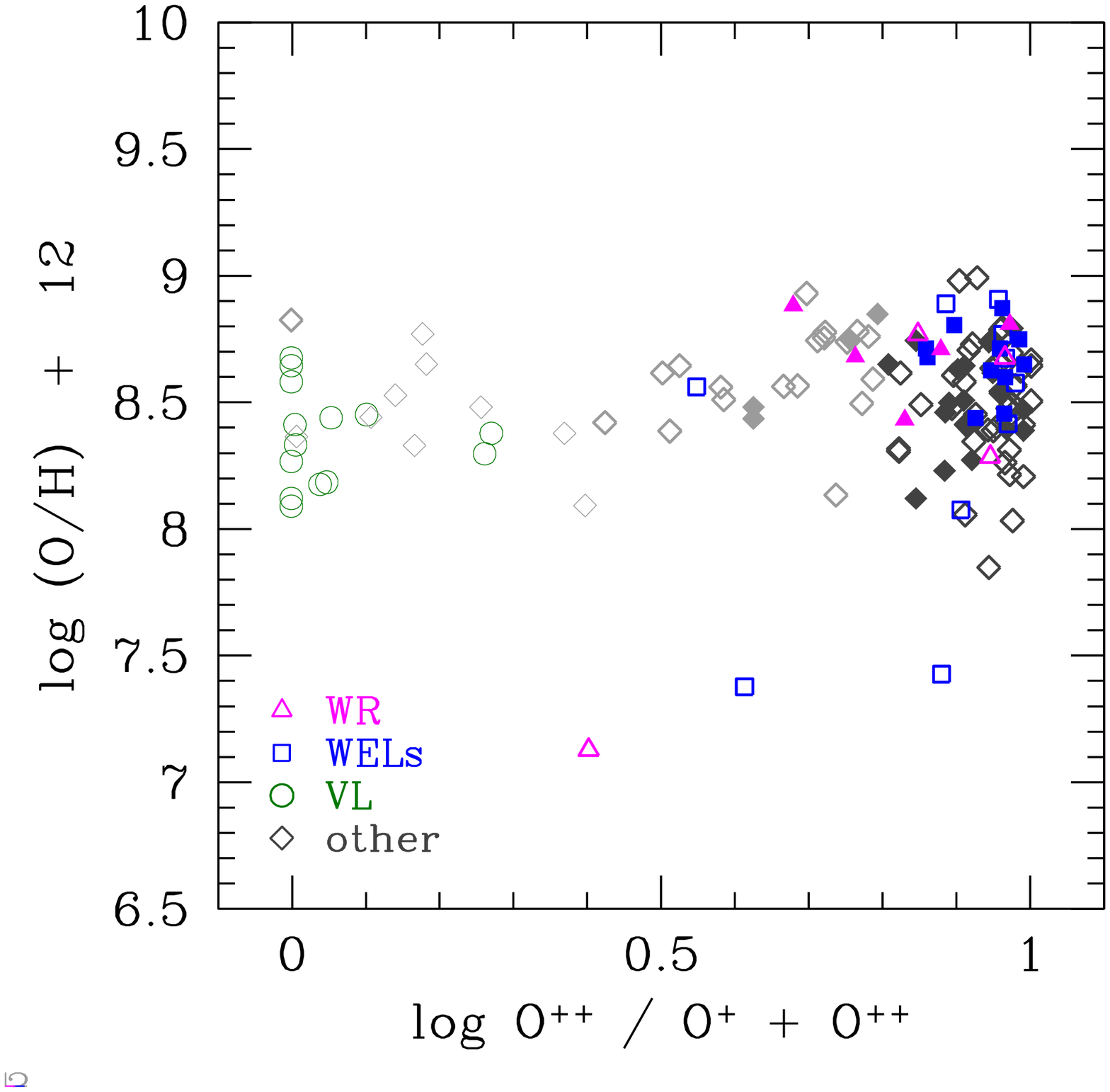}}   
\resizebox{0.32\hsize}{!}{\includegraphics{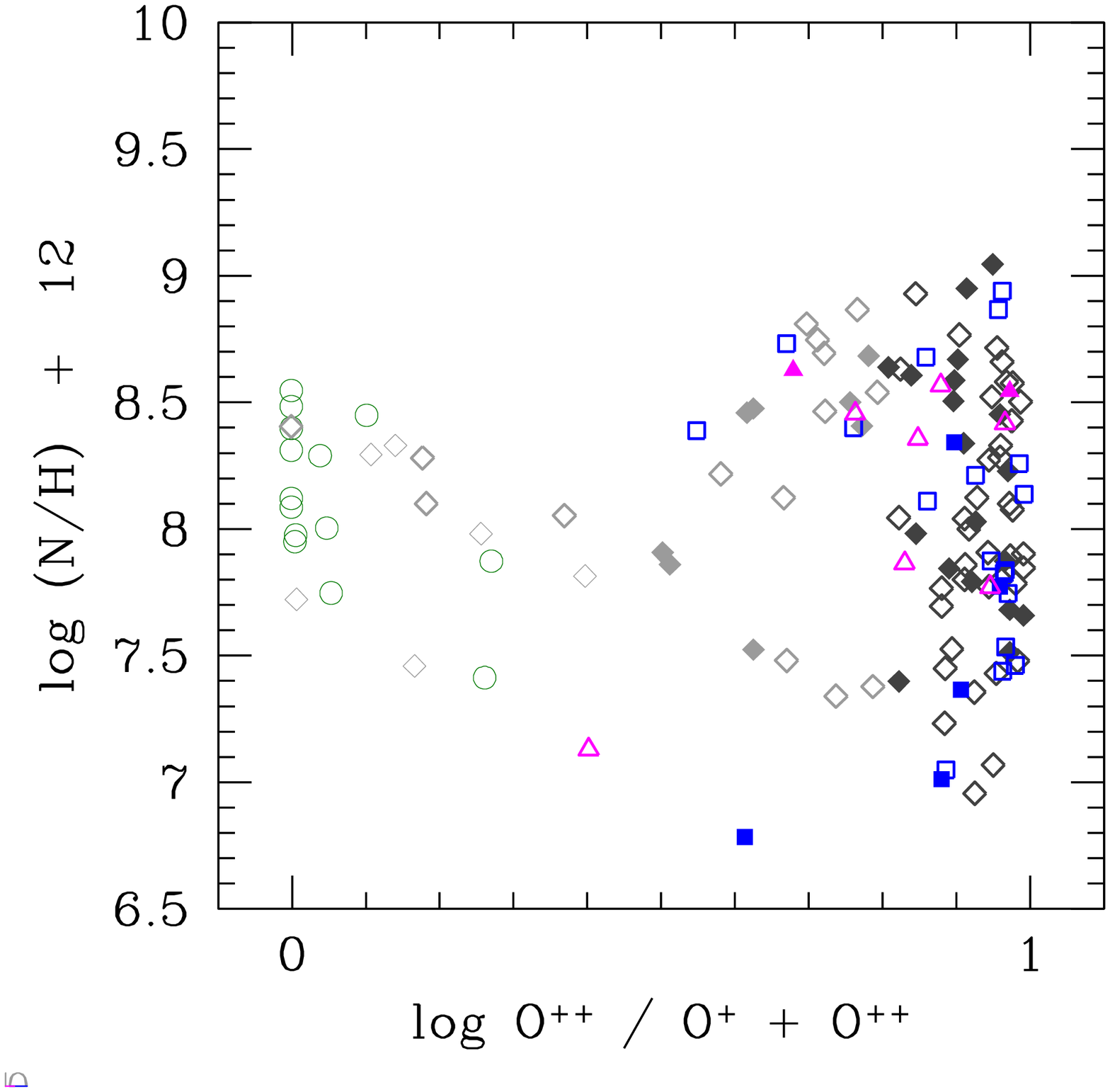}}   
\resizebox{0.32\hsize}{!}{\includegraphics{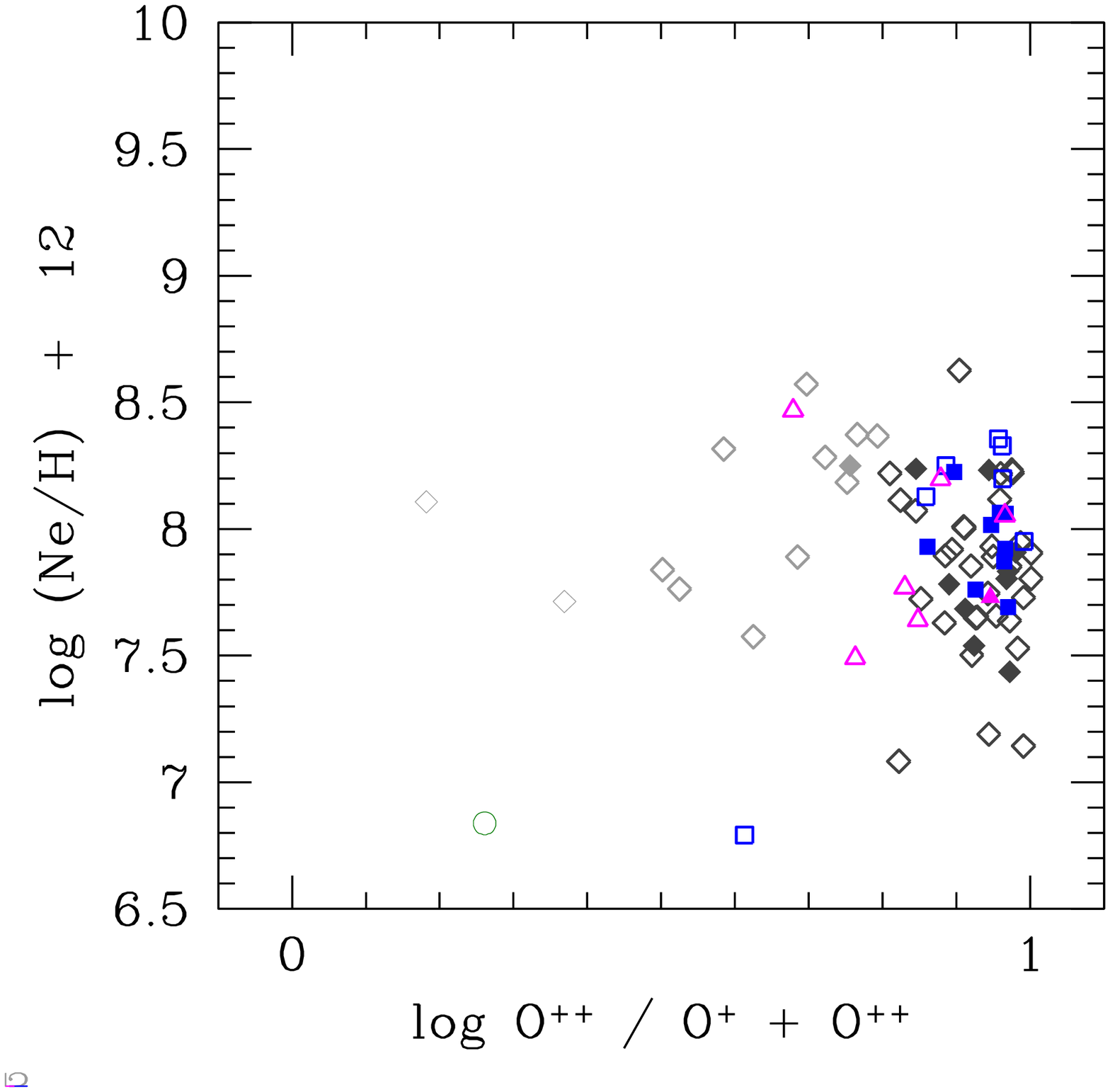}}   

\resizebox{0.32\hsize}{!}{\includegraphics{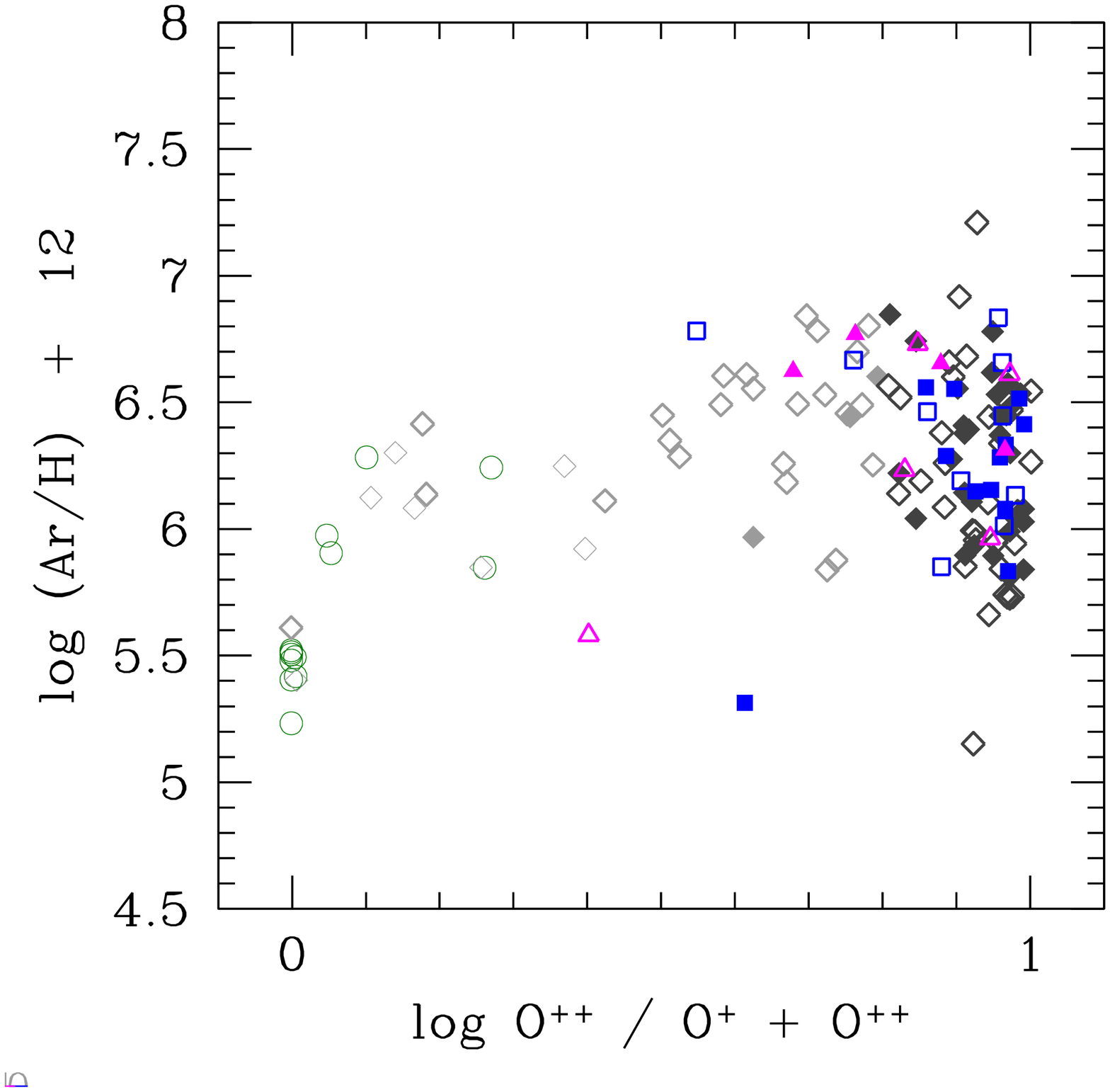}}   
\resizebox{0.32\hsize}{!}{\includegraphics{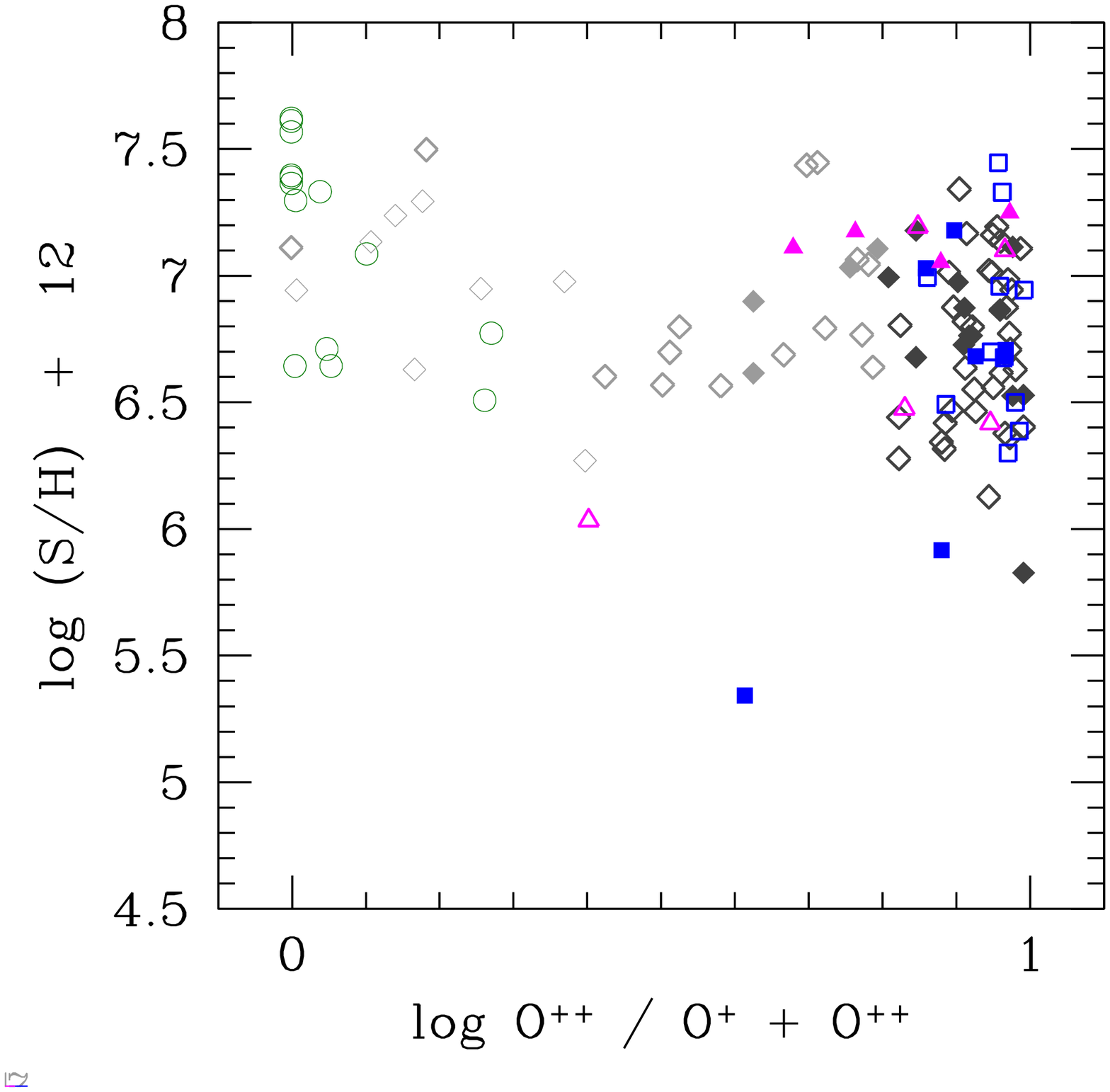}}   
\resizebox{0.32\hsize}{!}{\includegraphics{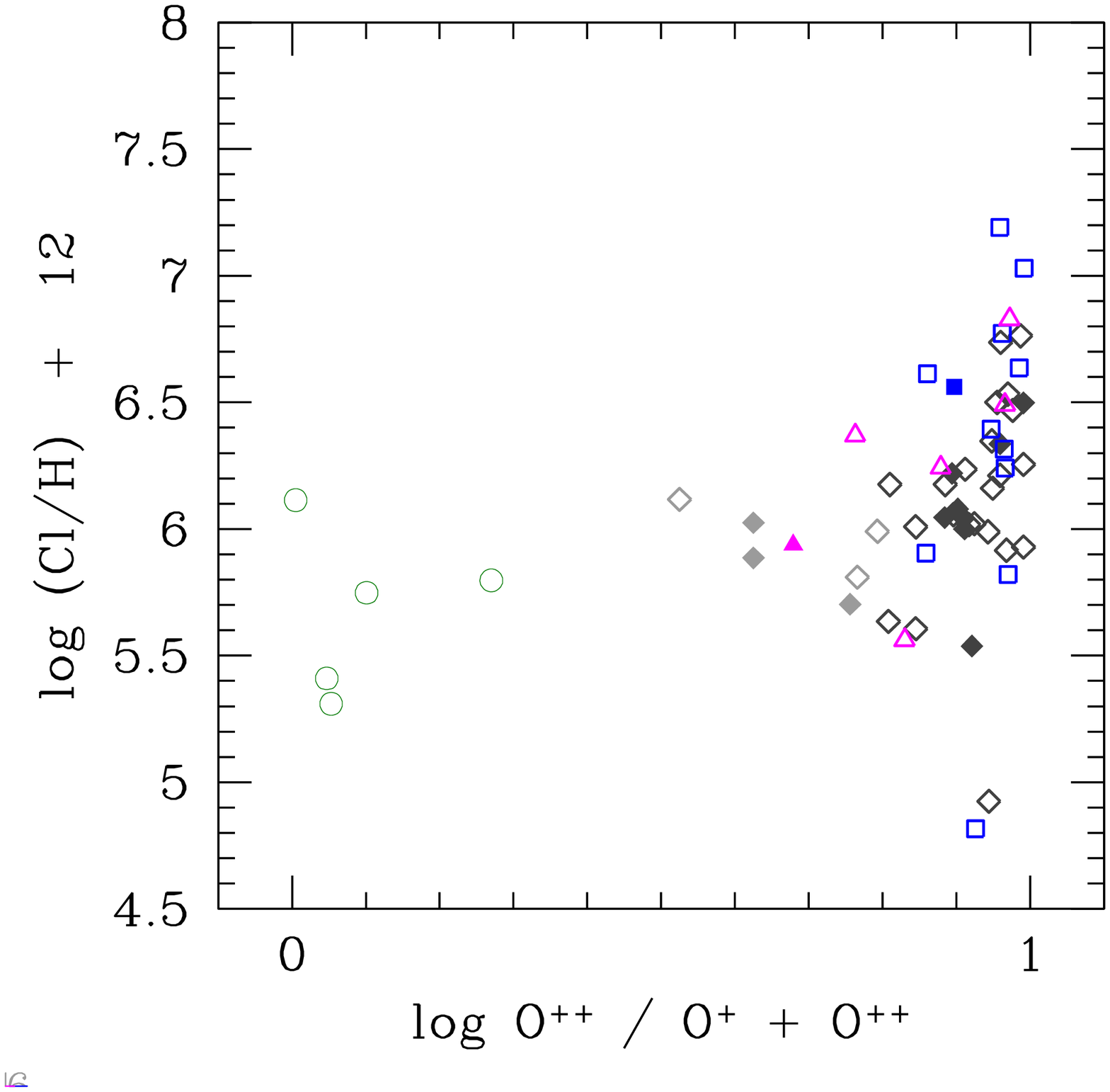}}   

\resizebox{0.32\hsize}{!}{\includegraphics{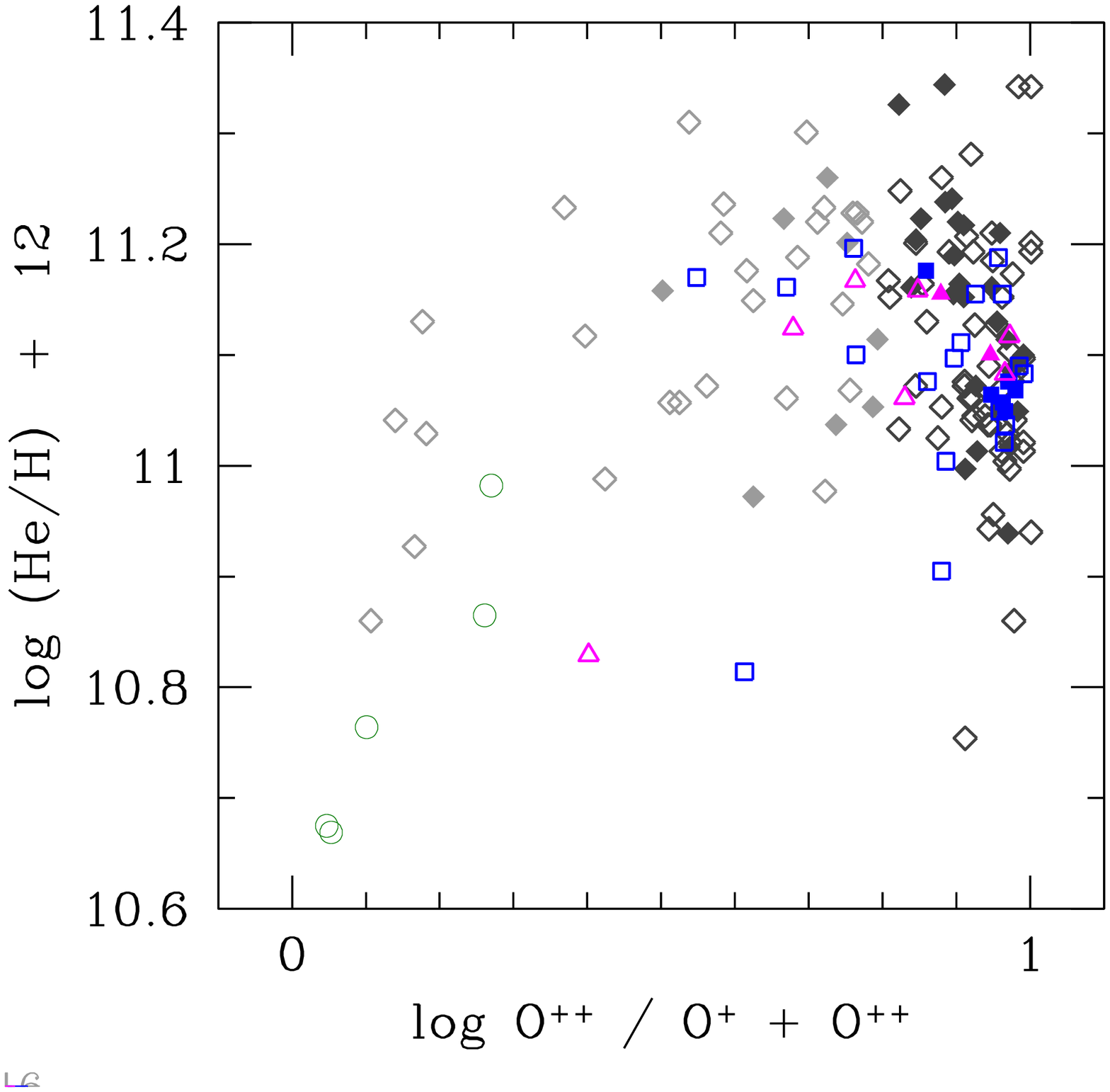}}   
\resizebox{0.32\hsize}{!}{\includegraphics{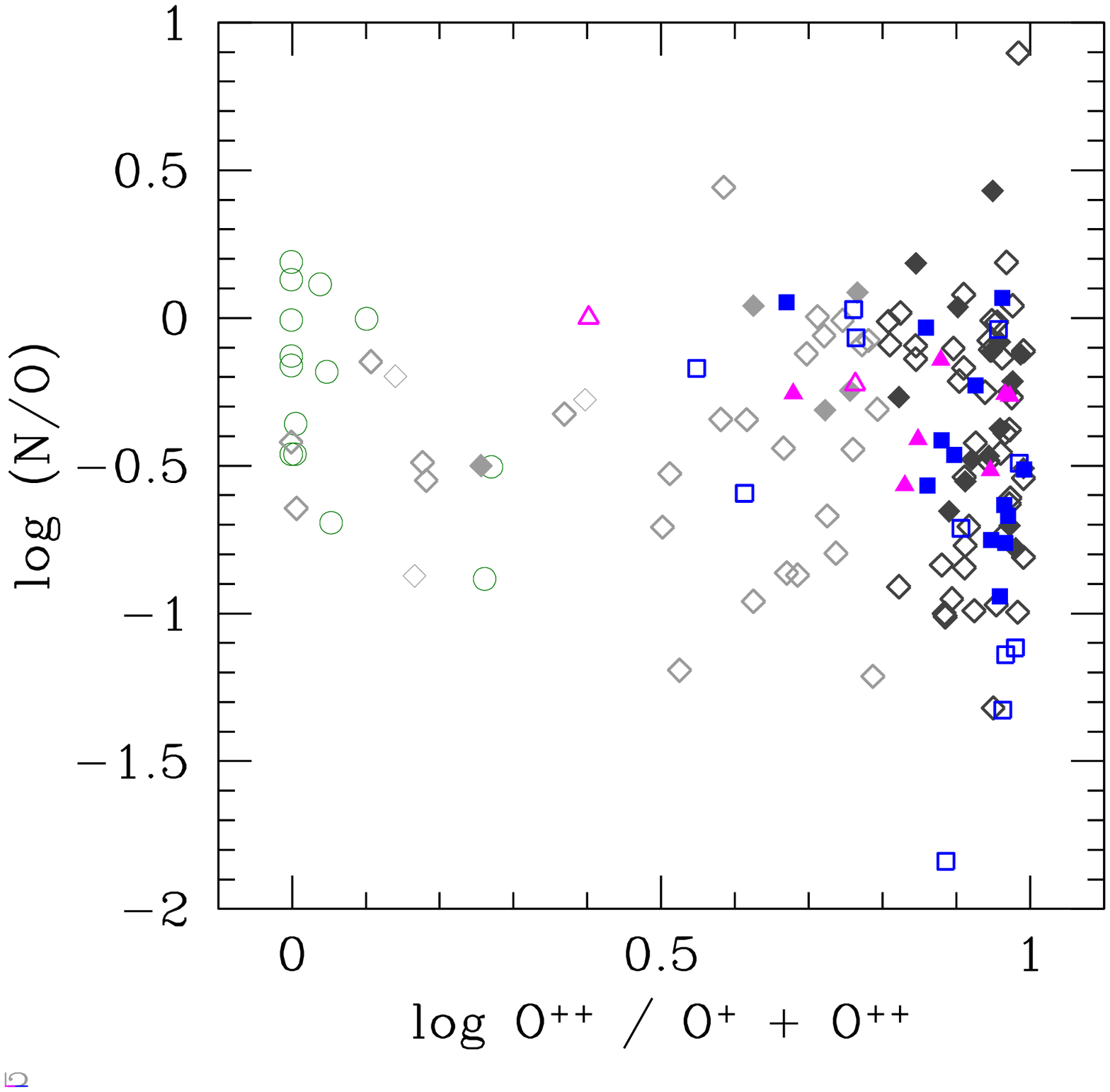}}   
\resizebox{0.32\hsize}{!}{\includegraphics{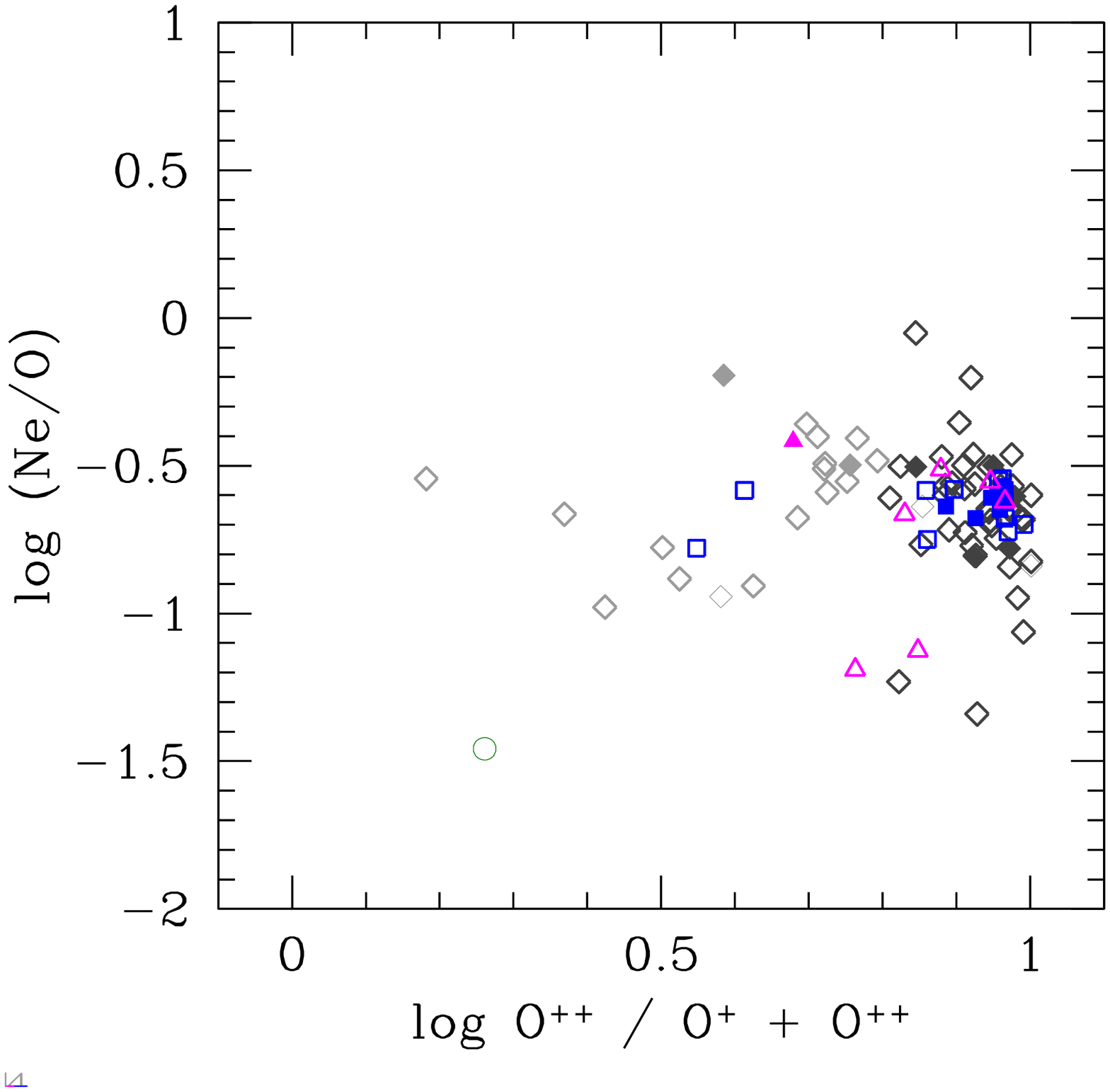}}   

\resizebox{0.32\hsize}{!}{\includegraphics{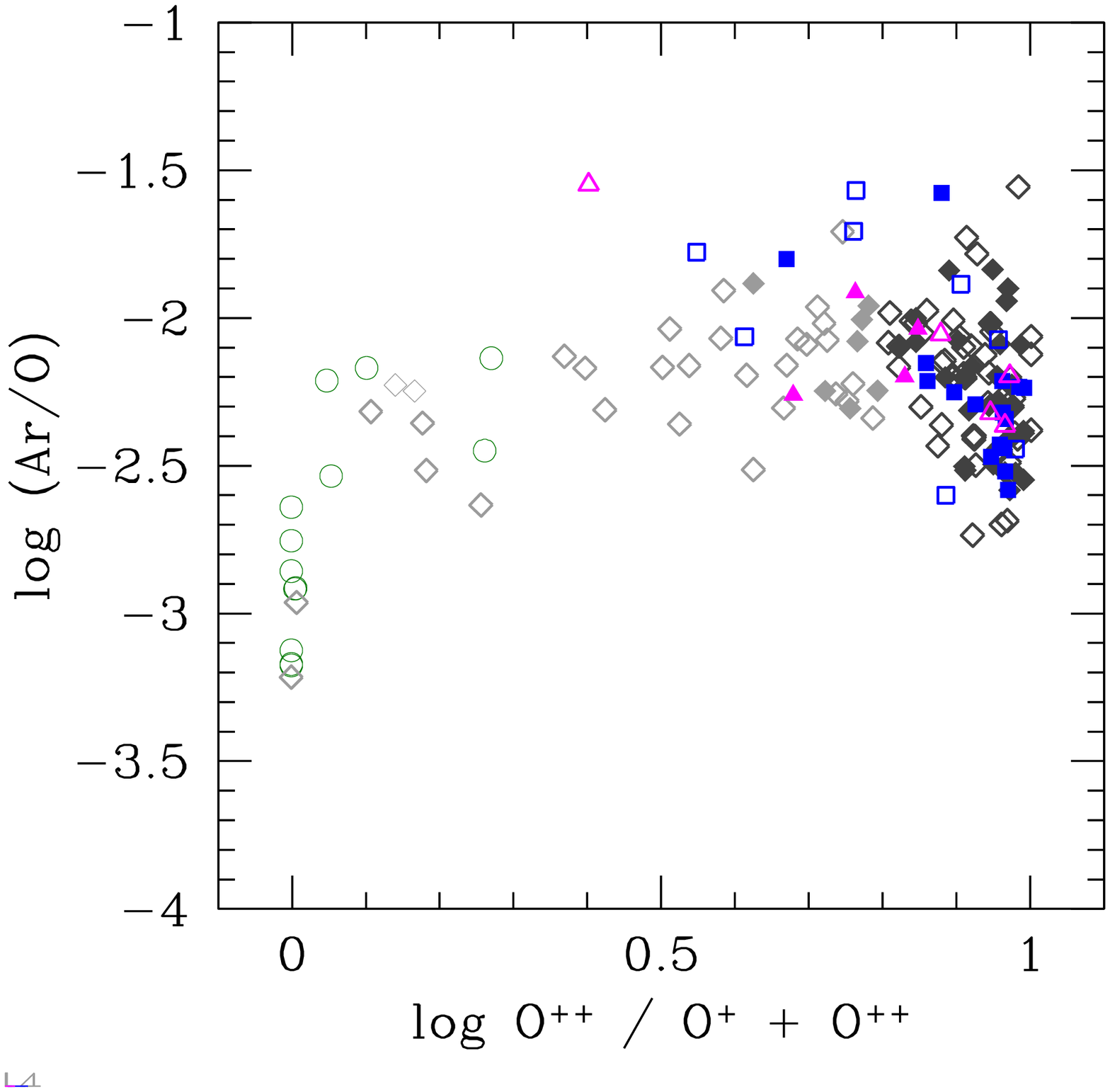}}   
\resizebox{0.32\hsize}{!}{\includegraphics{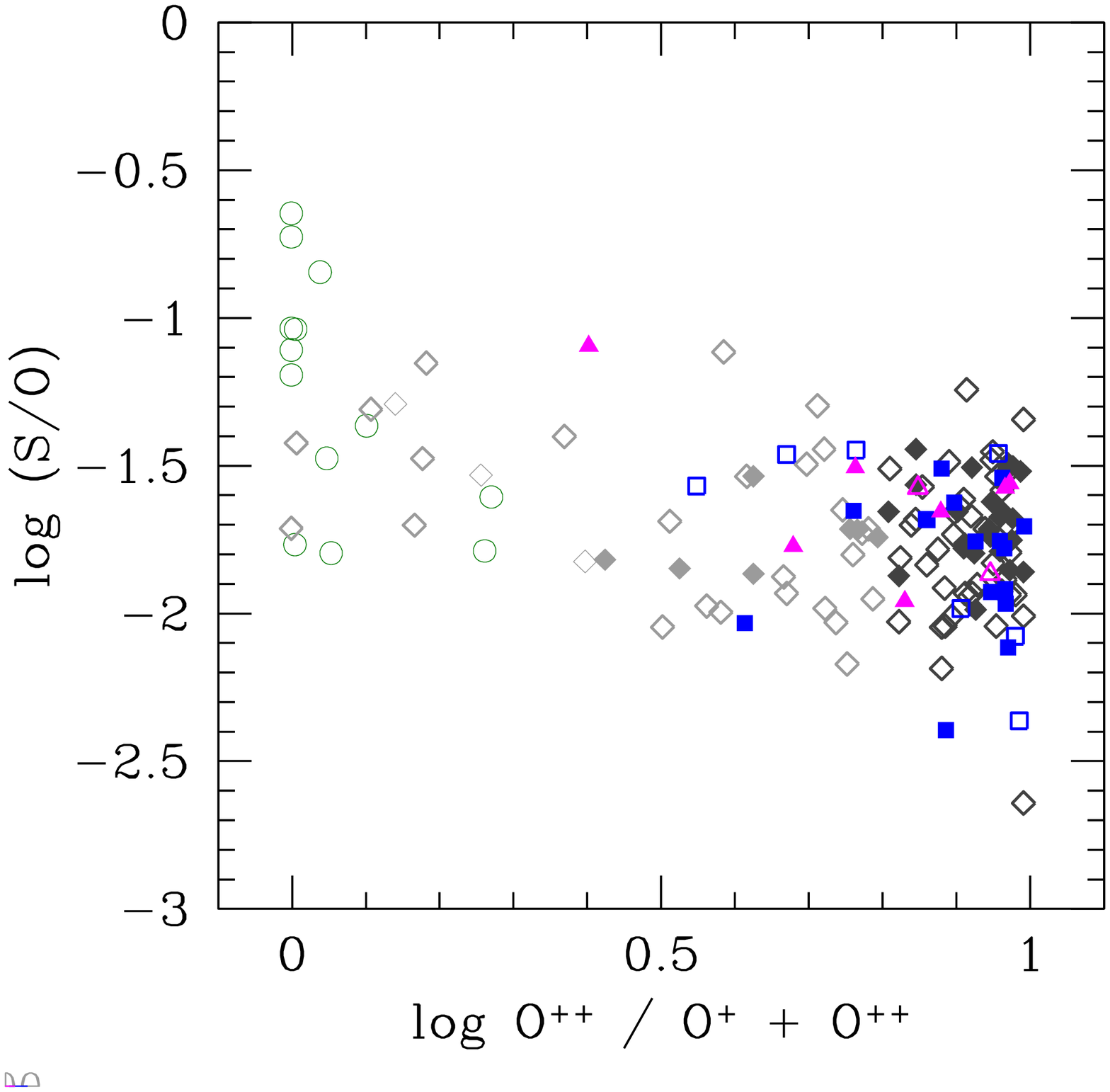}}   
\resizebox{0.32\hsize}{!}{\includegraphics{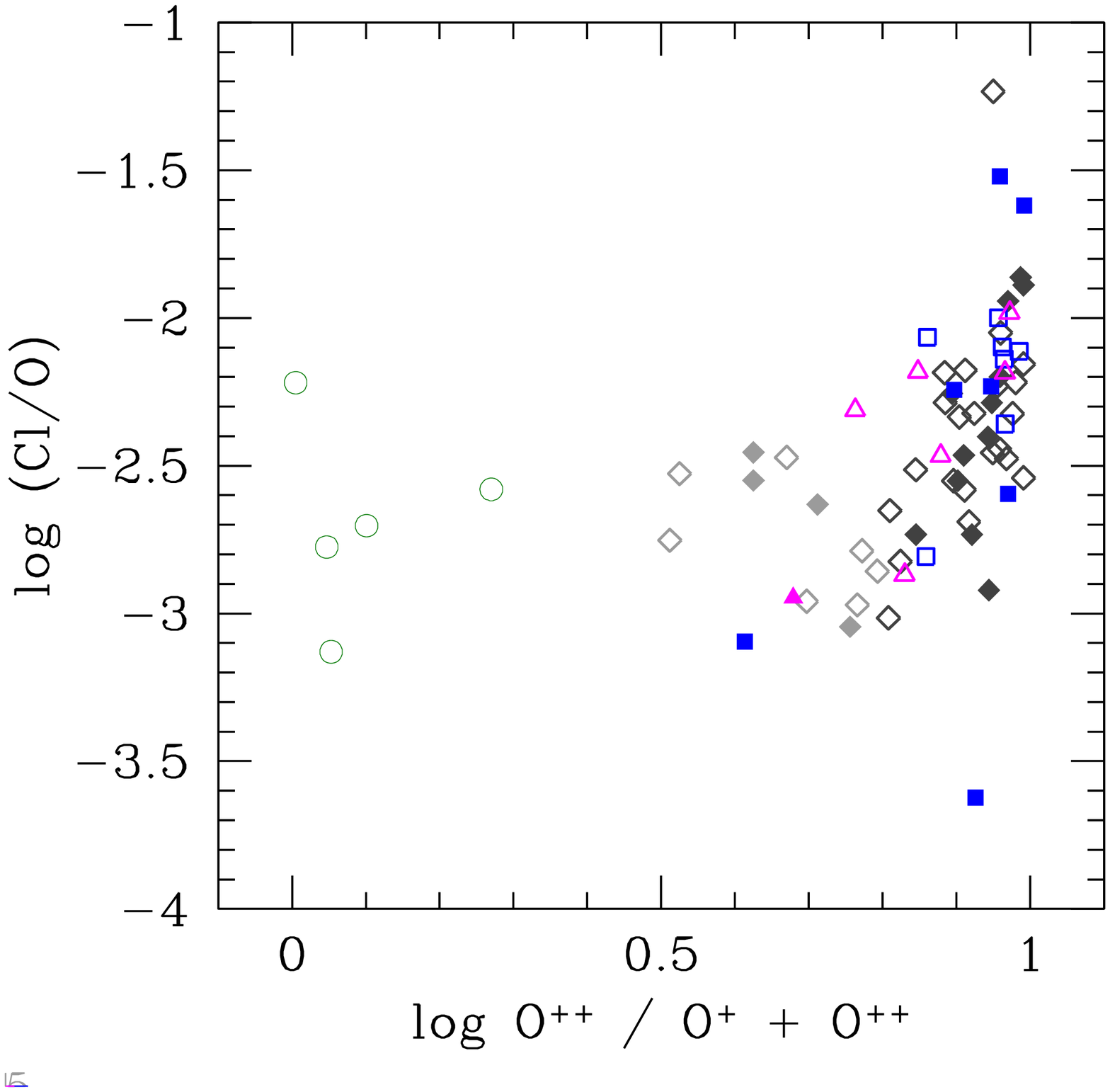}}   
\caption[]{
  The relation of different abundance ratios with ionization parameter
  \oppo\ for Galactic bulge PNe: \WRPNe --
  magenta triangles; \WELPNe\ -- blue squares; \VLPNe\ -- 
  green circles; normal PNe -- 
  black or grey
  diamonds. Filled symbols mark objects with best quality data. PNe with a
  quality of the derived abundance ratio above the adopted rejection limit
  (i.e. error$>$0.3\,dex) are represented with open thin-line symbols.
}
\label{oppo_abund}
\end{figure*}
}

\onlfig{22}{
\begin{figure*}
\resizebox{0.32\hsize}{!}{\includegraphics{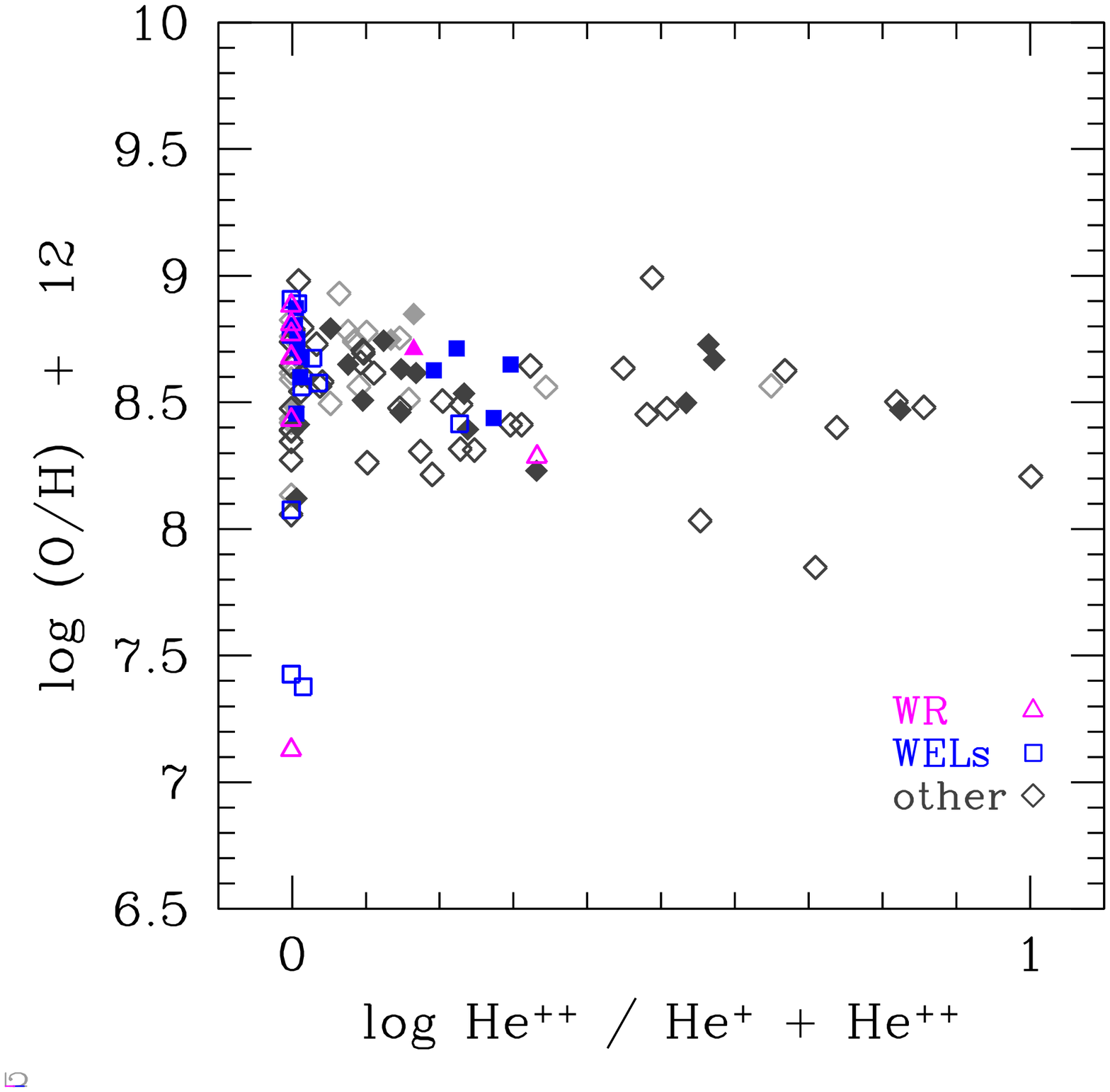}}   
\resizebox{0.32\hsize}{!}{\includegraphics{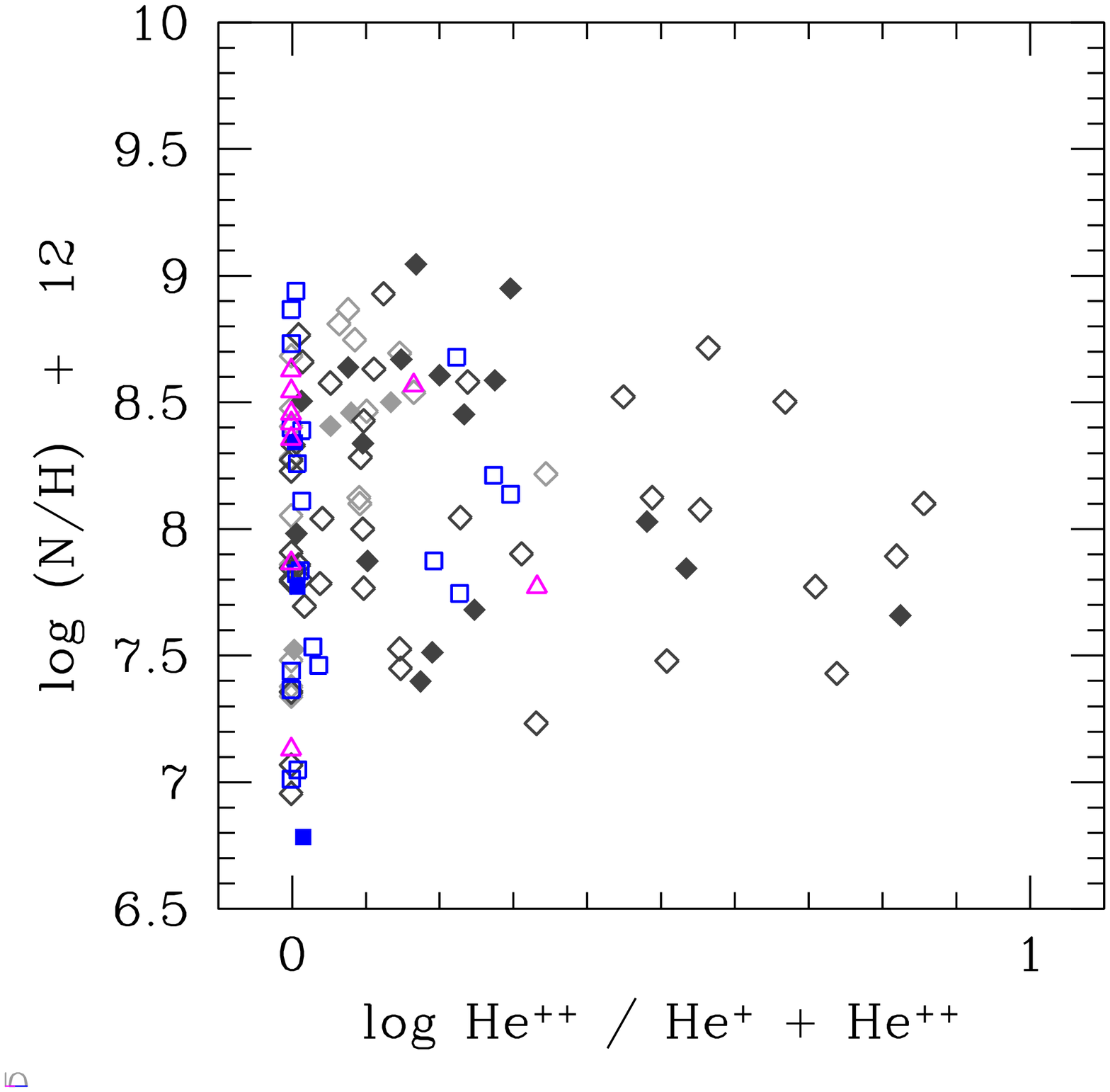}}   
\resizebox{0.32\hsize}{!}{\includegraphics{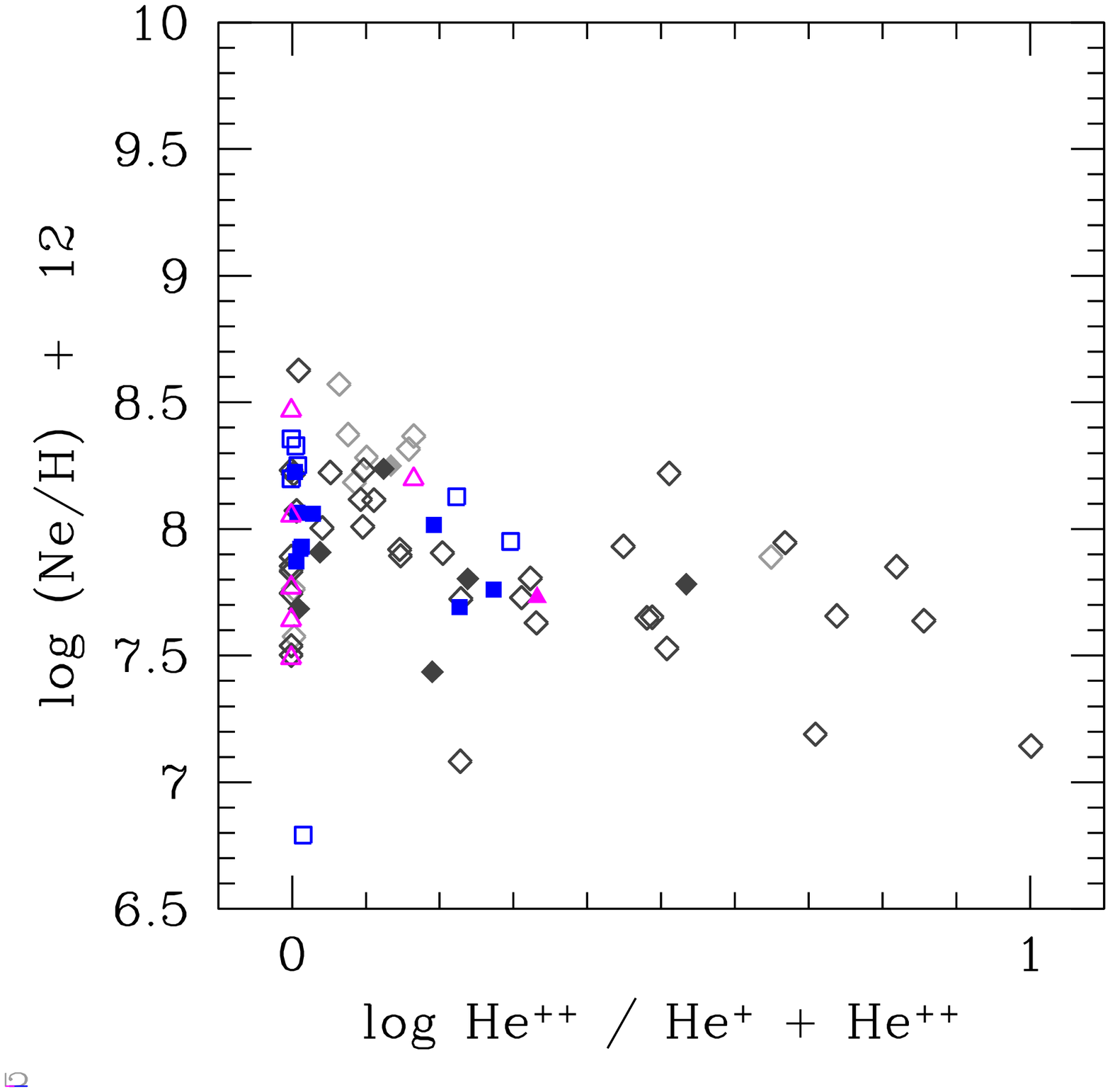}}   

\resizebox{0.32\hsize}{!}{\includegraphics{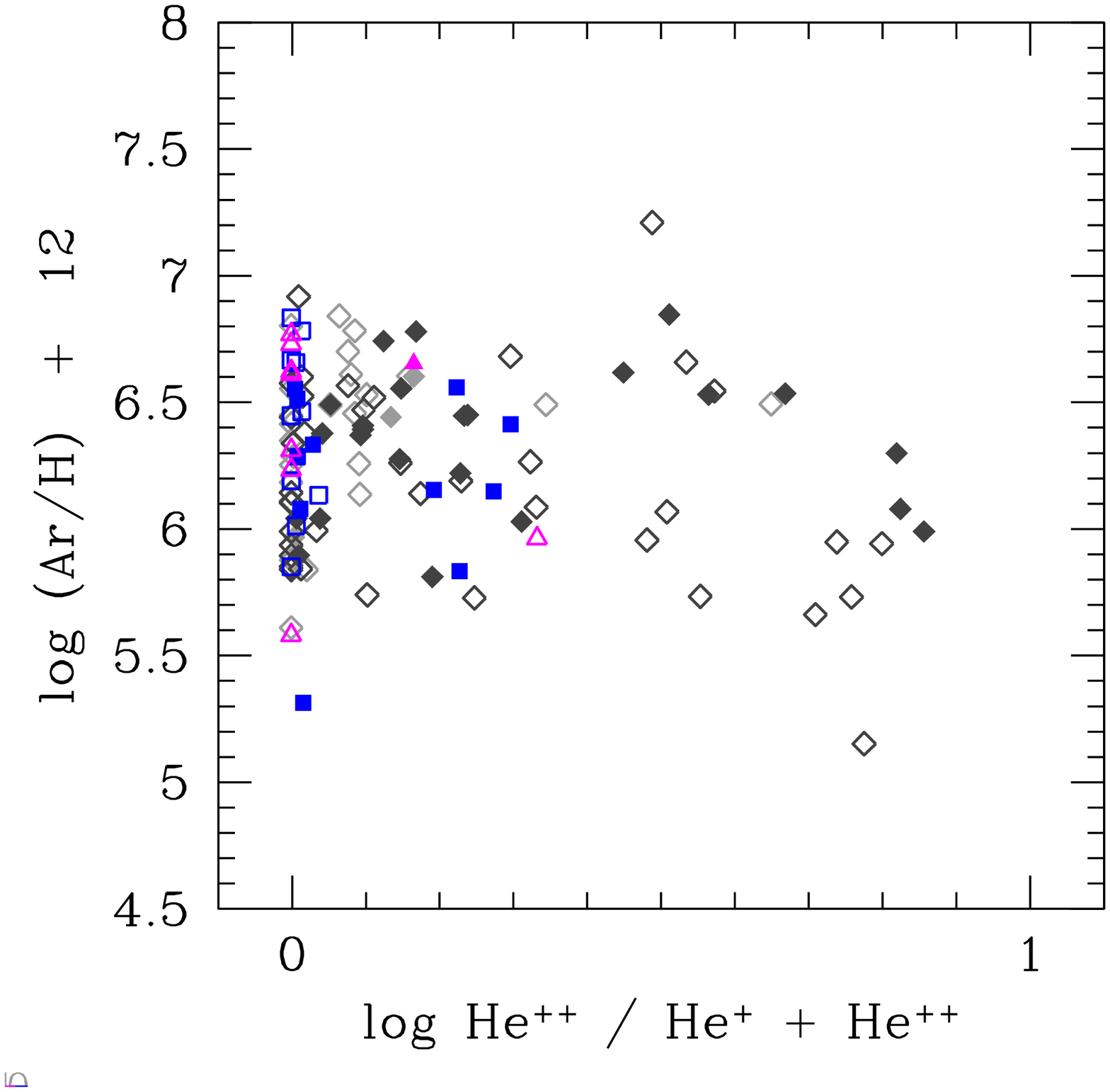}}   
\resizebox{0.32\hsize}{!}{\includegraphics{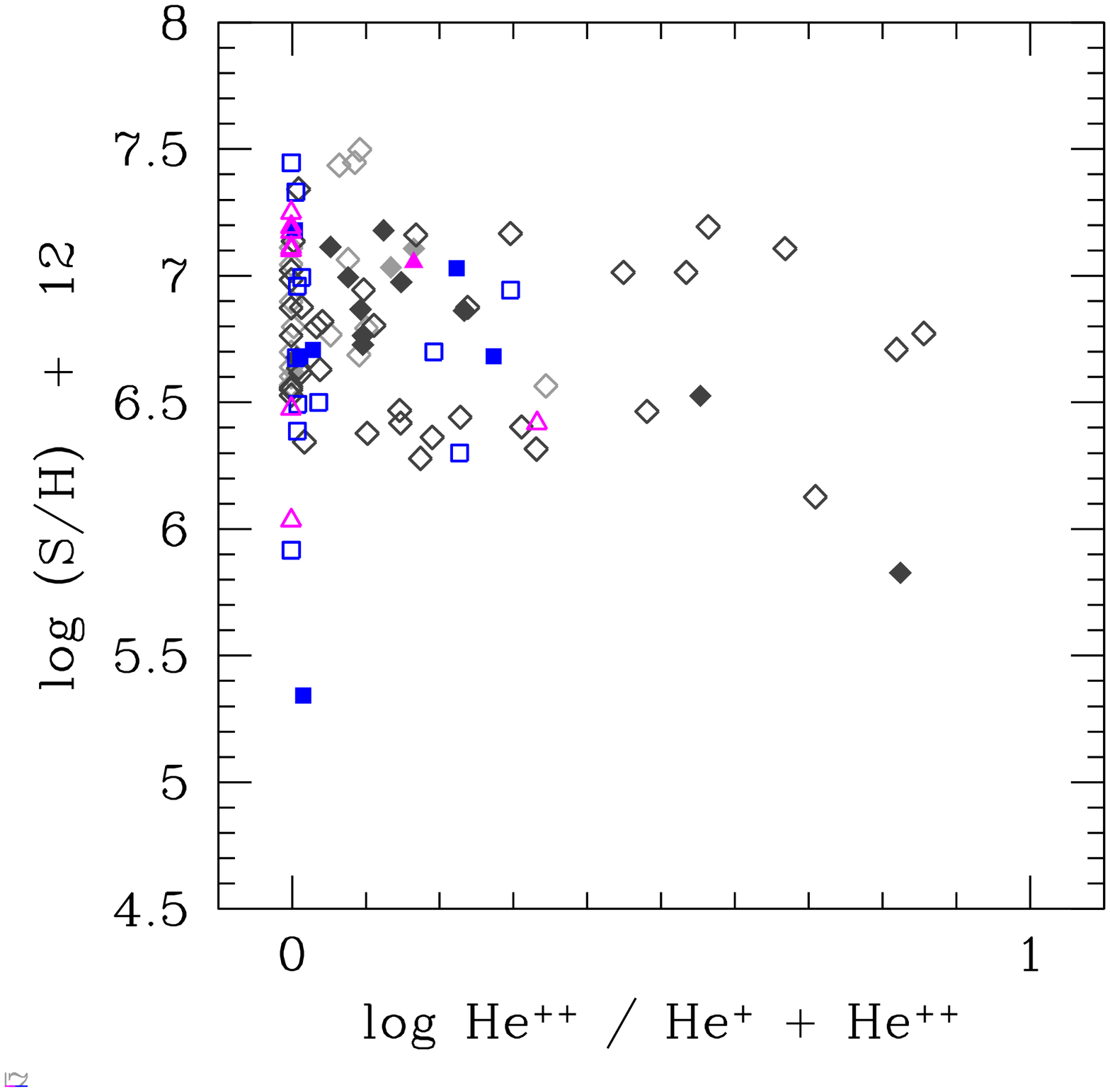}}   
\resizebox{0.32\hsize}{!}{\includegraphics{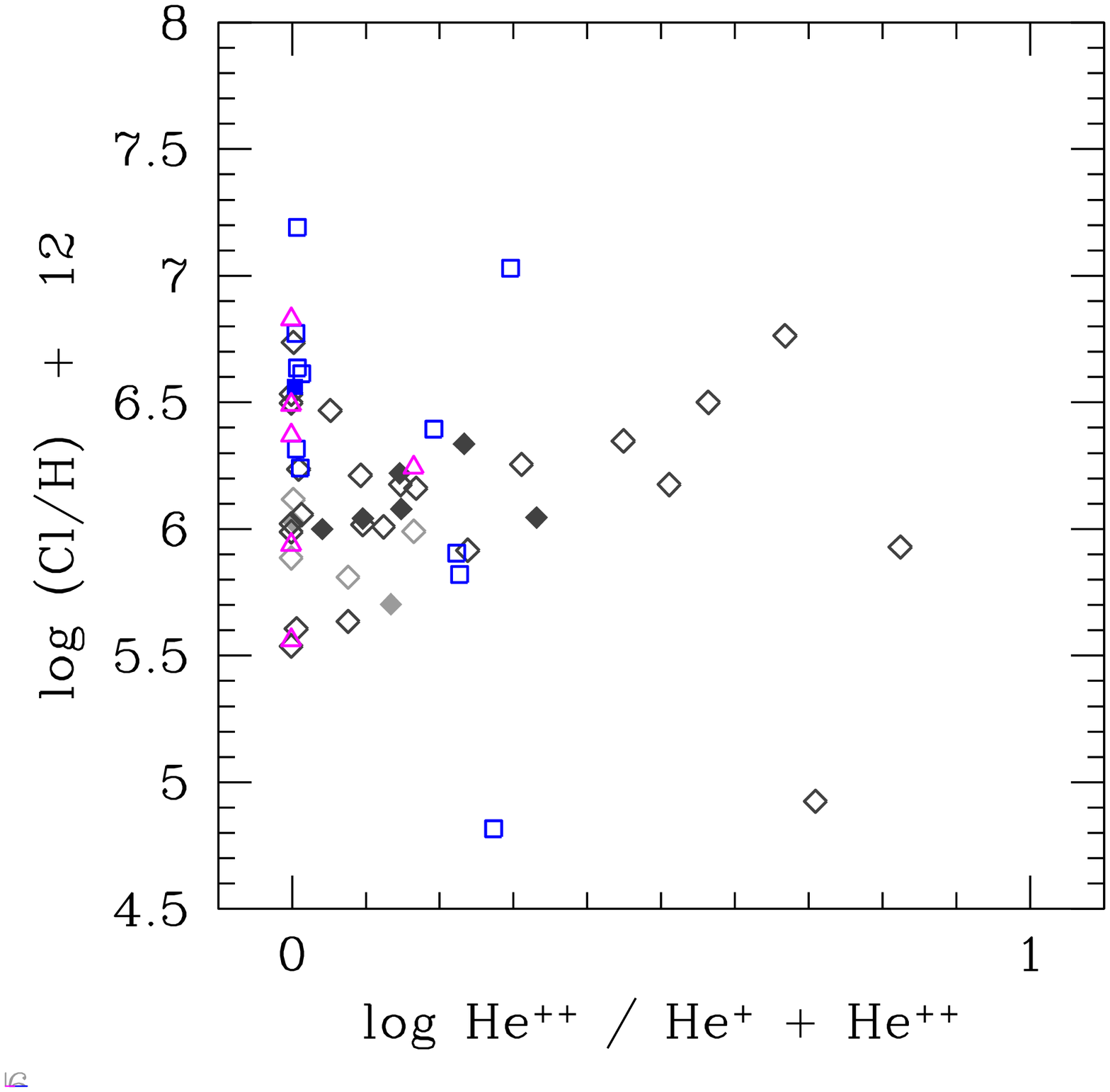}}   

\resizebox{0.32\hsize}{!}{\includegraphics{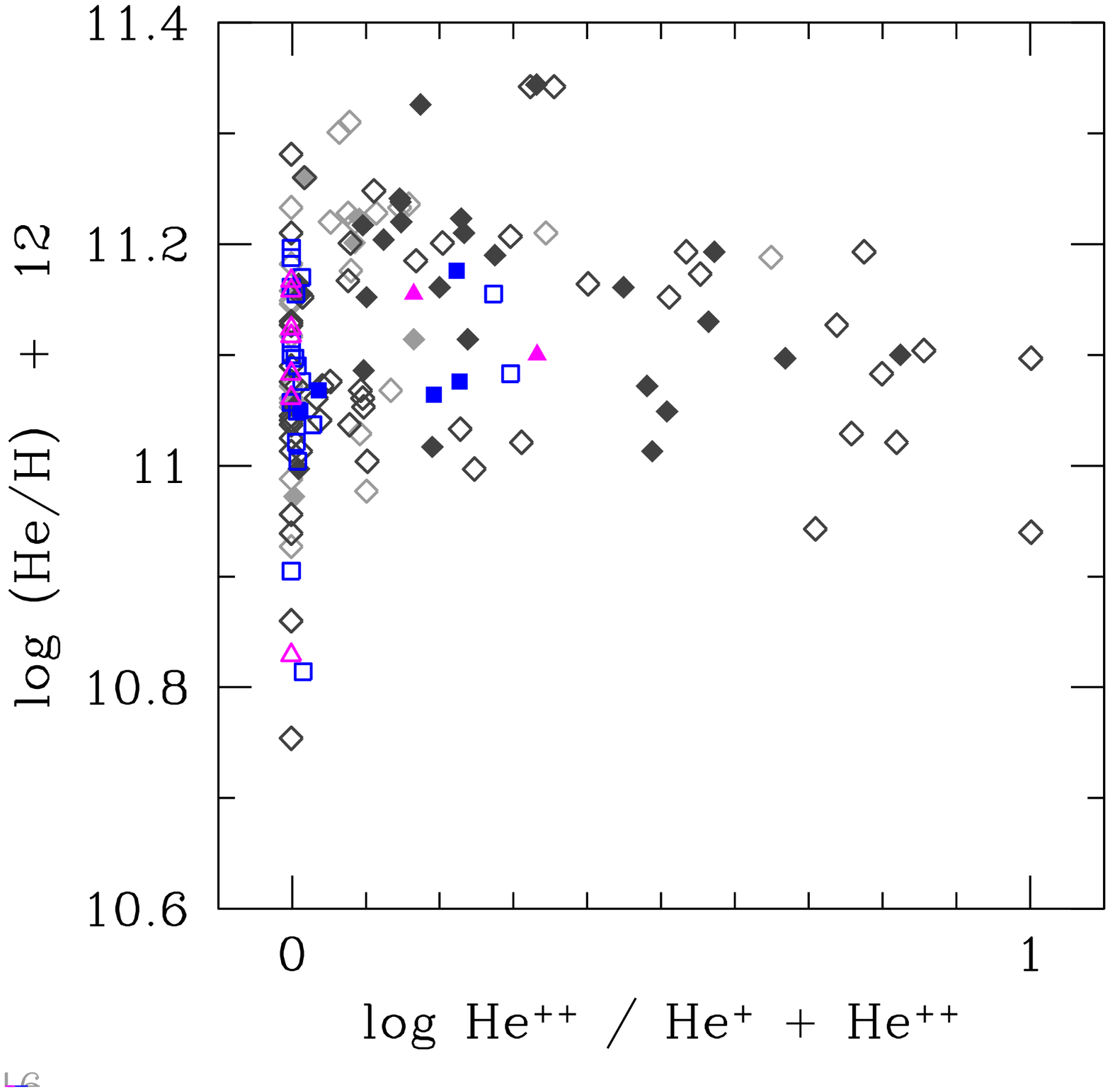}}   
\resizebox{0.32\hsize}{!}{\includegraphics{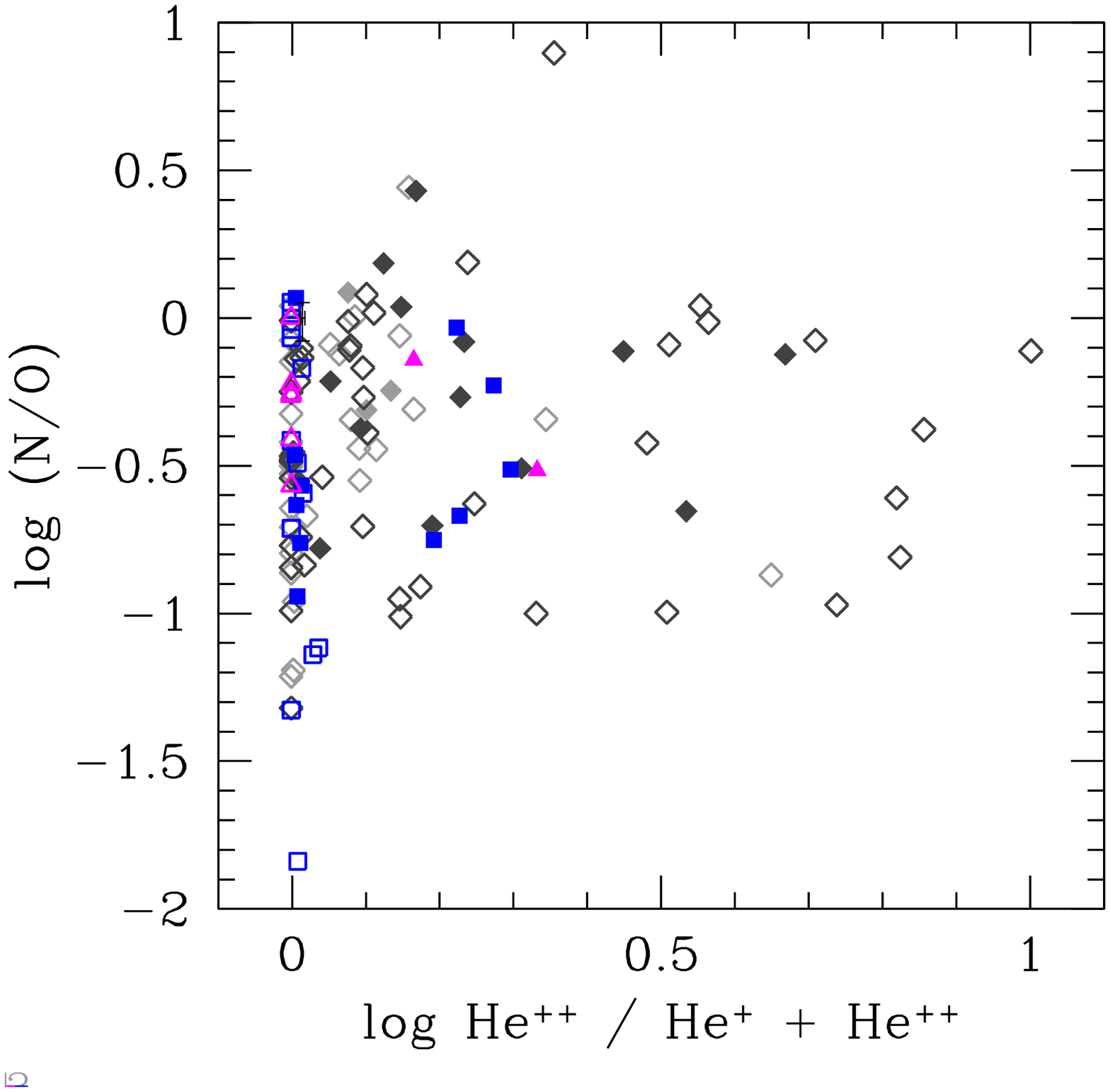}}   
\resizebox{0.32\hsize}{!}{\includegraphics{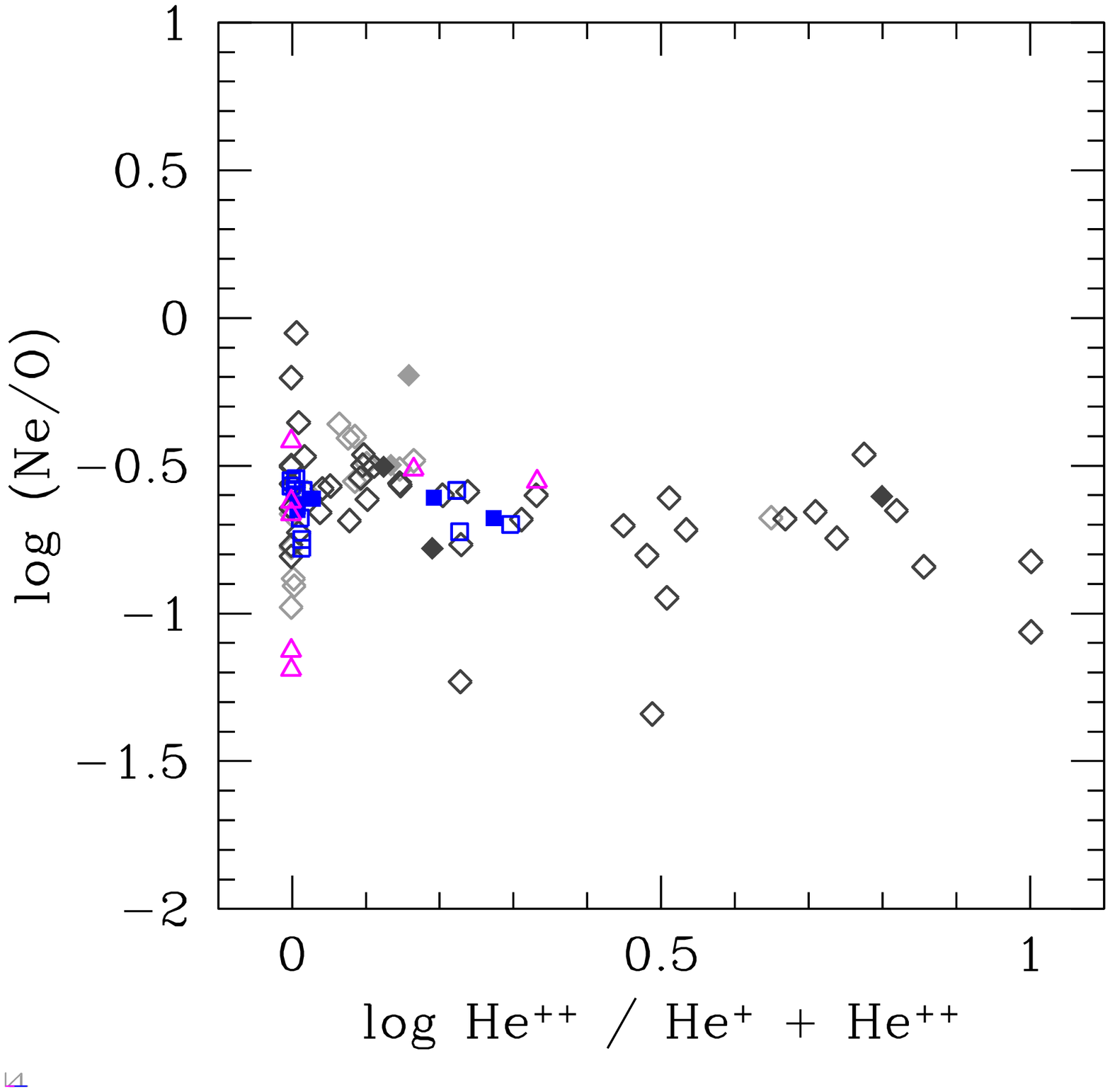}}   

\resizebox{0.32\hsize}{!}{\includegraphics{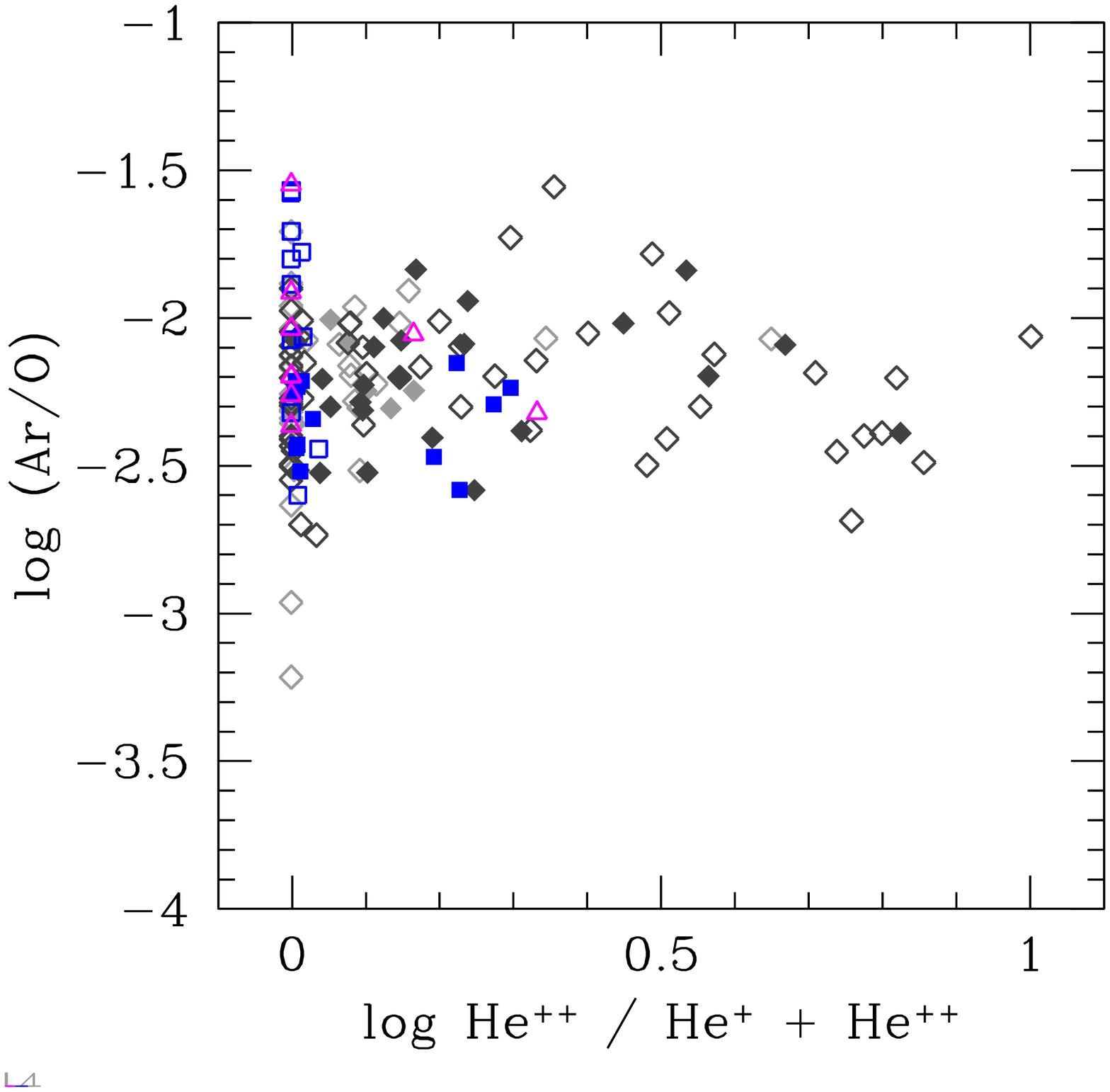}}   
\resizebox{0.32\hsize}{!}{\includegraphics{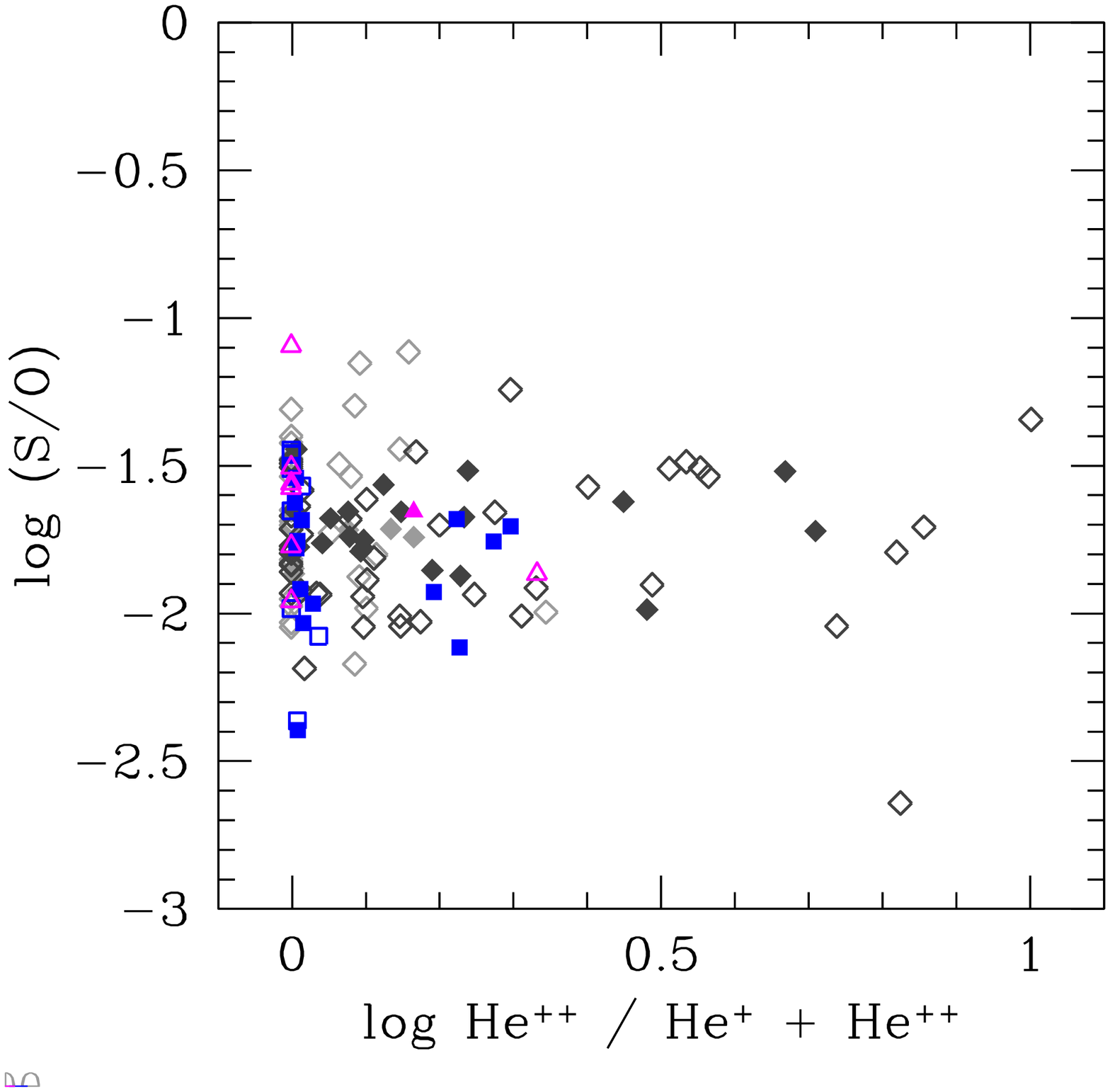}}   
\resizebox{0.32\hsize}{!}{\includegraphics{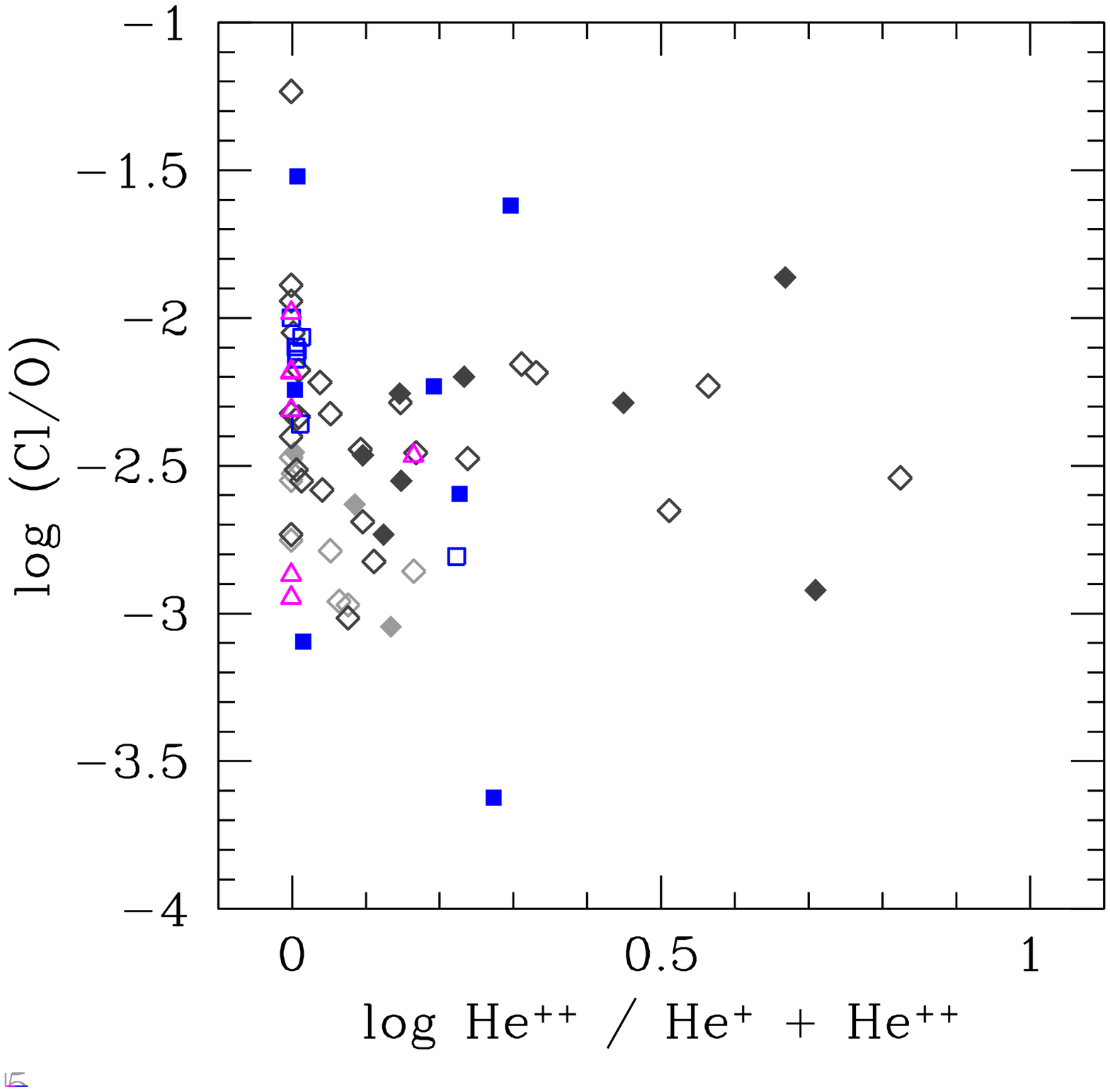}}   
\caption[]{
  The relation of different abundance ratios with ionization parameter
  \hepphe\ for Galactic bulge PNe. The same notation as in
  \figref{oppo_abund}.
}
\label{hepphe_abund}
\end{figure*}
}

The sample of Galactic bulge \WRPNe\ again does not seem numerous enough to
draw firm conclusions concerning the investigated abundances. The ranges of
all four abundance ratios would suggest similarity with normal PNe. On the
other hand, the derived median S/H and Ar/H abundances are approximately 0.3
dex larger for \WRPNe. While the median
log\,Ar/H=6.61 for \WRPNe\ there are only 14\% of normal PNe with that
argon abundance. The same holds for sulfur with median log\,S/H=7.10 for
\WRPNe\ and only 17\% of normal PNe lie above this abundance.  The nonparametric
tests do not allow to reject also the hypothesis that the distributions of
abundances of [WR] and normal PNe are identical.

In the case of the WEL PNe there seems to be even less doubt that there are no major
differences in distributions of Ar/H and S/H compared to the normal PNe.
If there is possibly any difference in abundances it seems the largest in
the case of neon  as the median log\,Ne/H is 0.15\,dex larger than
for normal PNe. The difference would be more pronounced if the WEL PNe were
compared with the normal PNe of \oppo$>$0.8 only\footnote{
  see upper-right panel of \figref{oppo_abund}
}. A similar situation holds for  the distribution of
Cl/H with a derived median abundance for WEL PNe larger by 0.36\,dex and
higher total proportion of WEL PNe with derived chlorine abundance (50\%) than
for normal PNe (25\%). But even in this case there remains a 5\% probability
that the difference in Cl/H is purely of random statistical origin.

\begin{figure*}
\resizebox{0.475\hsize}{!}{\includegraphics{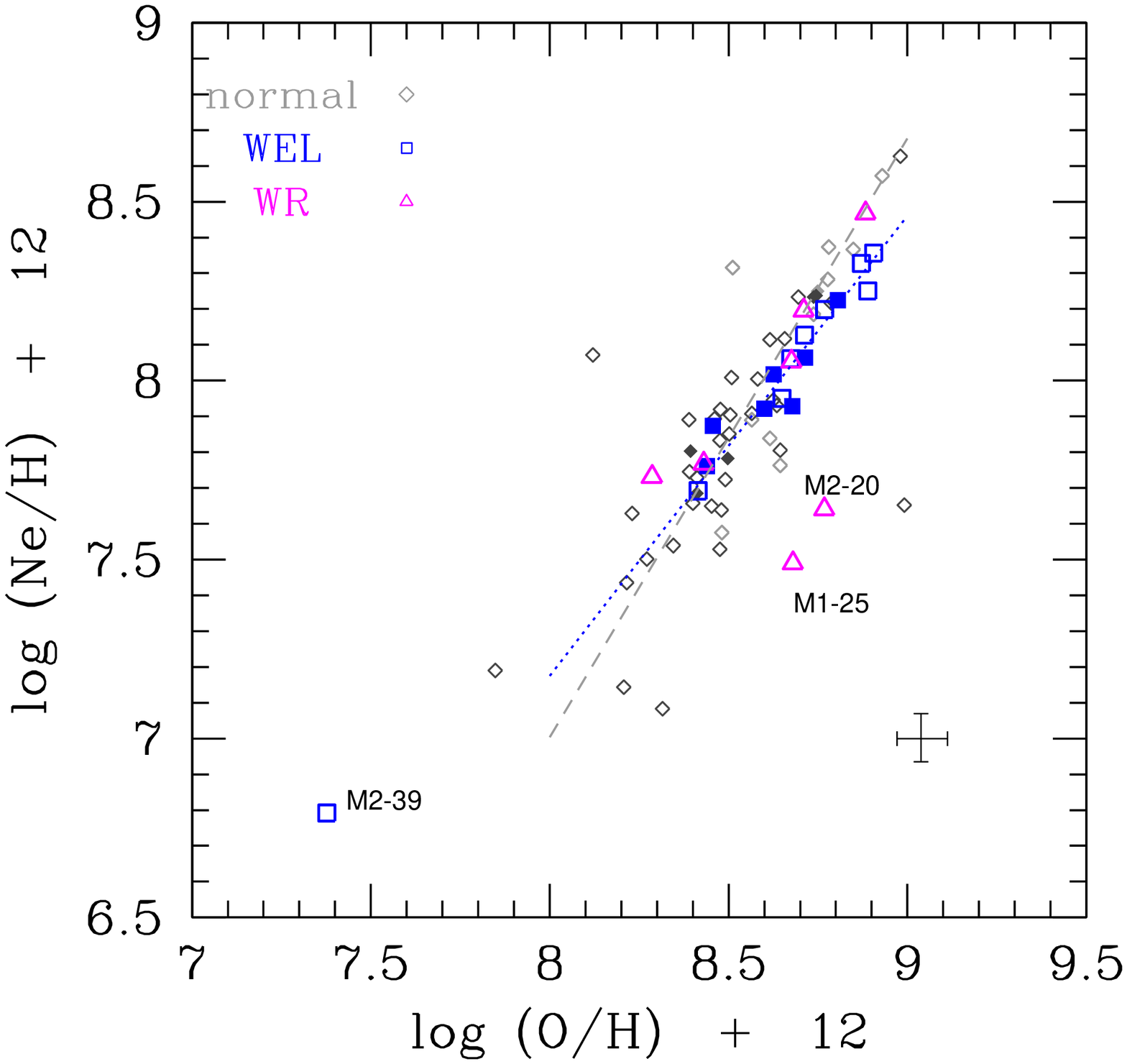}}  
\resizebox{0.475\hsize}{!}{\includegraphics{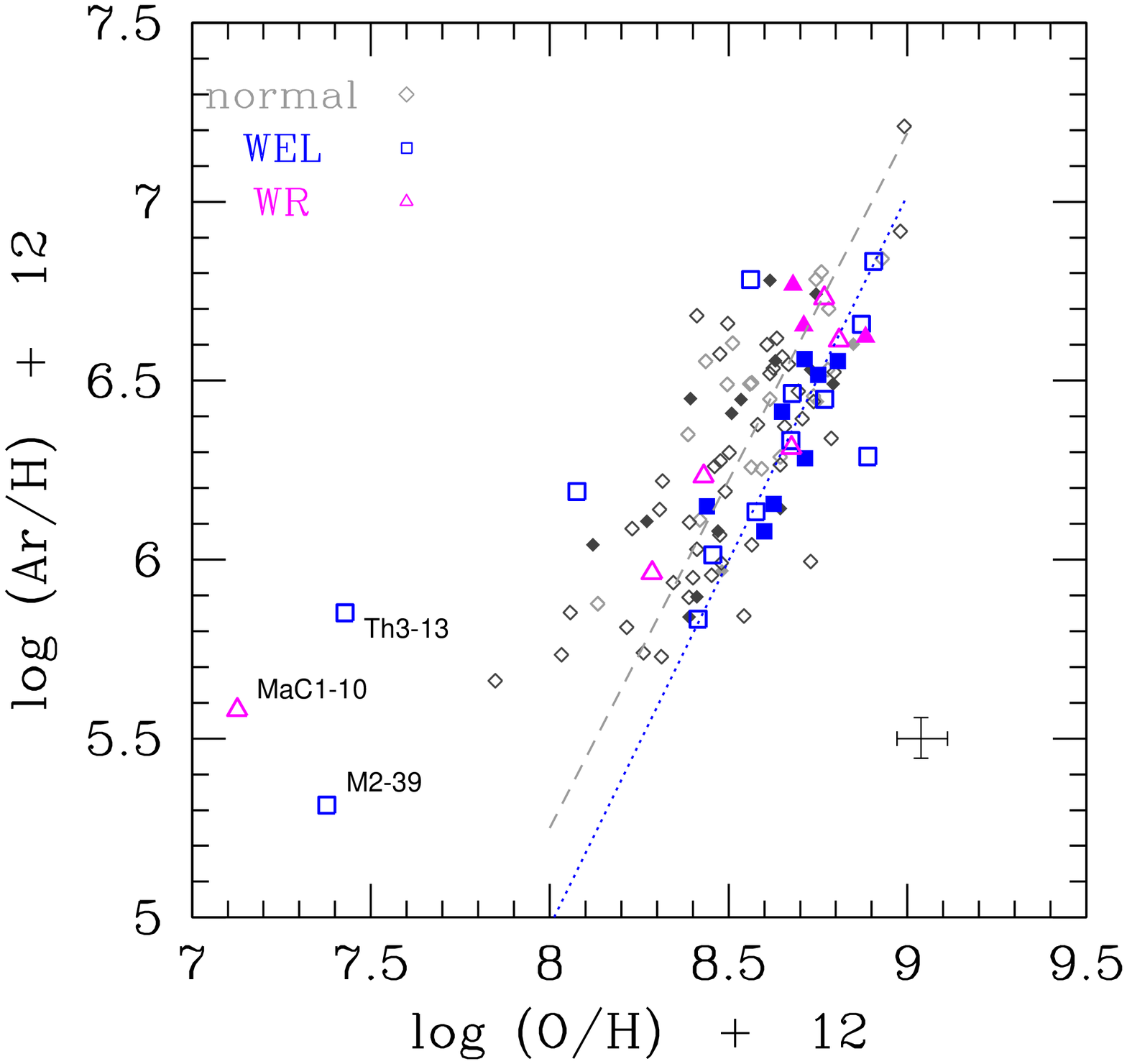}}  
\caption[]{
  Relations of Ne/H and Ar/H versus O/H for Galactic bulge [WR],
  WEL and normal PNe. The same notation of symbols as in \figref{wr_to_tn}.
  Lines are fits obtained by using WEL (short-dashed) and normal PNe
  (long-dashed) for objects with log\,O/H\,+12\,$>\,$8.
}
\label{oh_Xh}
\end{figure*}

In \figref{oh_Xh} the relations between oxygen abundance and the abundances
of neon and argon are presented for the three PNe
subsamples. For the WEL and normal PNe  we also overplot the lines
of fitted correlations between element abundances (dotted and dashed
respectively) derived for objects with log\,O/H+12$>$8.

Inspection of the left panel of \figref{oh_Xh} suggests that the same
tight relation between derived O and Ne abundances holds for WEL and normal
PNe. A closer analysis reveals, though, the already mentioned fact that the
neon abundances of the WEL PNe better match the values derived for normal PNe
with intermediate levels of oxygen ionization (grey symbols) than the highly
ionized PNe (black symbols). The slightly higher oxygen abundance of WEL
PNe is therefore apparently compensated by an equally increased neon abundance
preserving the relation between the two elements at practically the same level for
both WEL and normal PNe.

A different situation can be observed in the right panel of \figref{oh_Xh}
presenting the relation between O and Ar. The two correlation
lines have the same slope but are shifted by approximately 0.1\,dex
reflecting the difference in average oxygen abundance between the WEL and
normal PNe samples. In other words, if the difference in oxygen abundances
is real and not introduced by the method (see discussion above), then the
enhanced O abundance in WEL PNe is apparently not accompanied by an increase
in nebular Ar.

\subsubsection{Objects with peculiar abundances}

In Figures \ref{diag_heh_no} and \ref{oh_Xh} one can notice three PNe with
emission-line CSs distinguished by their low metallicity: MaC\,1-10, M\,2-39
and Th\,3-12.

The main reason for deriving low O/H ratio in these PNe is the unusually
high electron temperature of the nebular plasma. For the two
peculiar WEL PNe we determined:
\Te(O~{\sc iii})=26\,950K and \Te(N~{\sc ii})=26\,470K for Th\,3-12,  and
\Te(O~{\sc iii})=26\,960K and \Te(N~{\sc ii})=10\,270K for M\,2-39. 
For both nebulae we have two independent spectroscopic observations by
different observers confirming that the derived values are not due to incidental
errors.

In the case of  MaC\,1-10 -- the [WR] PN with peculiar nebular composition --
Te(OIII)=25\,870\,K while \Te(NII) cannot be derived as the
\rNii\ ratio 0.11  exceeds the usable range. These values are based on a single
spectroscopic observation in a set of sources we use in this paper. It finds
however confirmation in recently published lower quality spectra by Suarez
et~al. (2006).

Nonetheless, the MaC\,1-10 nebula is exceptional for other reasons. The
[WC\,8] spectral class of its central star makes it different from  the rest of
bulge \WRPNe\ and resembling more the [WR] type nuclei known in the Galactic
disk. Neither the distance nor radial velocity is available for this object
to judge wether MaC\,1-10 indeed belongs to the disk population. The total
nebular H${\beta}$ or radio flux are also unknown and the physical relation
of this nebula to the Galactic bulge was inferred from its apparent
location and diameter well below the 20\arcsec\ limit (MacConnell 1978).
 From the recent image published by Suarez et~al. (2006) it looks as the
nebula is of clearly bipolar morphology with the maximum extension measured
as 15x46\arcsec -- that makes the latter argument questionable. To add to
the mystery of the chemical composition and origin of this PN, in a recent survey
a H$_{2}$0 maser emission was discovered by Suarez et~al. (2007) in the
direction of MaC\,1-10.

The oxygen ionization level of MaC\,1-10 measured
by the \oppo\ ratio is only marginally above 0.4. We have chosen this
value (see above) as a minimum for the derived abundances to be accepted and
not seriously biased by the ICFs used. It is not excluded though that the
form of the ICFs adopted could add some additional contribution to the
apparent under-abundance of MaC\,1-10 in oxygen and other elements. However,
we can compare this nebula to PNe with a similar level of excitation (see
\figref{oppo_abund}) and therefore similar ICF factors. It is then evident
that the ratios He/H, O/H, N/H, Ar/H and S/H derived for MaC\,1-10 are the
lowest among them. On the other hand, abundance ratios N/O, Ar/O and S/O are
all larger than average in this group suggesting that oxygen is the most depleted
element in MaC\,1-10.

In the left panel of \figref{oh_Xh} presenting Ne/H abundance ratios as a
function of O/H another two \WRPNe\ stand out with their noticeably low neon
abundances. Combined with their oxygen abundance slightly above average
makes the Ne/O ratios of these objects unusually low. One of these
PNe is M\,1-25 reported for that property by Pe\~na \etal. (2001). The
other object is M\,2-20 sharing with the former also a similar [WC5-6]
spectral type.

\section{Discussion and conclusions}

We have presented a homogeneous set of  spectroscopic  measurements for a
sample of 90 PNe located in the direction of the Galactic bulge secured with
4-meter telescopes. For about a half of these objects, plasma parameters and
chemical abundances have been derived for the first time. We have discussed
the accuracy of the data based on internal tests of reliability and on
comparison with other recently published samples. We conclude that our data
is superior in quality to the data presented by Exter et al. (2004) and is
comparable to Wang \& Liu (2007) and G\'orny et al. (2004). We have
therefore merged our observations with the latter two sources into a large
common sample comprising 245 Galactic bulge and inner-disk PNe.

Using the new spectroscopic material, we performed a search for
emission-line CSs. We have discovered 8 new WEL CSs and 7 PNe with
very late (VL) emission-line CSs. The latter objects are associated with PNe
of extremely low ionization level. Despite the surprisingly large number of
objects discovered, it is not certain if the relative populations of WEL and
VL PNe are different in the bulge as compared to the inner-disk region of the
Galaxy. A similar extensive search for these types of objects in other
regions of the Milky Way would be desirable to draw firm conclusions.

Analyzing the spatial distribution of WEL, VL and classic \WRPNe\
belonging to the bulge we have discovered that there is a clear distinction
between the three groups in Galactic coordinates. WEL and VL PNe are
typically found at longitudes below 4\fdg5 from the center of the
Milky Way whereas \WRPNe\ are located further away. 

The properties of PNe surrounding emission-line stars have also been
analyzed. We have found the electron densities derived for these objects to
be larger than that of normal PNe. There seems also to be a difference in
electron temperatures with \Te(O~{\sc iii}) of \WELPNe\ and \WRPNe\ lower
 by 0.1\,dex on average than in normal PNe. The \Te(N~{\sc ii}) derived for
VL PNe also differs from normal PNe as their average temperature is of
6.760\,K only. The fact that the \Te(O~{\sc iii}) can be lower for \WRPNe\
and \WELPNe\ as compared to normal PNe has been noticed recently by Girard
et~al. (2007). The distribution of electron densities of disk \WRPNe\ have
seemed so far indistinguishable from their normal counterparts (G\'orny
\& Tylenda 2000, G\'orny 2001).

\WRPNe\ appear peculiar also by the fact that they are among the
brightest in the bulge. This explains why we were not successful in
discovering any new classic \WRPNe\ in our enlarged sample, which consists
mainly of fainter PNe.  On the other hand the same property can lead to
strong overestimation of the \WRPNe\ rate of occurrence in magnitude limited
samples of PNe, e.g. in other galaxies. The large luminosities of bulge
\WRPNe\ are likely due to the fact that their nebulae are more massive than
the average normal PNe. This can have important evolutionary implications.

The matter later ionized in the form of a planetary nebula is ejected
during late AGB stages of evolution. The higher masses of \WRPNe\ would
directly point to that stage of evolution as a period when the triggering
mechanism of a change from a normal intermediate-mass star into a future
hydrogen-depleted [WR]-type nucleus has started. It therefore favors the
hypothesis of a direct origin of \WRPNe\ from the AGB (either as a single star
or from a binary system) as opposied to a very late thermal pulse scenario
(VLTP) - see also  G\'orny \& Tylenda (2000). In the first case, the
creation of a more massive nebula and transformation of the surface chemical
composition of the star can naturally work together. In the VLTP case, on
the contrary, there is no possibility of increasing the mass of the nebula
considerably.

Investigating the present evolutionary state of the central stars and
surrounding nebulae we have found additional evidence that the \WELPNe,
\VLPNe\ and \WRPNe\ form three evolutionary unconnected manifestations of
enhanced mass loss phenomenon from PNe central stars.  With the help of comparison with simple model
calculations one can deduce from their locations in
evolutionary diagrams that the three groups differ in at least one
important property: the mass of the central star (closely related to
mass/age of the progenitor) or the mass of the surrounding nebula.  Both
factors may be working at the same time. From our study, it seems that the
VL PNe will not evolve into intermediate-type \WRPNe. In fact, although
the VL central stars satisfy the appropriate criteria and can be classified
as [WC\,11] objects\footnote{
  Similarly Pe\~na et al. (2001) were able to apply [WC] classification to
  some \WELCSs.}
they nevertheless seem different from the previously known
Galactic PNe with  central stars of similar spectral class.

From the analysis of our sample it looks as if CSs with VL type spectra
are quite common among Galactic bulge low-ionization PNe. In fact the
presence of such central stars may be a necessary condition for these PNe to
exist and be observable. Evolutionary time scales from model calculations
indicate that very low mass post-AGB stars would require an enhanced
mass-loss (manifested by emission lines in the spectra) to be able to reach
temperatures sufficient to ionize the surrounding matter before it
disperses. The range of masses of the stars from which the VL PNe originate may be
very narrow.

Finally, we analyzed the chemical abundances of the PNe with emission-line
CSs and compared them to those of normal PNe. We have found no significant
differences among the various groups.  Such a conclusion had previously been
reached by G\'orny \& Stasi\'nska (1995) and by Pe\~na \etal. (2001),
however, in those studies, the abundances were not obtained in a homogeneous
way. Recently Girard \etal. (2007) have published an analysis of new
observations of 30 \WRPNe\ (including some bulge and inner-disk objects) presenting
some indications that in these nebulae the N/O, Ar/H and S/H abundance
ratios are larger than in normal PNe. Interestingly, at least in the case of
the latter two ratios, our results would suggest a similar possibility.
On the other hand the O/H ratio may be higher for \WELPNe\ (and perhaps
\WRPNe) in comparison with other PNe. This would be consistent with the lower
\Te(O~{\sc iii}) of these PNe suggesting a larger number of effectively
cooling atoms in the nebular plasma. However the result may also be biased
by the method used to derive the abundances.

Three PNe with emission-line CSs: Th\,3-13,  M\,2-29 (WEL) and MaC\,1-10
([WR]), have been found to be extremely metal-poor compared to the rest.
These nebulae deserve further investigation. We have also confirmed the
unusually low Ne/O abundance in M\,1-25 reported by Pe\~na \etal. (2001) and
found another \WRPN\ with this property: M\,2-20.

A detailed analysis of the chemical abundances of the collected PNe sample and a
discussion of their consequences for our understanding of the formation and
evolution of the Galactic bulge itself and the PNe population in particular
are presented in Chiappini et~al. (2009).

\begin{acknowledgements}

C. C.  and F. C. thank CTIO and ESO staff in Chile and Pronex-Brazil for
partial support. C. C. would like to acknowledge the Swiss National Science
Foundation (SNF) for partial support. S.K.G. and G.S. acknowledge support
by the European Associated Laboratory "Astronomy Poland-France''.
Discussions with R. Walterbos and J. Danziger are greatly appreciated.

\end{acknowledgements}

\Online

\begin{appendix}
\section{Recovering correct intensities from second order contaminated
          spectra}

We describe here a procedure developed for our second order contaminated
spectra. This type of spectrum can be corrected if at least two different
standard stars were observed. The advantage of the presented approach is an
easy implementation within MIDAS routines requiring no labour-intensive manual
correcting or editing of every observed spectrum.

It is well known that the spectra at wavelengths longer than 5800\AA\ (double 
the atmospheric UV cutoff at 2900\AA) can be subject to
contamination from second order effects. However, the exact amount of this
contamination can vary depending on the site and  instrument
transmissivity to UV photons as well as the specific object properties.
Therefore this effect may be quite small or even negligible in many
cases. The use of order blocking filters is also not advised when one deals
with spectra covering a wide wavelength range and does not want to decrease the
sensitivity of its blue part. This was our case as we were equally
concerned with both an intrinsically weaker \oii\ $\lambda$7325 line at the red
end and the usually more poorly exposed lines at the blue end, especially
the crucial \oii\ $\lambda$3727 line.

Nevertheless after starting to reduce the spectra it became evident that the
order contamination problem of our CTIO observations is very prominent.
It first appears  when the flux calibration is attempted. Using the nomenclature of
MIDAS routines this is known as deriving the instrument response
function from standard star(s) observations and applying it to observed
objects. The response function at a given wavelength is defined as:

\[
R = \frac{I\;E}{t\;F},
\]

where I is the flux registered and measured from the CCD after flat-field
correction and bias \& background subtraction, F is the tabulated flux of the
star at the base of the atmosphere, t is the exposure time and E is a
correction for atmospheric extinction. E is computed from the actual airmass of
the observation and the atmospheric extinction constant A as:

\[
E(\lambda) = 10^{-0.4\;A(\lambda)\;airmass}.
\]

However in the case of second order contaminated spectra at wavelength
$\lambda$, the registered flux is partly due to light from the wavelength 
$\lambda$/2.
The amount of this contaminating light must be proportional to the absolute
flux but depends also on the airmass. If we represent the proportionality
factor by $X$, we can write for the actually registered intensity $I'$\,:

\begin{equation}
I'(\lambda) = I(\lambda) + X(\lambda)\;t\;F(\lambda/2)\;E(\lambda/2).
\end{equation}

Thus for the first observed but contaminated standard star to have a correct
response function we would have to solve the equation:

\[
R_1(\lambda) = \frac{(I'_1(\lambda) - 
X(\lambda)\;t_1\;F_1(\lambda/2)\;E_1(\lambda/2)\,)\;E_1(\lambda)}{t_1\;F_1(\lambda)}.
\]

X($\lambda$) should be a unique property of a spectrograph at a given setting
(neglecting some possible small variations with time as the properties
of the instrument may change with changing weather conditions or telescope
orientation). Therefore we can write a similar equation for the second star
with the same $X$($\lambda$):

\[
R_2(\lambda) = \frac{(I'_2(\lambda) - 
X(\lambda)\;t_2\;F_2(\lambda/2)\;E_2(\lambda/2)\,)\;E_2(\lambda)}{t_1\;F_2(\lambda)}
\]

\noindent
and solve them with the condition that the response functions from
observations of both stars should (ideally) give the same response solution:

\begin{equation}
R_1(\lambda) = R_2(\lambda)
\end{equation}

\noindent
or in full form as:

\[
\frac{I'_1(\lambda)\;E_1(\lambda)}{t_1\;F_1(\lambda)} - 
X(\lambda)\frac{t_1\;F_1(\lambda/2)\;E_1(\lambda/2)\;E_1(\lambda)}{t_1\;F_1(\lambda)} 
=
\]
\[
= \frac{I'_2(\lambda)\;E_2(\lambda)}{t_2\;F_2(\lambda)} - 
X(\lambda)\frac{t_2\;F_2(\lambda/2)\;E_2(\lambda/2)\;E_2(\lambda)}{t_2\;F_2(\lambda)}.
\]

\noindent
This can be noted also as

\[
R'_1(\lambda) - X(\lambda)\;R_1(\lambda/2) = R'_2(\lambda)  - 
X(\lambda)\;R_2(\lambda/2)
\]

\noindent
leading to:

\begin{equation}
X(\lambda) = \frac{R'_2(\lambda) - R'_1(\lambda)}{R_2(\lambda/2) - 
R_1(\lambda/2)}
\end{equation}

\noindent
where $R'_1(\lambda)$ and $R'_2(\lambda)$ are the response functions derived
with actually observed, contaminated intensities and $R_1(\lambda/2)$ and
$R_2(\lambda/2)$ are "response" functions derived with artificial standard
star spectra:

\[
  t\;F(\lambda/2)\;E(\lambda/2).
\]

$F(\lambda/2)$ is created with a MIDAS {\tt create/image} command from a
table comprising the original tabulated UV fluxes but assigned to doubled
wavelengths. On the other hand $E(\lambda/2)$ requires a modified extinction
table created from the  {\tt atmoexan.tbl} file of MIDAS by assigning the
extinction constants again to doubled wavelengths. This is possible with
another standard command {\tt create/table}.

The creation of the modified extinction table and separate $F(\lambda/2)$
frames for each standard star observed are the only steps requiring any
manual modifications of existing tables or data. Since the created files
follow the standard MIDAS image and table formats, the extinction correction
$E(\lambda/2)$ and $E(\lambda)$ can be executed with a standard MIDAS
procedure {\tt extinction/long} and the response function can be 
derived with the
commands {\tt integrate/long} and {\tt response/long}.

Once this is done it is easy to calculate the $X(\lambda)$ factors using
equation A.3 and then to derive the corrected response function.

Having observed more that two standard stars during a given night, we can
compute slightly different $X$ parameters using different standard star
pairs. We can also choose different stars to compute the final response:

\[
R_n(\lambda) = R'_n(\lambda) - R_n(\lambda/2).
\]

Our tests have shown however that due to the errors involved and the accuracy of the
computations it is best to correct the response function of the star with
the lowest UV/R light intensity ratio. On the other hand the $X$ factors
should be computed using all pairs including this star and stars with a high
UV/R ratio. In this way we are inferring the true $R$ by making only a small
correction to $R'$ and using $X$ that has been determined with high accuracy
from a few independent measurements\footnote{
  If two stars with similar and high UV/R are used it can lead to
  large errors or even  a false solution.
}.

The fact that the adopted procedure works correctly can be judged from
Figure\,A.1. In this figure,  black lines present the original registered
spectra of four standard stars observed during one night. Their corrected
shapes computed using equation A.1 are shown with green lines in this plot.

\begin{figure}
\resizebox{0.99\hsize}{!}{\includegraphics{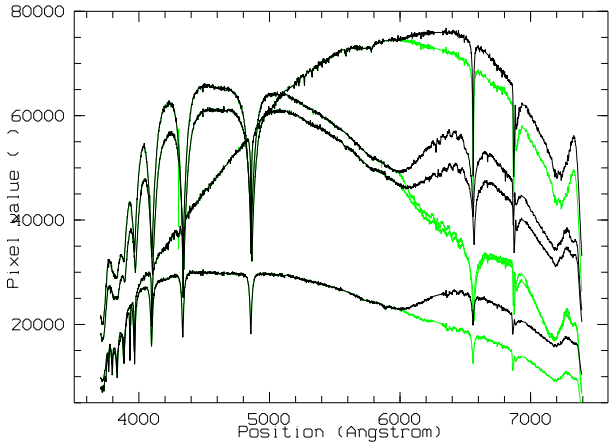}}
\caption[]{
Registered and atmospheric extinction corrected spectra of four standard
stars observed at CTIO on May 21st, 2002 (black lines) and after
removing second order contamination effects (green lines).
}
\label{}
\end{figure}

\begin{figure}
\resizebox{0.99\hsize}{!}{\includegraphics{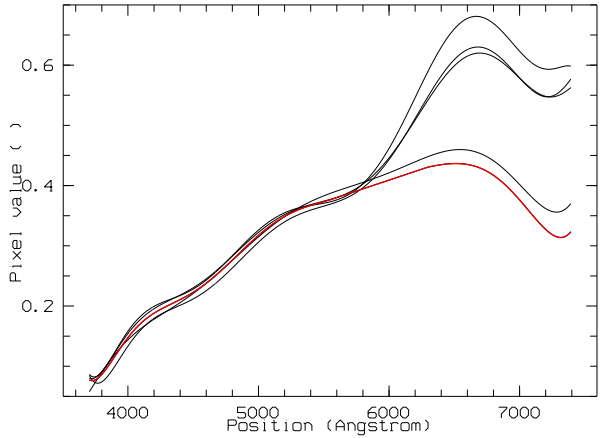}}
\caption[]{
Response functions derived from observations of four different standard
stars during the CTIO run on May 21st, 2002 (black lines) and the final
average response after removing second order contamination effects (red
line).
}
\label{}
\end{figure}

In Figure\,A.2 an average corrected response function is shown as a
red curve. The part between 5600A and 6400A has been derived by a linear
interpolation between the original function at shorter wavelengths and
the re-calculated one at longer wavelength. This was a necessary step since the
order contamination may still take place in this intermediate spectral
region but the necessary standard fluxes below 3200\AA\ are not available.

\end{appendix}

\clearpage

\setcounter{table}{1}
\begin{table*}
\caption{ 
  Logarithmic extinction C at \Hb\ and dereddened line intensities 
  on the scale H$\beta$=100.
}
\begin{tabular}{     l @{\hspace{-0.60cm}}
                     l @{\hspace{0.40cm}}
                     r r r r r r r r}
\hline
 PN G       &      &  000.1-02.3 &  000.2-04.6 &  000.3-04.6 &  000.4-01.9 &  000.4-02.9 &  000.5-03.1 &  000.7-02.7 &  000.7-03.7 \\
 Main Name  &      &  Bl3-10     &  Wray16-36  &  M2-28      &  M2-20      &  M3-19      &  KFL1       &  M2-21      &  M3-22      \\
 sample     &      &  CTIO'02    &  CTIO'02    &  CTIO'02    &  CTIO'02    &  CTIO'02    &  ESO'02     &  CTIO'02    &  CTIO'01    \\
            &      &             &             &             &             &             &             &             &             \\
 ext. C     &      &     1.62    &    1.88  ~~ &    1.39  ~~ &    1.72  ~~ &    1.44  ~~ &    1.46  ~~ &    1.18  ~~ &    1.03~~   \\
            &      &             &             &             &             &             &             &             &        ~~   \\
 3727 & {[O\ III]} &             &             &  101.94~c   &   59.78  ~~ &   11.90~c   &   10.10~c   &   20.21:~~  &        ~~   \\
 3869 & {[Ne III]} &  110.26  ~~ &             &   73.08~c   &   11.04  ~~ &   48.57~c   &   38.20~c   &   88.50  ~~ &        ~~   \\
\hline
\end{tabular}
\end{table*}

\setcounter{table}{1}
\begin{table*}
\caption{ 
  Continued
}
\begin{tabular}{     l @{\hspace{-0.60cm}}
                     l @{\hspace{0.40cm}}
                     r r r r r r r r}
\hline
 PN G       &      &  002.2-02.7 &  002.3-03.4 &  002.4-03.7 &  002.5-01.7 &  002.6+04.2 &  002.8-02.2 &  002.9-03.9 &  003.0-02.6 \\
 Main Name  &      &  M2-23   ~~ &  H2-37   ~~ &  M1-38      &  Pe2-11  ~~ &  Th3-27  ~~ &  Pe2-12  ~~ &  H2-39   ~~ &  KFL4       \\
 sample     &      &  CTIO'02    &  CTIO'02    &  CTIO'02    &  ESO'02  ~~ &  CTIO'01    &  ESO'02  ~~ &  CTIO'02    &  ESO'02     \\
            &      &             &             &             &             &             &             &             &             \\
 ext. C     &      &    1.05  ~~ &    1.33  ~~ &    1.08  ~~ &    2.15  ~~ &    2.47  ~~ &    1.70  ~~ &    1.30  ~~ &    1.25  ~~ \\
            &      &             &             &             &             &             &             &             &             \\
 3727 & {[O\ III]} &   14.54~c   &             &   98.91~c   &  129.24~c   &             &   42.95~c   &    8.84  ~~ &             \\
 3869 & {[Ne III]} &   99.64~c   &  202.43~c   &             &   61.93~c   &             &        ~c   &  104.00  ~~ &  103.81  ~~ \\
\hline
\end{tabular}
\end{table*}

\setcounter{table}{1}
\begin{table*}
\caption{ 
  Continued
}
\begin{tabular}{     l @{\hspace{-0.60cm}}
                     l @{\hspace{0.40cm}}
                     r r r r r r r r}
\hline
 PN G       &      &  003.9-03.1 &  004.1-03.8 &  004.2-03.2 &  004.2-04.3 &  004.8-05.0 &  004.9+04.9 &  004.9-04.9 &  005.0+03.0 \\
 Main Name  &      &  KFL7    ~~ &  KFL11   ~~ &  KFL-10  ~~ &  H1-60   ~~ &  M3-26   ~~ &  M1-25   ~~ &  M1-44   ~~ &  Pe1-9      \\
 sample     &      &  ESO'02  ~~ &  ESO'02  ~~ &  CTIO'02    &  CTIO'01    &  CTIO'02    &  CTIO'02    &  CTIO'01    &  ESO'02     \\
            &      &             &             &             &             &             &             &             &             \\
 ext. C     &      &    1.25  ~~ &    1.32  ~~ &    0.74  ~~ &    0.62  ~~ &    0.76  ~~ &    1.42  ~~ &    0.74  ~~ &    1.59  ~~ \\
            &      &             &             &             &             &             &             &             &             \\
 3727 & {[O\ III]} &             &   59.92~c   &    9.83:c   &             &   10.58~c   &   70.99  ~~ &             &   27.39  ~~ \\
 3869 & {[Ne III]} &   55.00~c   &   67.91~c   &   53.00~c   &             &   52.48~c   &    8.07  ~~ &             &   20.51~c   \\
\hline
\end{tabular}
\end{table*}

\setcounter{table}{1}
\begin{table*}
\caption{ 
  Continued
}
\begin{tabular}{     l @{\hspace{-0.60cm}}
                     l @{\hspace{0.40cm}}
                     r r r r r r r r}
\hline
 PN G       &      &  006.3+04.4 &  006.4+02.0 &  006.4-04.6 &  006.8+02.3 &  006.8-03.4 &  006.8+04.1 &  007.8-03.7 &  007.8-04.4 \\
 Main Name  &      &  H2-18   ~~ &  M1-31   ~~ &  Pe2-13  ~~ &  Th4-7   ~~ &  H2-45   ~~ &  M3-15   ~~ &  M2-34   ~~ &  H1-65      \\
 sample     &      &  CTIO'02    &  CTIO'02    &  CTIO'01    &  CTIO'02    &  CTIO'01    &  CTIO'01    &  CTIO'01    &  CTIO'01    \\
            &      &             &             &             &             &             &             &             &             \\
 ext. C     &      &    1.63  ~~ &    1.98  ~~ &    1.01  ~~ &    1.76  ~~ &    1.52  ~~ &    2.12  ~~ &    1.17  ~~ &    0.98  ~~ \\
            &      &             &             &             &             &             &             &             &             \\
 3727 & {[O\ III]} &   16.46~c   &   33.35  ~~ &             &   98.32~c   &             &             &             &             \\
 3869 & {[Ne III]} &   99.93~c   &   56.68  ~~ &             &   90.15~c   &             &             &             &             \\
\hline
\end{tabular}
\end{table*}

\setcounter{table}{1}
\begin{table*}
\caption{ 
  Continued
}
\begin{tabular}{     l @{\hspace{-0.60cm}}
                     l @{\hspace{0.40cm}}
                     r r r r r r r r}
\hline
 PN G       &      &  353.5-04.9 &  354.2+04.3 &  355.4-04.0 &  355.7-03.4 &  355.9+03.6 &  355.9+03.6 &  355.9-04.2 &  356.7-04.8 \\
 Main Name  &      &  H1-36   ~~ &  M2-10   ~~ &  Hf2-1   ~~ &  H2-23   ~~ &  H1-9    ~~ &  H1-9    ~~ &  M1-30   ~~ &  H1-41      \\
 sample     &      &  CTIO'02    &  CTIO'01    &  CTIO'01    &  CTIO'01    &  CTIO'02    &  CTIO'01    &  CTIO'01    &  CTIO'01    \\
            &      &             &             &             &             &             &             &             &             \\
 ext. C     &      &    1.09  ~~ &    1.32  ~~ &    0.90  ~~ &    1.23  ~~ &    1.71  ~~ &    1.75  ~~ &    1.04  ~~ &    6.03  ~~ \\
            &      &             &             &             &             &             &             &             &             \\
 3727 & {[O\ III]} &   55.47~c   &             &             &             &   63.93~c   &             &             &             \\
 3869 & {[Ne III]} &  163.71~c   &             &             &             &    2.60:c   &             &             &             \\
\hline
\end{tabular}
\end{table*}

\setcounter{table}{1}
\begin{table*}
\caption{ 
  Continued
}
\begin{tabular}{     l @{\hspace{-0.60cm}}
                     l @{\hspace{0.40cm}}
                     r r r r r r r r}
\hline
 PN G       &      &  357.5+03.1 &  357.5+03.2 &  357.6-03.3 &  357.9-03.8 &  358.5-04.2 &  358.5-04.2 &  359.0-04.1 &  359.0-04.8 \\
 Main Name  &      &  Th3-16  ~~ &  M3-42   ~~ &  H2-29   ~~ &  H2-30   ~~ &  H1-46   ~~ &  H1-46   ~~ &  M3-48   ~~ &  M2-25      \\
 sample     &      &  ESO'02  ~~ &  ESO'02  ~~ &  ESO'02  ~~ &  ESO'02  ~~ &  CTIO'02    &  CTIO'01    &  ESO'02  ~~ &  CTIO'01    \\
            &      &             &             &             &             &             &             &             &             \\
 ext. C     &      &    1.61  ~~ &    1.86  ~~ &    1.75  ~~ &    1.32  ~~ &    1.38  ~~ &    1.41  ~~ &    1.17  ~~ &    0.99  ~~ \\
            &      &             &             &             &             &             &             &             &             \\
 3727 & {[O\ III]} &  324.60  ~~ &  108.51  ~~ &   53.08~c   &             &   68.40~c   &             &  171.89~c   &             \\
 3869 & {[Ne III]} &             &  105.05  ~~ &    4.67:c   &   59.72~c   &   26.60~c   &             &   80.47~c   &             \\
\hline
\end{tabular}
\end{table*}

\setcounter{table}{2}
\begin{table*}
\caption{ Plasma parameters and abundances. The first row for each PN gives
 parameters computed from the nominal values of the observational data. The
 second and third row give the upper and lower limits respectively of these
 parameters. Column (1) gives the PNG number; Column (2) usual name; Column
 (3) electron density deduced from \rSii, columns (4) and (5) 
 electron temperatures from \rOiii\ and \rNii\ respectively (the
 value of \Te(N~{\sc ii}) is in parenthesis if \Te(O ~{\sc iii}) was chosen
 for all ions). Columns (6) to (12) give the
 He/H, N/H, O/H, Ne/H, S/H, Ar/H, Cl/H ratios, respectively. Column (13) gives the
 logarithmic extinction C at \Hb\ derived from our spectra.
}
\begin{tabular}{       @{\hspace{0.05cm}}
                     l @{\hspace{0.25cm}}
                     l @{\hspace{0.21cm}}
                     r @{\hspace{0.21cm}}
                     r @{\hspace{0.21cm}}
                     r @{\hspace{0.21cm}}
                     r @{\hspace{0.21cm}}
                     r @{\hspace{0.21cm}}
                     r @{\hspace{0.21cm}}
                     r @{\hspace{0.21cm}}
                     r @{\hspace{0.21cm}}
                     r @{\hspace{0.21cm}}
                     r @{\hspace{0.21cm}}
                     r } 
\hline
        PN G & Main Name & \Ne(S~{\sc ii})  &
                                      \Te(O~{\sc iii}) &
                                              \Te(N~{\sc ii})  &
                                                         He/H     &  N/H     &  O/H     &  Ne/H    &   S/H    &  Ar/H    &  Cl/H    &    ext. C \\
\hline
 000.1-02.3 & Bl 3-10    & 4.46E+02 & 12754 &      -~  & 2.20E-01 &        - & 4.41E-04 & 6.38E-05 &        - & 1.84E-06 &        - &  1.62E+00 \\
            &            & 8.46E+02 & 13440 &          & 2.72E-01 &          & 5.16E-04 & 9.41E-05 &          & 1.95E-06 &          &  1.71E+00 \\
            &            & 1.36E+02 & 12405 &          & 1.49E-01 &          & 3.60E-04 & 6.01E-05 &          & 1.51E-06 &          &  1.55E+00 \\
            &            &          &       &          &          &          &          &          &          &          &          &           \\
 000.3-04.6 & M 2-28     & 1.16E+03 &  8431 &   8190~  & 1.69E-01 & 7.34E-04 & 6.03E-04 & 2.36E-04 & 1.16E-05 & 5.02E-06 & 6.46E-07 &  1.39E+00 \\
            &            & 1.40E+03 &  8813 &   8399~  & 1.80E-01 & 9.14E-04 & 7.55E-04 & 3.12E-04 & 1.54E-05 & 6.13E-06 & 9.55E-07 &  1.49E+00 \\
            &            & 9.72E+02 &  7913 &   7921~  & 1.59E-01 & 6.01E-04 & 5.17E-04 & 1.92E-04 & 9.32E-06 & 4.19E-06 & 4.27E-07 &  1.27E+00 \\
            &            &          &       &          &          &          &          &          &          &          &          &           \\
 000.4-01.9 & M 2-20     & 3.90E+03 &  7835 &   8782~  & 1.44E-01 & 2.27E-04 & 5.86E-04 & 4.38E-05 & 1.56E-05 & 5.38E-06 & 3.85E-06 &  1.72E+00 \\
            &            & 5.06E+03 &  8203 &   8965~  & 1.53E-01 & 2.99E-04 & 7.54E-04 & 6.03E-05 & 2.33E-05 & 6.71E-06 & 7.40E-06 &  1.80E+00 \\
            &            & 3.11E+03 &  7419 &   8456~  & 1.33E-01 & 1.71E-04 & 4.89E-04 & 3.88E-05 & 1.14E-05 & 4.46E-06 & 2.45E-06 &  1.62E+00 \\
            &            &          &       &          &          &          &          &          &          &          &          &           \\
 000.4-02.9 & M 3-19     & 1.36E+03 &  8670 & (23458)  & 1.82E-01 & 4.95E-05 & 3.39E-04 & 1.15E-04 & 2.21E-06 & 2.40E-06 &        - &  1.44E+00 \\
            &            & 1.60E+03 &  9588 & (28027)  & 1.90E-01 & 3.57E-05 & 6.83E-04 & 2.42E-04 & 3.19E-06 & 3.51E-06 &          &  1.55E+00 \\
            &            & 1.09E+03 &  7496 & (18968)  & 1.68E-01 & 2.80E-05 & 2.15E-04 & 6.71E-05 & 1.01E-06 & 1.76E-06 &          &  1.35E+00 \\
            &            &          &       &          &          &          &          &          &          &          &          &           \\
 000.5-03.1 & KFL 1      & 2.22E+03 & 10250 &      -~  & 1.67E-01 & 1.33E-05 & 3.10E-04 & 5.30E-05 & 2.24E-06 & 1.55E-06 &        - &  1.46E+00 \\
            &            & 1.00E+05 & 10687 &       ~  & 1.74E-01 & 3.71E-05 & 3.99E-04 & 7.63E-05 & 4.63E-06 & 1.78E-06 &          &  1.57E+00 \\
            &            & 6.70E+02 &  9698 &       ~  & 1.56E-01 & 9.33E-06 & 2.52E-04 & 4.30E-05 & 1.67E-06 & 1.31E-06 &          &  1.38E+00 \\
            &            &          &       &          &          &          &          &          &          &          &          &           \\
 000.7-02.7 & M 2-21     & 4.83E+03 & 12685 & (16125)  & 1.19E-01 & 5.57E-05 & 2.60E-04 & 4.92E-05 & 2.00E-06 & 6.81E-07 & 6.60E-07 &  1.18E+00 \\
            &            & 6.57E+03 & 13181 & (17775)  & 1.25E-01 & 5.92E-05 & 2.94E-04 & 5.72E-05 & 2.27E-06 & 7.54E-07 & 8.17E-07 &  1.25E+00 \\
            &            & 3.68E+03 & 12306 & (14017)  & 1.12E-01 & 4.65E-05 & 2.18E-04 & 4.55E-05 & 1.64E-06 & 6.07E-07 & 5.07E-07 &  1.05E+00 \\
            &            &          &       &          &          &          &          &          &          &          &          &           \\
 000.7-03.7 & M 3-22     & 1.94E+03 & 14767 &      -~  & 1.26E-01 & 4.56E-05 & 2.95E-04 &        - & 6.72E-07 & 1.20E-06 & 8.49E-07 &  1.03E+00 \\
            &            & 2.39E+03 & 15195 &       ~  & 1.33E-01 & 5.32E-05 & 3.39E-04 &          & 7.55E-07 & 1.36E-06 & 1.08E-06 &  1.09E+00 \\
            &            & 1.65E+03 & 14121 &       ~  & 1.18E-01 & 4.00E-05 & 2.60E-04 &          & 6.21E-07 & 1.09E-06 & 6.35E-07 &  9.42E-01 \\
            &            &          &       &          &          &          &          &          &          &          &          &           \\
 000.9-02.0 & Bl 3-13    &        - &  8572 &      -~  & 1.14E-01 & 2.75E-05 & 5.86E-04 & 1.58E-04 &        - & 2.80E-06 &        - &  1.90E+00 \\
            &            &          &  8912 &       ~  & 1.19E-01 & 3.92E-05 & 7.98E-04 & 2.16E-04 &          & 3.54E-06 &          &  1.97E+00 \\
            &            &          &  7984 &       ~  & 1.07E-01 & 1.26E-05 & 4.71E-04 & 1.28E-04 &          & 2.44E-06 &          &  1.81E+00 \\
            &            &          &       &          &          &          &          &          &          &          &          &           \\
 000.9-04.8 & M 3-23     & 1.03E+03 & 13211 & (11000)  & 1.25E-01 & 3.18E-04 & 4.22E-04 & 8.83E-05 & 1.28E-05 & 3.43E-06 & 5.79E-06 &  1.33E+00 \\
            &            & 1.23E+03 & 13537 & (12697)  & 1.32E-01 & 4.00E-04 & 5.08E-04 & 1.00E-04 & 1.55E-05 & 3.89E-06 & 7.47E-06 &  1.44E+00 \\
            &            & 9.05E+02 & 12700 & ( 8431)  & 1.18E-01 & 2.83E-04 & 3.62E-04 & 7.57E-05 & 1.11E-05 & 3.08E-06 & 4.92E-06 &  1.26E+00 \\
            &            &          &       &          &          &          &          &          &          &          &          &           \\
 001.7-04.4 & H 1-55     & 2.51E+03 &     - &   6245~  & 5.81E-02 & 2.81E-04 & 2.83E-04 &        - & 1.22E-05 & 1.92E-06 & 5.60E-07 &  1.04E+00 \\
            &            & 2.99E+03 &       &   6427~  & 6.17E-02 & 3.22E-04 & 3.44E-04 &          & 1.42E-05 & 2.12E-06 & 6.73E-07 &  1.12E+00 \\
            &            & 2.08E+03 &       &   6085~  & 5.41E-02 & 2.31E-04 & 2.21E-04 &          & 1.04E-05 & 1.70E-06 & 4.43E-07 &  9.78E-01 \\
            &            &          &       &          &          &          &          &          &          &          &          &           \\
 001.7-04.6 & H 1-56     & 1.95E+03 &  7810 &  10146~  & 1.23E-01 & 1.81E-04 & 5.62E-04 &        - & 2.44E-06 & 3.28E-06 & 4.33E-06 &  6.69E-01 \\
            &            & 2.35E+03 &  7917 &  10947~  & 1.32E-01 & 2.50E-04 & 6.35E-04 &          & 3.02E-06 & 3.68E-06 & 5.81E-06 &  7.34E-01 \\
            &            & 1.67E+03 &  7632 &   9362~  & 1.17E-01 & 1.38E-04 & 5.02E-04 &          & 2.17E-06 & 2.94E-06 & 3.44E-06 &  6.08E-01 \\
            &            &          &       &          &          &          &          &          &          &          &          &           \\
 002.1-02.2 & M 3-20     & 4.06E+03 & 10494 & (13771)  & 1.12E-01 & 6.86E-05 & 3.98E-04 & 8.36E-05 & 4.82E-06 & 1.20E-06 & 1.74E-06 &  1.49E+00 \\
            &            & 5.15E+03 & 10774 & (15648)  & 1.19E-01 & 7.30E-05 & 4.40E-04 & 9.53E-05 & 5.25E-06 & 1.31E-06 & 1.97E-06 &  1.57E+00 \\
            &            & 3.13E+03 & 10245 & (12276)  & 1.06E-01 & 5.61E-05 & 3.43E-04 & 7.69E-05 & 4.10E-06 & 1.12E-06 & 1.32E-06 &  1.40E+00 \\
            &            &          &       &          &          &          &          &          &          &          &          &           \\
 002.1-04.2 & H 1-54     & 7.36E+03 &  9737 &  15602~  & 1.08E-01 & 6.24E-05 & 2.04E-04 & 3.33E-05 & 6.61E-06 & 1.36E-06 & 2.65E-06 &  1.13E+00 \\
            &            & 1.16E+04 & 10410 &  16912~  & 1.15E-01 & 8.09E-05 & 2.69E-04 & 4.32E-05 & 9.51E-06 & 1.70E-06 & 4.45E-06 &  1.21E+00 \\
            &            & 5.24E+03 &  9113 &  14351~  & 1.00E-01 & 4.54E-05 & 1.59E-04 & 2.29E-05 & 4.53E-06 & 1.11E-06 & 1.24E-06 &  1.04E+00 \\
            &            &          &       &          &          &          &          &          &          &          &          &           \\
 002.2-02.5 & KFL 2      &        - & 12522 &      -~  & 1.12E-01 & 3.01E-05 & 2.99E-04 & 3.38E-05 &        - & 1.17E-06 &        - &  1.39E+00 \\
            &            &          & 13262 &       ~  & 1.17E-01 & 4.17E-05 & 3.30E-04 & 4.14E-05 &          & 1.31E-06 &          &  1.50E+00 \\
            &            &          & 12313 &       ~  & 1.07E-01 & 1.75E-05 & 2.41E-04 & 2.85E-05 &          & 9.48E-07 &          &  1.30E+00 \\
            &            &          &       &          &          &          &          &          &          &          &          &           \\
 002.2-02.7 & M 2-23     & 1.23E+04 & 12237 & (25971)  & 1.05E-01 & 2.71E-05 & 2.05E-04 & 5.48E-05 & 1.40E-06 & 8.90E-07 & 3.86E-07 &  1.05E+00 \\
            &            & 2.43E+04 & 12698 & (29578)  & 1.13E-01 & 3.11E-05 & 2.42E-04 & 5.77E-05 & 2.33E-06 & 9.89E-07 & 5.24E-07 &  1.14E+00 \\
            &            & 8.35E+03 & 11903 & (19808)  & 9.72E-02 & 2.19E-05 & 1.80E-04 & 4.33E-05 & 1.01E-06 & 7.86E-07 & 3.02E-07 &  9.65E-01 \\
            &            &          &       &          &          &          &          &          &          &          &          &           \\
 002.3-03.4 & H 2-37     & 3.00E+01 & 14086 &      -~  & 1.91E-01 & 1.15E-04 & 1.14E-04 & 7.14E-05 & 2.44E-06 & 9.16E-07 &        - &  1.33E+00 \\
            &            & 3.33E+01 & 15892 &       ~  & 2.06E-01 & 2.07E-04 & 2.51E-04 & 8.61E-05 & 4.96E-06 & 1.40E-06 &          &  1.69E+00 \\
            &            & 3.00E+01 & 11344 &       ~  & 1.76E-01 & 8.08E-05 & 6.73E-05 & 4.16E-05 & 1.71E-06 & 7.74E-07 &          &  1.07E+00 \\
\hline
\end{tabular}
\end{table*}

\setcounter{table}{2}
\begin{table*}
\caption{ 
  Continued
}
\begin{tabular}{       @{\hspace{0.05cm}}
                     l @{\hspace{0.25cm}}
                     l @{\hspace{0.21cm}}
                     r @{\hspace{0.21cm}}
                     r @{\hspace{0.21cm}}
                     r @{\hspace{0.21cm}}
                     r @{\hspace{0.21cm}}
                     r @{\hspace{0.21cm}}
                     r @{\hspace{0.21cm}}
                     r @{\hspace{0.21cm}}
                     r @{\hspace{0.21cm}}
                     r @{\hspace{0.21cm}}
                     r @{\hspace{0.21cm}}
                     r } 
\hline
        PN G & Main Name & \Ne(S~{\sc ii})  &
                                      \Te(O~{\sc iii}) &
                                              \Te(N~{\sc ii})  &
                                                         He/H     &  N/H     &  O/H     &  Ne/H    &   S/H    &  Ar/H    &  Cl/H    &    ext. C \\
\hline
 002.4-03.7 & M 1-38     & 5.49E+03 &     - &   7414~  & 1.32E-02 & 8.91E-05 & 2.58E-04 &        - & 4.41E-06 & 3.12E-07 &        - &  1.08E+00 \\
            &            & 7.60E+03 &       &   7919~  & 1.41E-02 & 1.22E-04 & 5.00E-04 &          & 7.15E-06 & 4.31E-07 &          &  1.35E+00 \\
            &            & 4.15E+03 &       &   6769~  & 1.22E-02 & 6.68E-05 & 1.36E-04 &          & 3.20E-06 & 2.39E-07 &          &  8.62E-01 \\
            &            &          &       &          &          &          &          &          &          &          &          &           \\
 002.5-01.7 & Pe 2-11    & 1.94E+02 &     - &   8889~  & 1.71E-01 & 4.94E-04 & 5.67E-04 & 1.75E-04 & 2.04E-05 & 5.44E-06 &        - &  2.15E+00 \\
            &            & 1.11E+03 &       &   9240~  & 1.96E-01 & 6.27E-04 & 8.46E-04 & 3.74E-04 & 3.63E-05 & 9.61E-06 &          &  2.40E+00 \\
            &            & 3.00E+01 &       &   8230~  & 1.52E-01 & 3.90E-04 & 4.84E-04 & 1.57E-04 & 1.50E-05 & 3.94E-06 &          &  1.82E+00 \\
            &            &          &       &          &          &          &          &          &          &          &          &           \\
 002.6+04.2 & Th 3-27    & 5.09E+03 & 11084 &  11980~  & 1.53E-01 & 1.11E-03 & 4.13E-04 &        - & 1.45E-05 & 6.02E-06 & 1.45E-06 &  2.47E+00 \\
            &            & 7.53E+03 & 11336 &  12414~  & 1.62E-01 & 1.29E-03 & 4.63E-04 &          & 1.86E-05 & 6.49E-06 & 1.95E-06 &  2.54E+00 \\
            &            & 4.31E+03 & 10849 &  11374~  & 1.42E-01 & 9.62E-04 & 3.68E-04 &          & 1.15E-05 & 5.38E-06 & 9.75E-07 &  2.43E+00 \\
            &            &          &       &          &          &          &          &          &          &          &          &           \\
 002.8-02.2 & Pe 2-12    & 1.10E+03 &     - &   6897~  & 7.24E-02 & 1.97E-04 & 2.76E-04 &        - & 1.36E-05 & 1.33E-06 &        - &  1.70E+00 \\
            &            & 1.00E+05 &       &   7072~  & 7.94E-02 & 2.57E-03 & 1.23E-02 &          & 6.46E-04 & 5.89E-06 &          &  1.84E+00 \\
            &            & 4.07E+02 &       &   4401~  & 6.58E-02 & 1.76E-04 & 2.31E-04 &          & 1.09E-05 & 1.12E-06 &          &  1.52E+00 \\
            &            &          &       &          &          &          &          &          &          &          &          &           \\
 002.9-03.9 & H 2-39     & 2.38E+03 & 13197 & (14648)  & 1.05E-01 & 7.98E-05 & 2.58E-04 & 5.36E-05 & 2.53E-06 & 1.07E-06 & 1.80E-06 &  1.30E+00 \\
            &            & 2.90E+03 & 13727 & (18711)  & 1.12E-01 & 8.85E-05 & 2.97E-04 & 6.72E-05 & 3.04E-06 & 1.19E-06 & 2.16E-06 &  1.38E+00 \\
            &            & 1.96E+03 & 12763 & (10222)  & 1.01E-01 & 6.33E-05 & 2.21E-04 & 5.09E-05 & 1.76E-06 & 9.52E-07 & 1.06E-06 &  1.19E+00 \\
            &            &          &       &          &          &          &          &          &          &          &          &           \\
 003.0-02.6 & KFL 4      &        - & 11383 &      -~  & 1.59E-01 &        - & 3.20E-04 & 8.03E-05 &        - &        - &        - &  1.25E+00 \\
            &            &          & 11961 &       ~  & 1.72E-01 &          & 4.66E-04 & 1.02E-04 &          &          &          &  1.45E+00 \\
            &            &          & 10118 &       ~  & 1.44E-01 &          & 2.48E-04 & 5.01E-05 &          &          &          &  9.91E-01 \\
            &            &          &       &          &          &          &          &          &          &          &          &           \\
 003.2-04.4 & KFL 12     & 1.00E+05 &  8453 &      -~  & 1.01E-01 & 1.12E-05 & 7.74E-04 & 1.78E-04 & 3.11E-06 & 1.94E-06 &        - &  1.17E+00 \\
            &            & 1.00E+05 &  8681 &       ~  & 1.08E-01 & 1.47E-05 & 9.06E-04 & 1.94E-04 & 3.82E-06 & 2.16E-06 &          &  1.25E+00 \\
            &            & 1.00E+05 &  8242 &       ~  & 9.48E-02 & 9.51E-06 & 6.64E-04 & 1.53E-04 & 2.71E-06 & 1.73E-06 &          &  1.08E+00 \\
            &            &          &       &          &          &          &          &          &          &          &          &           \\
 003.3-04.6 & Ap 1-12    & 3.95E+03 &     - &   6074~  & 5.92E-03 & 1.95E-04 & 1.50E-04 &        - & 2.15E-05 &        - &        - &  7.02E-01 \\
            &            & 4.78E+03 &       &   6456~  & 6.94E-03 & 2.46E-04 & 2.38E-04 &          & 2.87E-05 &          &          &  7.81E-01 \\
            &            & 2.90E+03 &       &   5681~  & 4.39E-03 & 1.41E-04 & 8.22E-05 &          & 1.57E-05 &          &          &  6.03E-01 \\
            &            &          &       &          &          &          &          &          &          &          &          &           \\
 003.5-02.4 & IC 4673    & 1.02E+03 & 11089 & (11622)  & 8.69E-02 & 1.69E-04 & 2.99E-04 & 6.80E-05 & 9.64E-06 & 3.76E-06 & 3.41E-06 &  1.21E+00 \\
            &            & 1.16E+03 & 11538 & (12537)  & 9.18E-02 & 1.97E-04 & 3.42E-04 & 7.57E-05 & 1.08E-05 & 4.02E-06 & 4.19E-06 &  1.31E+00 \\
            &            & 8.48E+02 & 10802 & (10270)  & 8.24E-02 & 1.47E-04 & 2.49E-04 & 6.03E-05 & 8.06E-06 & 3.32E-06 & 2.48E-06 &  1.12E+00 \\
            &            &          &       &          &          &          &          &          &          &          &          &           \\
 003.6-02.3 & M 2-26     & 5.42E+02 &  9023 &   8565~  & 1.77E-01 & 4.28E-04 & 4.13E-04 & 1.30E-04 & 6.37E-06 & 3.30E-06 & 6.20E-07 &  1.64E+00 \\
            &            & 6.46E+02 &  9590 &   8809~  & 1.86E-01 & 5.28E-04 & 5.22E-04 & 1.75E-04 & 8.64E-06 & 3.95E-06 & 9.85E-07 &  1.72E+00 \\
            &            & 4.64E+02 &  8430 &   8365~  & 1.65E-01 & 3.08E-04 & 3.33E-04 & 1.12E-04 & 3.93E-06 & 2.80E-06 & 3.42E-07 &  1.54E+00 \\
            &            &          &       &          &          &          &          &          &          &          &          &           \\
 003.7-04.6 & M 2-30     & 3.10E+03 & 11042 &  12860~  & 1.21E-01 & 1.37E-04 & 4.46E-04 & 8.94E-05 & 8.78E-06 & 2.59E-06 & 1.07E-05 &  8.34E-01 \\
            &            & 3.70E+03 & 11245 &  13403~  & 1.25E-01 & 1.64E-04 & 5.25E-04 & 1.05E-04 & 1.06E-05 & 2.88E-06 & 1.48E-05 &  9.24E-01 \\
            &            & 2.52E+03 & 10650 &  12362~  & 1.13E-01 & 1.10E-04 & 3.99E-04 & 8.45E-05 & 7.42E-06 & 2.37E-06 & 8.26E-06 &  7.63E-01 \\
            &            &          &       &          &          &          &          &          &          &          &          &           \\
 003.8-04.3 & H 1-59     & 1.73E+03 & 11161 &  10360~  & 1.45E-01 & 3.33E-04 & 4.32E-04 & 8.54E-05 & 1.03E-05 & 4.15E-06 & 2.23E-06 &  8.74E-01 \\
            &            & 2.02E+03 & 11407 &  10726~  & 1.53E-01 & 4.14E-04 & 5.16E-04 & 1.05E-04 & 1.28E-05 & 4.69E-06 & 3.06E-06 &  9.43E-01 \\
            &            & 1.43E+03 & 10742 &  10027~  & 1.36E-01 & 2.67E-04 & 3.80E-04 & 8.49E-05 & 8.74E-06 & 3.86E-06 & 1.70E-06 &  7.81E-01 \\
            &            &          &       &          &          &          &          &          &          &          &          &           \\
 003.8-04.5 & H 2-41     & 4.87E+02 & 10384 & (21181)  & 2.12E-01 & 2.50E-05 & 2.03E-04 &        - & 1.90E-06 & 1.38E-06 &        - &  8.95E-01 \\
            &            & 6.24E+02 & 11033 & (24762)  & 2.25E-01 & 2.84E-05 & 2.78E-04 &          & 2.83E-06 & 1.83E-06 &          &  9.58E-01 \\
            &            & 4.10E+02 &  9579 & (17555)  & 1.98E-01 & 2.23E-05 & 1.60E-04 &          & 1.46E-06 & 1.20E-06 &          &  8.17E-01 \\
            &            &          &       &          &          &          &          &          &          &          &          &           \\
 003.9-03.1 & KFL 7      &        - & 13347 &      -~  & 8.70E-02 &        - & 1.41E-04 & 2.11E-05 &        - & 1.22E-06 &        - &  1.25E+00 \\
            &            &          & 16086 &       ~  & 9.11E-02 &          & 2.38E-04 & 3.57E-05 &          & 1.94E-06 &          &  1.32E+00 \\
            &            &          & 10748 &       ~  & 8.02E-02 &          & 8.48E-05 & 1.17E-05 &          & 6.64E-07 &          &  1.10E+00 \\
            &            &          &       &          &          &          &          &          &          &          &          &           \\
 004.1-03.8 & KFL 11     & 9.42E+02 & 10929 &  13580~  & 1.09E-01 & 8.07E-05 & 2.46E-04 & 5.57E-05 & 2.52E-06 & 1.27E-06 & 9.75E-07 &  1.32E+00 \\
            &            & 8.85E+03 & 11286 &  14286~  & 1.16E-01 & 9.63E-05 & 3.01E-04 & 6.89E-05 & 5.55E-06 & 1.48E-06 & 1.35E-06 &  1.47E+00 \\
            &            & 2.86E+02 & 10517 &  12271~  & 9.66E-02 & 6.16E-05 & 2.08E-04 & 5.06E-05 & 2.07E-06 & 1.10E-06 & 8.04E-07 &  1.18E+00 \\
            &            &          &       &          &          &          &          &          &          &          &          &           \\
 004.2-03.2 & KFL 10     & 1.19E+03 &  9681 & (24839)  & 1.74E-01 & 3.36E-05 & 3.00E-04 & 8.29E-05 & 2.94E-06 & 1.89E-06 & 1.66E-06 &  7.42E-01 \\
            &            & 1.45E+03 & 10017 & (26748)  & 1.83E-01 & 2.74E-05 & 3.66E-04 & 1.01E-04 & 3.06E-06 & 2.06E-06 & 1.75E-06 &  8.16E-01 \\
            &            & 1.01E+03 &  9461 & (22744)  & 1.63E-01 & 2.19E-05 & 2.66E-04 & 7.68E-05 & 2.34E-06 & 1.71E-06 & 1.30E-06 &  6.45E-01 \\
            &            &          &       &          &          &          &          &          &          &          &          &           \\
 004.2-04.3 & H 1-60     & 2.46E+03 &  9876 &      -~  & 1.17E-01 & 2.89E-05 & 3.78E-04 &        - & 3.16E-06 & 1.36E-06 &        - &  6.17E-01 \\
            &            & 6.20E+03 & 10011 &       ~  & 1.24E-01 & 4.05E-05 & 4.71E-04 &          & 4.07E-06 & 1.58E-06 &          &  6.75E-01 \\
            &            & 1.30E+03 &  9513 &       ~  & 1.10E-01 & 2.19E-05 & 3.45E-04 &          & 2.39E-06 & 1.21E-06 &          &  5.44E-01 \\
\hline
\end{tabular}
\end{table*}

\setcounter{table}{2}
\begin{table*}
\caption{ 
  Continued
}
\begin{tabular}{       @{\hspace{0.05cm}}
                     l @{\hspace{0.25cm}}
                     l @{\hspace{0.21cm}}
                     r @{\hspace{0.21cm}}
                     r @{\hspace{0.21cm}}
                     r @{\hspace{0.21cm}}
                     r @{\hspace{0.21cm}}
                     r @{\hspace{0.21cm}}
                     r @{\hspace{0.21cm}}
                     r @{\hspace{0.21cm}}
                     r @{\hspace{0.21cm}}
                     r @{\hspace{0.21cm}}
                     r @{\hspace{0.21cm}}
                     r } 
\hline
        PN G & Main Name & \Ne(S~{\sc ii})  &
                                      \Te(O~{\sc iii}) &
                                              \Te(N~{\sc ii})  &
                                                         He/H     &  N/H     &  O/H     &  Ne/H    &   S/H    &  Ar/H    &  Cl/H    &    ext. C \\
\hline
 004.8-05.0 & M 3-26     & 1.03E+03 &  9818 & (26251)  & 1.73E-01 & 2.82E-05 & 2.89E-04 & 7.83E-05 & 2.62E-06 & 1.82E-06 & 1.50E-06 &  7.63E-01 \\
            &            & 1.25E+03 & 10197 & (28483)  & 1.84E-01 & 2.43E-05 & 3.31E-04 & 9.52E-05 & 2.97E-06 & 1.95E-06 & 1.70E-06 &  8.61E-01 \\
            &            & 9.06E+02 &  9697 & (24251)  & 1.64E-01 & 1.95E-05 & 2.49E-04 & 7.02E-05 & 1.84E-06 & 1.58E-06 & 9.53E-07 &  6.65E-01 \\
            &            &          &       &          &          &          &          &          &          &          &          &           \\
 004.9+04.9 & M 1-25     & 4.85E+03 &  8058 &   8672~  & 1.47E-01 & 2.85E-04 & 4.79E-04 & 3.09E-05 & 1.49E-05 & 5.86E-06 & 2.34E-06 &  1.42E+00 \\
            &            & 6.50E+03 &  8267 &   8873~  & 1.55E-01 & 3.11E-04 & 5.49E-04 & 4.08E-05 & 1.73E-05 & 6.36E-06 & 2.87E-06 &  1.54E+00 \\
            &            & 3.63E+03 &  7903 &   8342~  & 1.35E-01 & 2.27E-04 & 4.35E-04 & 2.89E-05 & 1.29E-05 & 5.16E-06 & 1.55E-06 &  1.34E+00 \\
            &            &          &       &          &          &          &          &          &          &          &          &           \\
 005.0+03.0 & Pe 1- 9    & 4.73E+02 &     - &  11084~  & 1.82E-01 & 1.76E-05 & 8.21E-05 & 2.11E-05 & 1.30E-06 & 6.91E-07 &        - &  1.59E+00 \\
            &            & 2.79E+03 &       &  11841~  & 1.92E-01 & 2.81E-05 & 1.31E-04 & 4.10E-05 & 2.47E-06 & 9.90E-07 &          &  1.66E+00 \\
            &            & 4.55E+01 &       &   9512~  & 1.70E-01 & 1.64E-05 & 5.76E-05 & 1.57E-05 & 9.68E-07 & 5.60E-07 &          &  1.50E+00 \\
            &            &          &       &          &          &          &          &          &          &          &          &           \\
 005.5-02.5 & M 3-24     & 1.18E+03 &     - &   9793~  & 1.65E-01 & 2.18E-04 & 3.22E-04 & 1.02E-04 & 5.33E-06 & 2.56E-06 & 1.10E-06 &  1.53E+00 \\
            &            & 1.45E+03 &       &  10118~  & 1.71E-01 & 2.48E-04 & 3.55E-04 & 1.18E-04 & 5.94E-06 & 2.86E-06 & 1.24E-06 &  1.63E+00 \\
            &            & 9.69E+02 &       &   9515~  & 1.56E-01 & 2.00E-04 & 2.80E-04 & 8.09E-05 & 4.69E-06 & 2.26E-06 & 9.57E-07 &  1.44E+00 \\
            &            &          &       &          &          &          &          &          &          &          &          &           \\
 005.6-04.7 & KFL 16     & 8.68E+02 & 12471 &   8685~  & 1.54E-01 & 4.97E-05 & 3.67E-04 & 7.76E-05 & 3.75E-06 & 3.12E-06 &        - &  6.83E-01 \\
            &            & 8.79E+03 & 12857 &   8982~  & 1.60E-01 & 8.25E-05 & 4.81E-04 & 9.13E-05 & 8.38E-06 & 3.64E-06 &          &  8.41E-01 \\
            &            & 2.77E+02 & 11500 &   8227~  & 1.42E-01 & 4.74E-05 & 3.20E-04 & 4.83E-05 & 3.49E-06 & 2.70E-06 &          &  5.12E-01 \\
            &            &          &       &          &          &          &          &          &          &          &          &           \\
 005.7-03.6 & KFL 13     & 1.62E+02 &  8232 &   8423~  & 1.69E-01 & 1.75E-04 & 4.87E-04 &        - & 7.70E-06 & 2.91E-06 & 3.65E-07 &  1.24E+00 \\
            &            & 2.13E+02 &  8861 &   8949~  & 1.79E-01 & 3.32E-04 & 9.11E-04 &          & 1.82E-05 & 4.65E-06 & 1.60E-06 &  1.30E+00 \\
            &            & 1.11E+02 &  7106 &   7869~  & 1.57E-01 & 1.36E-04 & 3.76E-04 &          & 5.30E-06 & 2.35E-06 & 1.84E-07 &  1.17E+00 \\
            &            &          &       &          &          &          &          &          &          &          &          &           \\
 006.0+02.8 & Th 4- 3    & 6.78E+03 &     - &   8679~  & 1.09E-02 & 5.27E-05 & 2.32E-04 &        - & 8.77E-06 & 2.53E-07 &        - &  1.71E+00 \\
            &            & 1.00E+05 &       &   9478~  & 1.20E-02 & 8.18E-04 & 2.19E-02 &          & 4.28E-04 & 1.32E-06 &          &  1.83E+00 \\
            &            & 1.24E+03 &       &   5212~  & 1.01E-02 & 3.41E-05 & 1.55E-04 &          & 3.84E-06 & 1.78E-07 &          &  1.55E+00 \\
            &            &          &       &          &          &          &          &          &          &          &          &           \\
 006.0+03.1 & M 1-28     & 1.64E+02 & 11094 &   9844~  & 1.72E-01 & 9.00E-04 & 3.24E-04 & 2.07E-04 & 2.49E-05 & 4.03E-06 &        - &  9.02E-01 \\
            &            & 1.97E+03 & 11276 &  10134~  & 1.87E-01 & 1.38E-03 & 4.82E-04 & 2.86E-04 & 5.02E-05 & 5.37E-06 &          &  1.12E+00 \\
            &            & 3.00E+01 & 10443 &   9173~  & 1.46E-01 & 7.85E-04 & 3.00E-04 & 1.80E-04 & 1.97E-05 & 3.21E-06 &          &  7.03E-01 \\
            &            &          &       &          &          &          &          &          &          &          &          &           \\
 006.0-03.6 & M 2-31     & 3.85E+03 &  9687 &  12238~  & 1.24E-01 & 3.85E-04 & 4.55E-04 &        - & 1.38E-05 & 2.60E-06 & 3.25E-06 &  1.37E+00 \\
            &            & 4.94E+03 &  9875 &  12718~  & 1.33E-01 & 4.66E-04 & 5.20E-04 &          & 1.60E-05 & 2.89E-06 & 4.14E-06 &           \\
            &            & 2.88E+03 &  9467 &  11844~  & 1.16E-01 & 3.35E-04 & 4.07E-04 &          & 1.23E-05 & 2.37E-06 & 2.58E-06 &           \\
            &            &          &       &          &          &          &          &          &          &          &          &           \\
 006.2-03.7 & KFL 15     & 4.92E+02 & 14141 &      -~  & 1.05E-01 & 7.81E-05 & 3.18E-04 & 7.09E-05 & 5.12E-06 & 1.99E-06 &        - &  1.49E+00 \\
            &            & 4.01E+03 & 14588 &       ~  & 1.11E-01 & 1.08E-04 & 4.40E-04 & 8.72E-05 & 7.17E-06 & 2.27E-06 &          &  1.60E+00 \\
            &            & 3.00E+01 & 13197 &       ~  & 9.59E-02 & 5.92E-05 & 2.69E-04 & 5.50E-05 & 2.97E-06 & 1.76E-06 &          &  1.39E+00 \\
            &            &          &       &          &          &          &          &          &          &          &          &           \\
 006.3+03.3 & H 2-22     & 4.91E+02 &     - &   6630~  & 1.10E-01 & 2.14E-04 & 3.37E-04 &        - & 1.73E-05 & 2.00E-06 &        - &  2.04E+00 \\
            &            & 2.41E+03 &       &   7497~  & 1.21E-01 & 5.11E-04 & 1.80E-03 &          & 4.17E-05 & 4.65E-06 &          &  2.18E+00 \\
            &            & 7.83E+01 &       &   5486~  & 9.91E-02 & 1.46E-04 & 1.43E-04 &          & 1.10E-05 & 1.35E-06 &          &  1.89E+00 \\
            &            &          &       &          &          &          &          &          &          &          &          &           \\
 006.3+04.4 & H 2-18     & 2.26E+03 & 10928 & (13871)  & 1.10E-01 & 6.10E-05 & 3.67E-04 & 8.10E-05 & 4.27E-06 & 1.10E-06 & 2.22E-06 &  1.63E+00 \\
            &            & 2.72E+03 & 11201 & (16891)  & 1.18E-01 & 7.03E-05 & 4.36E-04 & 8.76E-05 & 5.14E-06 & 1.25E-06 & 3.35E-06 &  1.72E+00 \\
            &            & 1.87E+03 & 10564 & (10749)  & 1.04E-01 & 5.10E-05 & 3.19E-04 & 7.17E-05 & 3.32E-06 & 1.01E-06 & 1.42E-06 &  1.51E+00 \\
            &            &          &       &          &          &          &          &          &          &          &          &           \\
 006.4+02.0 & M 1-31     & 8.66E+03 &  7639 &  10738~  & 1.54E-01 & 7.37E-04 & 8.05E-04 & 2.27E-04 & 2.80E-05 & 6.80E-06 & 8.05E-06 &  1.98E+00 \\
            &            & 1.40E+04 &  7980 &  11414~  & 1.67E-01 & 9.09E-04 & 1.06E-03 & 3.20E-04 & 3.89E-05 & 8.09E-06 & 1.37E-05 &  2.07E+00 \\
            &            & 5.93E+03 &  7264 &   9533~  & 1.45E-01 & 4.77E-04 & 6.54E-04 & 1.95E-04 & 1.95E-05 & 5.80E-06 & 3.81E-06 &  1.87E+00 \\
            &            &          &       &          &          &          &          &          &          &          &          &           \\
 006.4-04.6 & Pe 2-13    & 2.02E+03 & 11814 &      -~  & 1.56E-01 &        - & 4.66E-04 &        - &        - & 3.51E-06 &        - &  1.01E+00 \\
            &            & 2.36E+03 & 12196 &       ~  & 1.65E-01 &          & 5.23E-04 &          &          & 3.94E-06 &          &  1.08E+00 \\
            &            & 1.71E+03 & 11540 &       ~  & 1.48E-01 &          & 4.00E-04 &          &          & 2.96E-06 &          &  9.31E-01 \\
            &            &          &       &          &          &          &          &          &          &          &          &           \\
 006.8+02.3 & Th 4- 7    & 2.07E+03 & 14514 & (15076)  & 1.18E-01 & 1.07E-04 & 2.84E-04 & 4.46E-05 & 2.91E-06 & 9.04E-07 &        - &  1.76E+00 \\
            &            & 2.48E+03 & 15294 & (16954)  & 1.25E-01 & 1.26E-04 & 3.12E-04 & 4.80E-05 & 3.30E-06 & 1.08E-06 &          &  1.87E+00 \\
            &            & 1.78E+03 & 14165 & (12944)  & 1.11E-01 & 9.33E-05 & 2.30E-04 & 3.52E-05 & 2.43E-06 & 6.79E-07 &          &  1.66E+00 \\
            &            &          &       &          &          &          &          &          &          &          &          &           \\
 006.8-03.4 & H 2-45     & 1.06E+04 & 11654 & (17005)  & 1.03E-01 & 7.02E-05 & 2.45E-04 &        - & 3.37E-06 & 6.92E-07 & 3.15E-06 &  1.52E+00 \\
            &            & 1.76E+04 & 11866 & (21259)  & 1.10E-01 & 9.33E-05 & 2.81E-04 &          & 3.91E-06 & 7.71E-07 & 3.64E-06 &  1.60E+00 \\
            &            & 7.17E+03 & 11302 & (10342)  & 9.68E-02 & 6.15E-05 & 2.22E-04 &          & 3.06E-06 & 6.45E-07 & 2.73E-06 &  1.46E+00 \\
            &            &          &       &          &          &          &          &          &          &          &          &           \\
 006.8+04.1 & M 3-15     & 5.40E+03 &  8431 &  10644~  & 1.31E-01 & 3.51E-04 & 6.44E-04 &        - & 1.77E-05 & 4.09E-06 & 6.73E-06 &  2.12E+00 \\
            &            & 8.14E+03 &  8584 &  11056~  & 1.38E-01 & 4.08E-04 & 7.17E-04 &          & 2.10E-05 & 5.11E-06 & 8.57E-06 &  2.18E+00 \\
            &            & 4.07E+03 &  8235 &  10119~  & 1.21E-01 & 3.10E-04 & 5.79E-04 &          & 1.53E-05 & 3.36E-06 & 4.90E-06 &  2.06E+00 \\
\hline
\end{tabular}
\end{table*}

\setcounter{table}{2}
\begin{table*}
\caption{ 
  Continued
}
\begin{tabular}{       @{\hspace{0.05cm}}
                     l @{\hspace{0.25cm}}
                     l @{\hspace{0.21cm}}
                     r @{\hspace{0.21cm}}
                     r @{\hspace{0.21cm}}
                     r @{\hspace{0.21cm}}
                     r @{\hspace{0.21cm}}
                     r @{\hspace{0.21cm}}
                     r @{\hspace{0.21cm}}
                     r @{\hspace{0.21cm}}
                     r @{\hspace{0.21cm}}
                     r @{\hspace{0.21cm}}
                     r @{\hspace{0.21cm}}
                     r } 
\hline
        PN G & Main Name & \Ne(S~{\sc ii})  &
                                      \Te(O~{\sc iii}) &
                                              \Te(N~{\sc ii})  &
                                                         He/H     &  N/H     &  O/H     &  Ne/H    &   S/H    &  Ar/H    &  Cl/H    &    ext. C \\
\hline
 007.8-03.7 & M 2-34     & 3.08E+02 &  8424 &   8726~  & 1.78E-01 & 5.46E-04 & 4.73E-04 &        - & 1.31E-05 & 4.16E-06 & 9.28E-07 &  1.17E+00 \\
            &            & 3.82E+02 &  8830 &   8966~  & 1.90E-01 & 7.25E-04 & 5.91E-04 &          & 1.92E-05 & 4.94E-06 & 1.49E-06 &  1.23E+00 \\
            &            & 2.38E+02 &  7937 &   8489~  & 1.68E-01 & 4.41E-04 & 3.98E-04 &          & 9.75E-06 & 3.58E-06 & 5.56E-07 &  1.10E+00 \\
            &            &          &       &          &          &          &          &          &          &          &          &           \\
 007.8-04.4 & H 1-65     & 6.07E+03 &     - &   5804~  & 9.06E-03 & 2.50E-04 & 1.85E-04 &        - & 4.19E-05 & 3.26E-07 &        - &  9.81E-01 \\
            &            & 9.38E+03 &       &   5946~  & 9.66E-03 & 3.45E-04 & 2.98E-04 &          & 6.80E-05 & 4.13E-07 &          &  1.05E+00 \\
            &            & 4.16E+03 &       &   5532~  & 8.57E-03 & 2.03E-04 & 1.40E-04 &          & 3.09E-05 & 2.84E-07 &          &  8.80E-01 \\
            &            &          &       &          &          &          &          &          &          &          &          &           \\
 008.4-03.6 & H 1-64     & 3.19E+02 &     - &   6562~  & 1.35E-01 & 1.91E-04 & 5.89E-04 &        - & 1.97E-05 & 2.60E-06 &        - &  1.41E+00 \\
            &            & 1.32E+03 &       &   6703~  & 1.43E-01 & 2.77E-04 & 1.00E-03 &          & 3.17E-05 & 3.56E-06 &          &  1.57E+00 \\
            &            & 3.00E+01 &       &   6232~  & 1.22E-01 & 1.61E-04 & 5.51E-04 &          & 1.58E-05 & 2.27E-06 &          &  1.25E+00 \\
            &            &          &       &          &          &          &          &          &          &          &          &           \\
 009.0+04.1 & Th 4- 5    & 1.56E+03 & 10399 &  10878~  & 1.62E-01 & 2.84E-04 & 3.43E-04 &        - & 7.28E-06 & 2.80E-06 & 2.17E-06 &  1.73E+00 \\
            &            & 1.91E+03 & 10637 &  11176~  & 1.71E-01 & 3.28E-04 & 3.76E-04 &          & 8.41E-06 & 3.04E-06 & 2.60E-06 &  1.82E+00 \\
            &            & 1.33E+03 & 10187 &  10570~  & 1.53E-01 & 2.49E-04 & 3.07E-04 &          & 6.28E-06 & 2.53E-06 & 1.77E-06 &  1.67E+00 \\
            &            &          &       &          &          &          &          &          &          &          &          &           \\
 009.3+04.1 & Th 4- 6    & 8.38E+03 & 10053 &      -~  & 1.15E-01 & 1.00E-04 & 3.68E-04 &        - & 5.32E-06 & 9.76E-07 &        - &  1.35E+00 \\
            &            & 1.65E+04 & 10265 &       ~  & 1.22E-01 & 1.36E-04 & 4.29E-04 &          & 6.14E-06 & 1.07E-06 &          &  1.41E+00 \\
            &            & 6.08E+03 &  9794 &       ~  & 1.05E-01 & 8.67E-05 & 3.28E-04 &          & 4.79E-06 & 8.87E-07 &          &  1.29E+00 \\
            &            &          &       &          &          &          &          &          &          &          &          &           \\
 010.4+04.5 & M 2-17     & 1.49E+03 &  9094 &  12549~  & 1.66E-01 & 1.71E-04 & 2.80E-04 &        - & 5.87E-06 & 2.31E-06 & 2.13E-06 &  1.05E+00 \\
            &            & 1.81E+03 &  9248 &  13050~  & 1.78E-01 & 1.93E-04 & 3.08E-04 &          & 6.79E-06 & 2.60E-06 & 2.60E-06 &  1.08E+00 \\
            &            & 1.31E+03 &  8882 &  12096~  & 1.57E-01 & 1.41E-04 & 2.46E-04 &          & 4.93E-06 & 2.15E-06 & 1.68E-06 &  9.64E-01 \\
            &            &          &       &          &          &          &          &          &          &          &          &           \\
 010.6+03.2 & Th 4-10    & 2.13E+03 &  7198 &   8405~  & 1.33E-01 & 3.93E-04 & 4.99E-04 &        - & 1.40E-05 & 4.18E-06 & 2.37E-06 &  1.34E+00 \\
            &            & 2.61E+03 &  7840 &   8635~  & 1.42E-01 & 5.74E-04 & 7.73E-04 &          & 2.65E-05 & 5.89E-06 & 6.47E-06 &  1.41E+00 \\
            &            & 1.66E+03 &  6541 &   8168~  & 1.25E-01 & 2.82E-04 & 3.58E-04 &          & 8.98E-06 & 3.18E-06 & 1.14E-06 &  1.26E+00 \\
            &            &          &       &          &          &          &          &          &          &          &          &           \\
 350.9+04.4 & H 2- 1     & 9.10E+03 & 14419 &  11656~  & 4.16E-02 & 1.45E-05 & 5.80E-05 &        - & 9.25E-07 & 2.61E-07 & 8.13E-09 &  1.01E+00 \\
            &            & 1.34E+04 & 15946 &  12155~  & 4.43E-02 & 1.78E-05 & 7.66E-05 &          & 1.21E-06 & 3.13E-07 & 1.11E-08 &  1.10E+00 \\
            &            & 6.39E+03 & 13347 &  10435~  & 3.81E-02 & 1.19E-05 & 4.11E-05 &          & 6.98E-07 & 2.05E-07 & 5.03E-09 &  9.10E-01 \\
            &            &          &       &          &          &          &          &          &          &          &          &           \\
 352.0-04.6 & H 1-30     & 3.11E+03 &  9848 &   9471~  & 1.60E-01 & 8.49E-04 & 5.54E-04 & 1.73E-04 & 1.51E-05 & 5.52E-06 & 1.02E-06 &  1.53E+00 \\
            &            & 3.98E+03 & 10074 &   9766~  & 1.70E-01 & 1.00E-03 & 6.34E-04 & 1.87E-04 & 1.80E-05 & 6.32E-06 & 1.32E-06 &  1.62E+00 \\
            &            & 2.57E+03 &  9567 &   9125~  & 1.51E-01 & 7.02E-04 & 4.98E-04 & 1.54E-04 & 1.32E-05 & 5.15E-06 & 8.22E-07 &  1.43E+00 \\
            &            &          &       &          &          &          &          &          &          &          &          &           \\
 353.5-04.9 & H 1-36     & 9.87E+03 & 36563 &      -~  & 8.78E-02 & 5.92E-05 & 7.04E-05 & 1.55E-05 & 1.34E-06 & 4.59E-07 & 8.42E-08 &  1.09E+00 \\
            &            & 1.22E+04 & 38094 &       ~  & 9.13E-02 & 6.75E-05 & 8.46E-05 & 1.80E-05 & 1.62E-06 & 5.39E-07 & 1.10E-07 &  1.19E+00 \\
            &            & 6.66E+03 & 31360 &       ~  & 8.10E-02 & 4.73E-05 & 6.01E-05 & 1.30E-05 & 1.14E-06 & 4.21E-07 & 6.50E-08 &  9.82E-01 \\
            &            &          &       &          &          &          &          &          &          &          &          &           \\
 354.2+04.3 & M 2-10     & 1.27E+03 &     - &   7291~  & 1.41E-01 & 2.99E-04 & 2.73E-04 &        - & 7.93E-06 & 3.58E-06 & 7.69E-07 &  1.32E+00 \\
            &            & 1.51E+03 &       &   7439~  & 1.50E-01 & 3.33E-04 & 3.18E-04 &          & 9.01E-06 & 4.13E-06 & 8.78E-07 &  1.38E+00 \\
            &            & 1.07E+03 &       &   7126~  & 1.32E-01 & 2.68E-04 & 2.36E-04 &          & 6.96E-06 & 3.15E-06 & 6.66E-07 &  1.26E+00 \\
            &            &          &       &          &          &          &          &          &          &          &          &           \\
 355.4-04.0 & Hf 2-1     & 6.96E+02 & 11710 &  10514~  & 1.35E-01 & 5.19E-04 & 5.36E-04 &        - & 1.56E-05 & 3.40E-06 & 3.16E-06 &  9.03E-01 \\
            &            & 8.25E+02 & 12064 &  11501~  & 1.43E-01 & 7.01E-04 & 5.99E-04 &          & 1.80E-05 & 3.65E-06 & 3.98E-06 &  9.54E-01 \\
            &            & 5.74E+02 & 11414 &   9579~  & 1.28E-01 & 3.51E-04 & 4.74E-04 &          & 1.28E-05 & 3.07E-06 & 2.12E-06 &  8.40E-01 \\
            &            &          &       &          &          &          &          &          &          &          &          &           \\
 355.7-03.4 & H 2-23     & 3.41E+03 & 10061 &      -~  & 1.19E-01 & 6.32E-05 & 4.41E-04 &        - & 7.46E-06 & 1.39E-06 &        - &  1.23E+00 \\
            &            & 4.75E+03 & 10350 &       ~  & 1.27E-01 & 8.70E-05 & 4.97E-04 &          & 8.71E-06 & 1.55E-06 &          &  1.31E+00 \\
            &            & 2.88E+03 &  9855 &       ~  & 1.11E-01 & 5.11E-05 & 3.82E-04 &          & 6.38E-06 & 1.24E-06 &          &  1.18E+00 \\
            &            &          &       &          &          &          &          &          &          &          &          &           \\
 355.9+03.6 & H 1- 9     & 1.00E+05 & 11230 & (10189)  & 7.32E-02 & 2.59E-05 & 1.98E-04 & 6.88E-06 & 3.23E-06 & 7.03E-07 &        - &  1.71E+00 \\
            &            & 1.00E+05 & 12653 & (18659)  & 7.71E-02 & 2.95E-05 & 2.41E-04 & 8.10E-06 & 3.84E-06 & 7.94E-07 &          &  1.81E+00 \\
            &            & 2.70E+04 & 10649 & ( 9974)  & 6.74E-02 & 1.47E-05 & 8.26E-05 & 2.45E-06 & 2.00E-06 & 5.21E-07 &          &  1.63E+00 \\
            &            &          &       &          &          &          &          &          &          &          &          &           \\
 355.9-04.2 & M 1-30     & 3.61E+03 &  6494 &   7064~  & 1.45E-01 & 5.40E-04 & 4.79E-04 &        - & 1.65E-05 & 7.59E-06 & 2.81E-06 &  1.04E+00 \\
            &            & 5.08E+03 &  6915 &   7217~  & 1.54E-01 & 7.75E-04 & 7.61E-04 &          & 3.34E-05 & 1.14E-05 & 8.63E-06 &  1.11E+00 \\
            &            & 2.73E+03 &  5822 &   6857~  & 1.33E-01 & 4.23E-04 & 3.90E-04 &          & 1.08E-05 & 5.89E-06 & 1.49E-06 &  9.54E-01 \\
            &            &          &       &          &          &          &          &          &          &          &          &           \\
 356.7-04.8 & H 1-41     & 7.43E+02 & 10052 &  10851~  & 1.36E-01 & 1.28E-04 & 3.13E-04 &        - & 4.90E-06 & 1.47E-06 & 1.96E-06 &  6.03E-01 \\
            &            & 8.71E+02 & 10238 &  11269~  & 1.42E-01 & 1.50E-04 & 3.59E-04 &          & 5.54E-06 & 1.61E-06 & 2.36E-06 &  6.63E-01 \\
            &            & 6.01E+02 &  9848 &  10584~  & 1.29E-01 & 1.13E-04 & 2.81E-04 &          & 4.38E-06 & 1.37E-06 & 1.61E-06 &  5.41E-01 \\
            &            &          &       &          &          &          &          &          &          &          &          &           \\
 356.9+04.4 & M 3-38     & 3.06E+03 & 14015 &  17800~  & 1.30E-01 & 3.81E-04 & 2.47E-04 & 6.36E-05 & 7.50E-06 & 2.82E-06 & 8.25E-07 &  2.06E+00 \\
            &            & 3.53E+03 & 14579 &  18900~  & 1.38E-01 & 5.02E-04 & 2.88E-04 & 7.19E-05 & 9.32E-06 & 3.14E-06 & 1.16E-06 &  2.16E+00 \\
            &            & 2.45E+03 & 13310 &  16697~  & 1.22E-01 & 3.41E-04 & 2.13E-04 & 5.77E-05 & 6.36E-06 & 2.56E-06 & 6.20E-07 &  1.97E+00 \\
\hline
\end{tabular}
\end{table*}

\setcounter{table}{2}
\begin{table*}
\caption{ 
  Continued
}
\begin{tabular}{       @{\hspace{0.05cm}}
                     l @{\hspace{0.25cm}}
                     l @{\hspace{0.21cm}}
                     r @{\hspace{0.21cm}}
                     r @{\hspace{0.21cm}}
                     r @{\hspace{0.21cm}}
                     r @{\hspace{0.21cm}}
                     r @{\hspace{0.21cm}}
                     r @{\hspace{0.21cm}}
                     r @{\hspace{0.21cm}}
                     r @{\hspace{0.21cm}}
                     r @{\hspace{0.21cm}}
                     r @{\hspace{0.21cm}}
                     r } 
\hline
        PN G & Main Name & \Ne(S~{\sc ii})  &
                                      \Te(O~{\sc iii}) &
                                              \Te(N~{\sc ii})  &
                                                         He/H     &  N/H     &  O/H     &  Ne/H    &   S/H    &  Ar/H    &  Cl/H    &    ext. C \\
\hline
 356.9+04.5 & M 2-11     & 2.15E+03 & 14427 &  14610~  & 1.43E-01 & 1.63E-04 & 2.75E-04 & 5.77E-05 & 4.81E-06 & 1.41E-06 & 6.56E-08 &  1.26E+00 \\
            &            & 2.55E+03 & 14976 &  15254~  & 1.51E-01 & 2.00E-04 & 3.20E-04 & 6.57E-05 & 5.72E-06 & 1.55E-06 & 8.54E-08 &  1.37E+00 \\
            &            & 1.77E+03 & 13905 &  13893~  & 1.33E-01 & 1.37E-04 & 2.43E-04 & 5.36E-05 & 4.07E-06 & 1.27E-06 & 5.06E-08 &  1.17E+00 \\
            &            &          &       &          &          &          &          &          &          &          &          &           \\
 357.1+03.6 & M 3- 7     & 4.11E+03 &  7645 &   8900~  & 1.30E-01 & 2.39E-04 & 4.50E-04 &        - & 1.29E-05 & 4.26E-06 & 6.72E-06 &  1.54E+00 \\
            &            & 5.34E+03 &  7792 &   9205~  & 1.39E-01 & 2.67E-04 & 5.11E-04 &          & 1.60E-05 & 4.70E-06 & 8.58E-06 &  1.61E+00 \\
            &            & 3.16E+03 &  7498 &   8646~  & 1.22E-01 & 2.04E-04 & 3.99E-04 &          & 1.07E-05 & 3.89E-06 & 4.92E-06 &  1.48E+00 \\
            &            &          &       &          &          &          &          &          &          &          &          &           \\
 357.1-04.7 & H 1-43     & 8.86E+03 &     - &   6137~  & 1.50E-02 & 2.05E-04 & 1.32E-04 &        - & 2.49E-05 & 3.03E-07 &        - &  1.07E+00 \\
            &            & 1.52E+04 &       &   6474~  & 1.61E-02 & 2.80E-04 & 2.03E-04 &          & 4.66E-05 & 3.73E-07 &          &  1.14E+00 \\
            &            & 5.42E+03 &       &   5580~  & 1.42E-02 & 1.47E-04 & 8.43E-05 &          & 1.53E-05 & 2.36E-07 &          &  9.87E-01 \\
            &            &          &       &          &          &          &          &          &          &          &          &           \\
 357.2-04.5 & H 1-42     & 4.13E+03 &  9828 & (12984)  & 1.11E-01 & 7.39E-05 & 4.64E-04 & 1.12E-04 & 5.84E-06 & 1.48E-06 & 2.03E-06 &  8.53E-01 \\
            &            & 5.93E+03 & 10172 & (14192)  & 1.16E-01 & 8.21E-05 & 5.09E-04 & 1.17E-04 & 6.51E-06 & 1.58E-06 & 2.16E-06 &  9.13E-01 \\
            &            & 3.28E+03 &  9684 & (11872)  & 1.01E-01 & 6.28E-05 & 3.84E-04 & 9.47E-05 & 4.90E-06 & 1.33E-06 & 1.52E-06 &  7.51E-01 \\
            &            &          &       &          &          &          &          &          &          &          &          &           \\
 357.3+03.3 & M 3-41     & 2.14E+03 &     - &   7267~  & 4.73E-02 & 1.01E-04 & 1.53E-04 &        - & 5.14E-06 & 9.41E-07 & 2.57E-07 &  1.66E+00 \\
            &            & 2.67E+03 &       &   7427~  & 4.98E-02 & 1.15E-04 & 1.92E-04 &          & 5.96E-06 & 1.06E-06 & 3.77E-07 &  1.73E+00 \\
            &            & 1.72E+03 &       &   7103~  & 4.39E-02 & 9.17E-05 & 1.34E-04 &          & 4.64E-06 & 8.52E-07 & 1.49E-07 &  1.58E+00 \\
            &            &          &       &          &          &          &          &          &          &          &          &           \\
 357.4-04.6 & M 2-22     & 1.37E+03 &  9661 &   9097~  & 1.66E-01 & 4.68E-04 & 4.28E-04 &        - & 9.44E-06 & 3.59E-06 & 1.20E-06 &  1.22E+00 \\
            &            & 1.69E+03 &  9857 &   9282~  & 1.76E-01 & 5.50E-04 & 4.85E-04 &          & 1.08E-05 & 3.97E-06 & 1.45E-06 &  1.31E+00 \\
            &            & 1.17E+03 &  9430 &   8826~  & 1.56E-01 & 3.98E-04 & 3.85E-04 &          & 8.31E-06 & 3.19E-06 & 1.02E-06 &  1.15E+00 \\
            &            &          &       &          &          &          &          &          &          &          &          &           \\
 357.5+03.2 & M 3-42     & 1.03E+03 & 10862 &   9312~  & 1.42E-01 & 5.47E-04 & 6.73E-04 & 1.66E-04 & 2.08E-05 & 7.03E-06 & 1.50E-06 &  1.86E+00 \\
            &            & 1.71E+04 & 11063 &   9552~  & 1.50E-01 & 9.44E-04 & 1.11E-03 & 2.45E-04 & 8.27E-05 & 8.05E-06 & 2.07E-06 &  2.01E+00 \\
            &            & 2.68E+02 & 10087 &   8701~  & 1.31E-01 & 4.47E-04 & 6.37E-04 & 1.30E-04 & 1.73E-05 & 6.48E-06 & 1.18E-06 &  1.69E+00 \\
            &            &          &       &          &          &          &          &          &          &          &          &           \\
 357.6-03.3 & H 2-29     & 2.17E+02 &     - &   7550~  & 1.71E-01 & 1.13E-04 & 2.39E-04 & 5.18E-05 & 9.48E-06 & 1.77E-06 &        - &  1.75E+00 \\
            &            & 1.29E+03 &       &   7874~  & 1.85E-01 & 1.61E-04 & 5.44E-04 & 1.33E-04 & 1.68E-05 & 2.71E-06 &          &  1.86E+00 \\
            &            & 3.00E+01 &       &   6700~  & 1.57E-01 & 9.42E-05 & 1.67E-04 & 3.34E-05 & 7.65E-06 & 1.43E-06 &          &  1.54E+00 \\
            &            &          &       &          &          &          &          &          &          &          &          &           \\
 358.5-04.2 & H 1-46     & 3.75E+03 &  9992 &  15848~  & 1.10E-01 & 6.20E-05 & 1.87E-04 & 3.18E-05 & 5.81E-06 & 1.28E-06 & 3.45E-07 &  1.38E+00 \\
            &            & 4.72E+03 & 10243 &  16352~  & 1.17E-01 & 7.26E-05 & 2.10E-04 & 3.36E-05 & 6.63E-06 & 1.38E-06 & 4.30E-07 &  1.48E+00 \\
            &            & 3.09E+03 &  9781 &  14847~  & 1.03E-01 & 5.42E-05 & 1.65E-04 & 2.53E-05 & 4.92E-06 & 1.13E-06 & 2.67E-07 &  1.28E+00 \\
            &            &          &       &          &          &          &          &          &          &          &          &           \\
 359.0-04.1 & M 3-48     & 5.21E+02 &  8884 &   9266~  & 1.66E-01 & 5.58E-04 & 5.55E-04 & 2.20E-04 & 2.80E-05 & 6.07E-06 & 1.30E-06 &  1.17E+00 \\
            &            & 2.59E+03 &  9161 &   9624~  & 1.79E-01 & 7.84E-04 & 7.69E-04 & 3.44E-04 & 4.14E-05 & 8.11E-06 & 2.01E-06 &  1.35E+00 \\
            &            & 6.85E+01 &  8117 &   8779~  & 1.53E-01 & 4.75E-04 & 5.09E-04 & 2.10E-04 & 2.40E-05 & 5.10E-06 & 1.12E-06 &  1.02E+00 \\
            &            &          &       &          &          &          &          &          &          &          &          &           \\
 359.0-04.8 & M 2-25     & 2.76E+02 &  9528 &  10009~  & 1.47E-01 & 4.35E-04 & 4.47E-04 &        - & 9.87E-06 & 3.68E-06 & 4.32E-07 &  9.94E-01 \\
            &            & 3.49E+02 &  9643 &  10302~  & 1.57E-01 & 5.18E-04 & 5.10E-04 &          & 1.14E-05 & 4.56E-06 & 5.56E-07 &  1.07E+00 \\
            &            & 2.08E+02 &  9286 &   9695~  & 1.41E-01 & 4.09E-04 & 4.21E-04 &          & 9.12E-06 & 2.93E-06 & 3.49E-07 &  9.29E-01 \\
            &            &          &       &          &          &          &          &          &          &          &          &           \\
 359.1-01.7 & M 1-29     & 2.72E+03 & 10911 &   9824~  & 1.53E-01 & 5.24E-04 & 4.72E-04 & 1.38E-04 & 1.05E-05 & 4.12E-06 & 7.96E-07 &  2.12E+00 \\
            &            & 3.19E+03 & 11153 &  10033~  & 1.62E-01 & 6.83E-04 & 5.58E-04 & 1.51E-04 & 1.28E-05 & 4.59E-06 & 1.10E-06 &  2.23E+00 \\
            &            & 2.19E+03 & 10540 &   9503~  & 1.42E-01 & 4.45E-04 & 4.22E-04 & 1.21E-04 & 8.94E-06 & 3.84E-06 & 6.46E-07 &  2.04E+00 \\
            &            &          &       &          &          &          &          &          &          &          &          &           \\
 359.1-02.9 & M 3-46     & 5.04E+02 &  7425 &   8160~  & 2.00E-01 & 6.45E-04 & 8.52E-04 & 3.73E-04 & 2.73E-05 & 6.94E-06 & 9.34E-07 &  1.72E+00 \\
            &            & 1.48E+03 &  7762 &   8387~  & 2.17E-01 & 8.98E-04 & 1.19E-03 & 5.49E-04 & 3.89E-05 & 9.12E-06 & 1.67E-06 &  1.87E+00 \\
            &            & 6.34E+01 &  6965 &   7839~  & 1.79E-01 & 5.43E-04 & 7.47E-04 & 2.96E-04 & 2.28E-05 & 5.80E-06 & 5.50E-07 &  1.53E+00 \\
            &            &          &       &          &          &          &          &          &          &          &          &           \\
 359.4-03.4 & H 2-33     & 9.64E+02 &  8636 & (21816)  & 1.59E-01 & 1.75E-05 & 5.47E-04 & 1.53E-04 & 3.69E-06 & 2.86E-06 &        - &  1.75E+00 \\
            &            & 1.00E+05 &  8845 & (26143)  & 1.63E-01 & 6.72E-05 & 7.34E-04 & 1.91E-04 & 6.79E-06 & 3.41E-06 &          &  1.85E+00 \\
            &            & 3.13E+02 &  8104 & (11324)  & 1.49E-01 & 1.56E-05 & 4.50E-04 & 1.29E-04 & 3.30E-06 & 2.59E-06 &          &  1.66E+00 \\
            &            &          &       &          &          &          &          &          &          &          &          &           \\
 359.6-04.8 & H 2-36     & 2.93E+02 & 12336 &      -~  & 2.21E-01 & 1.71E-05 & 1.70E-04 & 4.26E-05 & 2.07E-06 & 1.22E-06 & 1.11E-06 &  1.08E+00 \\
            &            & 3.74E+02 & 12660 &       ~  & 2.31E-01 & 2.19E-05 & 1.98E-04 & 4.51E-05 & 2.70E-06 & 1.45E-06 & 1.41E-06 &  1.19E+00 \\
            &            & 2.31E+02 & 11917 &       ~  & 2.11E-01 & 1.45E-05 & 1.47E-04 & 3.49E-05 & 1.69E-06 & 1.05E-06 & 8.90E-07 &  9.70E-01 \\
            &            &          &       &          &          &          &          &          &          &          &          &           \\
 359.7-04.4 & KFL 3      & 1.47E+02 &  9676 &   8503~  & 1.67E-01 & 1.33E-04 & 3.66E-04 &        - & 4.87E-06 & 1.81E-06 &        - &  1.06E+00 \\
            &            & 2.03E+02 & 10174 &   8666~  & 1.76E-01 & 1.77E-04 & 4.59E-04 &          & 6.85E-06 & 2.45E-06 &          &  1.12E+00 \\
            &            & 9.12E+01 &  9021 &   8301~  & 1.56E-01 & 1.12E-04 & 3.14E-04 &          & 4.05E-06 & 1.07E-06 &          &  9.91E-01 \\
            &            &          &       &          &          &          &          &          &          &          &          &           \\
 359.9-04.5 & M 2-27     & 4.07E+03 &  8325 &  10504~  & 1.50E-01 & 8.39E-04 & 6.70E-04 & 2.37E-04 & 2.09E-05 & 5.78E-06 & 5.86E-06 &  1.66E+00 \\
            &            & 5.72E+03 &  8817 &  10740~  & 1.59E-01 & 1.02E-03 & 7.88E-04 & 2.80E-04 & 2.69E-05 & 6.69E-06 & 8.51E-06 &  1.74E+00 \\
            &            & 3.02E+03 &  8021 &   9985~  & 1.40E-01 & 6.67E-04 & 5.31E-04 & 1.66E-04 & 1.48E-05 & 4.84E-06 & 3.49E-06 &  1.55E+00 \\
\hline
\end{tabular}
\end{table*}


\begin{thebibliography}{}

       \bibitem{}
Acker, A., Neiner, C., 2003, A\&A, 403, 659
       \bibitem{}
Barlow, M.J., 1987, MNRAS, 227, 161
       \bibitem{}
Beaulieu S.F., Freeman K.C., Kalnajs A.J., et al., 2000, AJ 120, 855
       \bibitem{}
Bl\"ocker, T., 1995, A\&A, 229, 755
       \bibitem{}
Chiappini, C., G\'orny. S.K., Stasi\'nska. G., Barbuy, B., 2009, 
A\&A, 494, 591, astro-ph/0812.0558
       \bibitem{}
Crowther, P.A., De Marco, O., Barlow, M.J., 1998, MNRAS 296, 367
       \bibitem{}
Cuisinier, F., Maciel, W.J., K\"oppen, J., et al., 2000, A\&A 353, 543
       \bibitem{}
Durand, S., Acker, A., Zijlstra, A., 1998, A\&AS 132, 13
       \bibitem{}
Escudero, A.V., Costa, R.D.D., 2001, A\&A, 380, 300
       \bibitem{}
Escudero, A. V., Costa, R. D. D., Maciel, W. J., 2004, A\&A, 414, 211
       \bibitem{}
Exter, K.M., Barlow, M.J., Walton, N.A., 2004, MNRAS, 349, 1291
       \bibitem{}
Gesicki, K., Zijlstra, A.A., Acker, A., et al., 2006, A\&A, 451, 925
       \bibitem{}
Girard, P., K\"oppen, J., Acker, A., 2007, A\&A, 463, 265
       \bibitem{}
G\'orny, S.K., 1996, Ap\&SS, 238, 79
       \bibitem{}
G\'orny, S.K., 2001, Ap\&SS, 275, 67
       \bibitem{}
G\'orny, S.K., Stasi\'nska, G., 1995, A\&A, 303, 893
       \bibitem{}
G\'orny, S.K., Stasi\'nska, G., Escudero, A.V., Costa, R.D.D., 2004, 
A\&A, 427, 231
       \bibitem{}
G\'orny S.K., Stasi\'nska G., Tylenda R., 1997, A\&A, 318, 256
       \bibitem{}
G\'orny, S.K., Tylenda, R., 2000, A\&A 362, 1008
       \bibitem{}
G\'orny, S.K., Tylenda, R., Szczerba, R, 1994, A\&A 284, 949
       \bibitem{}
Hu, J.Y., Bibo, E.A., 1990, A\&A, 234, 435
       \bibitem{}
Kingsburgh, R.L., Barlow, M.J., 1994, MNRAS, 275, 605
       \bibitem{}
Liu, X.-W., 2006, in IAU Symp. 234: Planetary Nebulae in our Galaxy and
  Beyond, eds. Michel J. Barlow and Roberto H. M\'endez., 
  Cambridge University Pres, p.219
       \bibitem{} 
Liu, X.-W., Storey, P.J., Barlow, M.J., et al., 2000, MNRAS 312, 585
       \bibitem{}
MacConnell, D.J., 1978, A\&ASS, 33, 219
       \bibitem{}
Paczy\'nski, B., 1971, Acta Astron., 21, 417
       \bibitem{}
Pe\~na, M., 2005, RevMexA\&A, 41, 423
       \bibitem{}
Pe\~na, M., Stasi\'nska, G., Medina, S., 2001, A\&A 367, 983
       \bibitem{}
Porter R.~L., Ferland G.~J., MacAdam K.~B., 2007, ApJ, 657, 327
       \bibitem{}
Sch\"onberner, D., Tylenda, R., 1990, A\&A, 234, 439
       \bibitem{}
Seaton, M.J., 1979, MNRAS 187, 73
       \bibitem{}
Shklovsky, I. 1956, AZh, 33, 222
       \bibitem{}
Stasi\'nska, G., 2005, A\&A, 434, 507
       \bibitem{}
Stasi\'nska, G., 2007, in StellarNucleosynthesis: 50 years after B2FH,
       C. Charbonnel \& J.-P. Zahn (eds.), EAS PublicationsSeries, in press
       \bibitem{}
Stasi\'nska, G., Tylenda, R., 1994, A\&A, 289, 225
       \bibitem{}
Storey, P.J., Zeippen, C.J., 2000, MNRAS, 312, 813
       \bibitem{}
Su\'aez, O., Garc\'ia-Lario, P., Manchado, A., et~al., 2006, A\&A 458, 173
       \bibitem{}
Su\'aez, O., Gomez, J.F., Morata, O., 2007, A\&A, 467, 1085
       \bibitem{}
       Tayal S.~S., 2007, ApJS, 171, 331 Tylenda, R., Acker, A., 
Stenholm, B., 1993, A\&AS, 102, 595
       \bibitem{}
Wang, W., Liu X.-W., 2007, MNRAS, 381, 669
       \bibitem{}
Zijlstra, A.A., Gesicki, K., Walsh, J.R., P\'equignot, D., van Hoof, P.A.M., 
  Minniti, D., 2006, MNRAS, 369, 875

\end{thebibliography}
\end{document}